\RequirePackage{fix-cm}
\documentclass[smallextended]{svjour3}       % onecolumn (second format)
\smartqed  % flush right qed marks, e.g. at end of proof
\usepackage{graphicx,epstopdf,url}

\usepackage[lined,ruled,boxed,commentsnumbered,linesnumbered]{algorithm2e}

\usepackage{subfigure}

\usepackage{multirow}
\usepackage{amsmath}

\journalname{Journal of Network and Systems Management}
\begin{document}

\title{Reliability and Survivability Analysis of\\ Data Center Network Topologies}


\author{Rodrigo de Souza Couto         \and \\
        Stefano Secci \and \\
        Miguel Elias Mitre Campista \and \\
        Lu\'is Henrique Maciel Kosmalski Costa
}


\institute{Rodrigo S. Couto, Miguel Elias M. Campista and Lu\'is Henrique M. K. Costa \at
              Universidade Federal do Rio de Janeiro - COPPE/PEE/GTA - POLI/DEL\\
              P.O. Box 68504 – CEP 21941-972, Rio de Janeiro, RJ, Brazil\\
              Tel.: +55 21 3938 8635\\
              Fax: +55  21 3938 8628\\
              \email{souza@gta.ufrj.br,miguel@gta.ufrj.br,luish@gta.ufrj.br}           \\ \\
             \emph{Present address} of Rodrigo S. Couto \at
              Universidade do Estado do Rio de Janeiro - FEN/DETEL/PEL\\
              CEP 20550-013, Rio de Janeiro, RJ, Brazil\\
              Tel.: +55 21 2334 0027\\
              \email{rodrigo.couto@uerj.br}
           \and
           Stefano Secci \at
              Sorbonne Universit\'es, UPMC Univ Paris 06, UMR 7606, LIP6 \\ 
              F-75005, Paris, France \\
              Tel.: +33 (0)1 4427 3678\\
              Fax:  +33 (0)1 4427 8783\\
              \email{stefano.secci@umpc.fr} 
}

\date{Received: date / Accepted: date}

\date{Neither the entire paper nor any part of its content has been published or 
has been accepted for publication elsewhere. 
It has not been submitted to any other journal.}

\date{The final publication is available at Springer via\\ http://dx.doi.org/10.1007/s10922-015-9354-8}

\maketitle

\begin{abstract}
The architecture of several data centers have been proposed as alternatives to the conventional three-layer one.
Most of them employ commodity equipment for cost reduction. Thus, robustness to failures becomes even more important, because commodity equipment is more failure-prone. Each architecture has a different network topology design with a specific level of redundancy. In this work, we aim at analyzing the benefits of different data center topologies taking the reliability and survivability requirements into account. We consider the topologies of three alternative data center architecture: Fat-tree, BCube, and DCell. Also, we compare these topologies with a conventional three-layer data center topology. 
Our analysis is independent of specific equipment, traffic patterns, or network protocols, for the sake of generality. We derive closed-form formulas for the Mean Time To Failure of each topology. The results allow us to indicate the best topology for each failure scenario. In particular, we conclude that BCube is more robust to link failures than the other topologies, whereas DCell has the most robust topology when considering switch failures. 
Additionally, we show that all considered alternative 
topologies outperform a three-layer topology for both types of failures. We also determine to which extent the robustness of BCube and DCell is influenced by the number of network interfaces per server. 
\keywords{Data center networks \and cloud networks \and survivability \and reliability \and robustness}
\end{abstract}

\section{Introduction}
\label{sec:introduction}

Data center networking has been receiving a lot of attention in the last few years as it plays an essential role in cloud computing and big data applications. In particular, as data center (DC) sizes steadily increase, operational expenditures (OPEX) and capital expenditures (CAPEX) become more and more important in the choice of the DC network (DCN) architecture~\cite{jennings2014Cloud}. Conventional DCN architecture employing high-end equipment suffers from prohibitive costs for large network sizes~\cite{al2008scalable}. As a consequence, a variety of alternative DCN architecture has been proposed to better meet cost efficiency, scalability, and communication requirements. Among the most cited alternative DCN architecture, we can mention Fat-tree~\cite{al2008scalable}, BCube~\cite{guo2009bcube}, and DCell~\cite{guo2008dcell}. These architecture have different topologies but share the goal of providing a modular infrastructure using low-cost equipment. The conventional DC topologies and Fat-tree are switch-centric, 
where only 
switches forward packets, whereas BCube and DCell are server-centric topologies, where servers also participate in packet forwarding.

Although the utilization of low-cost network elements reduces the CAPEX of a DC, it also likely makes the network more failure prone~\cite{al2008scalable,guo2009bcube,greenber2009cost}. Hence, in the medium to long term, low-cost alternatives would incur in OPEX increase, caused by the need to restore the network. The tradeoff between CAPEX and OPEX can be more significant if we consider the increasing deployment of DCs in environments with difficult access for maintenance, e.g., within a sealed shipping container (e.g., Modular Data Center)~\cite{guo2009bcube}. In this case, repairing or replacing failed elements can be very cumbersome. Therefore, the DCN needs to be robust,
i.e., it should survive as long as possible without going through maintenance procedures. Network robustness is thus an important concern in the design of low-cost DCN architecture.

DCN robustness depends on the physical topology and on the ability of protocols to react to failures. In this work, we focus on the first aspect, by analyzing the performance of recently proposed DCN topologies under failure conditions. Although fault-tolerant network protocols are mandatory to guarantee network robustness, in the long term the topological organization of DCNs plays a major role. In current literature, alternative DCN topologies have been analyzed in terms of cost~\cite{popa2010cost}, scalability~\cite{danLi2011Scalable}, and network capacity~\cite{guo2009bcube}. Guo~\textit{et. al}~\cite{guo2009bcube} also addresses DCN robustness, by comparing Fat-tree, BCube, and DCell alternatives when there are switch or server failures. Nevertheless, as this comparison is not the primary focus of~\cite{guo2009bcube}, the topologies are analyzed with respect to only one robustness criterion. Also, the conclusions of Guo~\textit{et. al} are bound to specific traffic patterns and routing protocols.

In this work, we provide a generic, protocol-, hardware-, and traffic-agnostic analysis of DCN robustness, focusing on topological characteristics. Our motivation is that as commodity equipment is increasingly employed in DCNs, DC designers have a wide and heterogeneous vendor choice. Hence, we do not limit our analysis to specific vendors. Also, as a DCN topology might be employed by different applications, its robustness analysis should be independent of the traffic matrix. We analyze robustness aspects of Fat-tree, BCube, and DCell. As detailed later, we focus on these representative topologies because they have been receiving a lot of attention in recent literature and because they are conceived to be based on low-cost equipment. Also, we compare the alternative topologies with a conventional three-layer DCN topology. In summary, the contributions of this article are as follows:
\begin{itemize}
\item We point out the characteristics that make the analyzed topologies more vulnerable or robust to certain types of failures. We show that BCube and DCell outperform Fat-tree both on link and switch failures. In a Fat-tree, when a given fraction of the total links or switches fail, the number of reachable servers is reduced by the same fraction. BCube topology is the most robust against link failures, maintaining at least 84\% of its servers connected when 40\% of its links are down, while in DCell this lower bound is 74\% of servers. On the other hand, DCell is the best one for switch failures, maintaining 100\% of its servers for a period up to 12 times longer than BCube. We also observe that the robustness to failures grows proportionally to the number of server network interfaces in BCube and DCell. Finally, we show that all alternative DCN topologies outperform a three-layer topology in terms of both link and switch failures. 
\item We characterize and analyze the DCN, both analytically and by simulation, against each failure type (i.e., link, switch, or server) separately. Our proposed methodology relies on the MTTF (Mean Time To Failure) and on other metrics regarding the path length and DCN reachability. In particular, we provide closed-form formulas to model the MTTF of the considered topologies, and to predict server disconnections, thus helping to estimate DCN maintenance periods. 
\end{itemize}

This article is organized as follows. Section~\ref{sec:topologies} details the topologies used in this work. Section~\ref{sec:methodology} describes our proposed methodology. The evaluation, as well as the description of metrics, are presented in Sections~\ref{sec:reliableTime}~and~\ref{sec:survival}. Section~\ref{sec:qualitative} summarizes the obtained results with a qualitative evaluation of DCN topologies. Section~\ref{sec:edpAnalysis} addresses the sensibility of the used metrics according to the choice of DCN gateways. Section~\ref{sec:het} complements the evaluation, considering that the DCN is composed of heterogeneous equipment. Finally, Section~\ref{sec:related} discusses related work and Section~\ref{sec:conclusion} concludes this article and presents future directions. 

\section{Data Center Network Topologies}
\label{sec:topologies}

DCN topologies can be structured or unstructured.
Structured topologies have a deterministic formation rule and are built by connecting basic modules. They can be copper-only topologies, employing exclusively copper connections (e.g., Gigabit Ethernet), as conventional three-layer DC topologies, Fat-tree, BCube, and DCell; or can be hybrid, meaning that they also use optical links to improve energy efficiency and network capacity, as C-Through and Helios~\cite{christoforos2012Optical}.
On the other hand, unstructured topologies do not have a deterministic formation rule. These topologies can be built by using a stochastic algorithm (e.g., Jellyfish~\cite{singla2012jellyfish}) or the output of an optimization problem (e.g., REWIRE~\cite{curtisrewire}). The advantage of unstructured topologies is that they are easier to scale up, as they do not have a rigid structure.
In this work, we focus on structured copper-only topologies, since they are receiving major attention in literature~\cite{raiciu2011improving,meng2010improving}. Next, we detail the topologies analyzed in this work.

\subsection{Three-layer}
\label{conventional}

\begin{figure}
\centering
\includegraphics[width=0.7\textwidth]{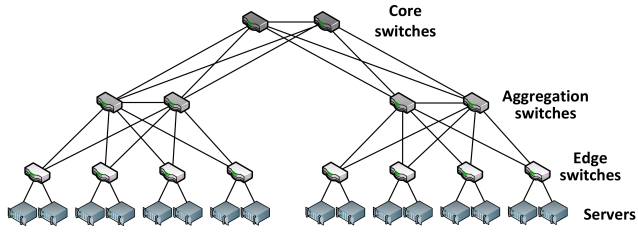}
\caption{Three-layer topology with 2 edge ports ($n_e=2$) and 4 aggregation ports ($n_a=4$).}
\label{fig:conventional}
\end{figure}

Most of today's commercial DCNs employ a conventional hierarchical topology, composed of three layers: the edge, the aggregation, and the core~\cite{scotland}.
There is no unique definition in the literature for a conventional three-layer DC topology, since it highly depends on DC design decisions and commercial equipment specifications.
Hence, we define our \textit{conventional} Three-layer topology based on a DCN architecture recommended by Cisco in~\cite{scotland}. 
In the Three-layer topology, the core layer is composed of two switches directly connected between each other, which act as DC gateways. Each core switch is connected to all aggregation switches. The aggregation switches are organized in pairs, where in each pair the aggregation switches are directly connected to each other, as in Figure~\ref{fig:conventional}. Each aggregation switch in a pair is connected to the same group of $n_a$ edge switches. Each edge switch has $n_e$ ports to connect directly to the servers. Hence, each pair of aggregation switches provides connectivity to $n_a*n_e$ servers and we need $\frac{|\mathcal{S}|}{n_a*n_e}$ pairs to build a DC with $|\mathcal{S}|$ servers. A \textit{module} is a group of servers in Three-layer where the connectivity is maintained by the same pair of aggregation switches.
Figure~\ref{fig:conventional} shows an example of a Three-layer topology with 16 servers, $n_a=4$ and $n_e=2$.

In commercial DCNs, edge switches are generally connected to the servers using 1 Gbps Ethernet ports. The ports that connect the aggregation switches to the core and edge switches are generally 10 Gbps Ethernet.
Hence, as can be noted, three-layer topologies employ high capacity equipment in the core and aggregation layers. The alternative DC architecture propose topological enhancements to enable the utilization of commodity switches throughout the network, as we describe next.

\subsection{Fat-tree}
\label{Fat-tree}

\begin{figure}
\centering
\includegraphics[width=0.7\textwidth]{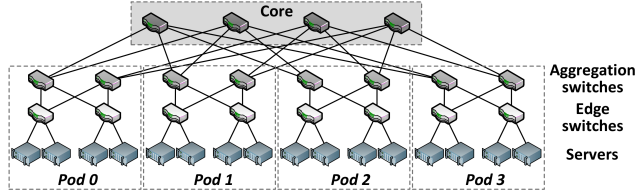}
\caption{Fat-tree with 4-port switches ($n=4$).}
\label{fig:fatTree}
\end{figure}

We refer to Fat-tree as the DCN topology proposed in~\cite{al2008scalable}, designed using the concept of ``fat-tree'', a special case of a Clos network. VL2~\cite{greenberg2009vl2} also uses a Clos network but is not considered in our analysis because it is very similar to the Fat-tree. 
As shown in Figure~\ref{fig:fatTree}, the Fat-tree topology has two sets of elements: core and pods.
The first set is composed of switches that interconnect the pods. Pods are composed of aggregation switches, edge switches, and servers. Each port of each switch in the core is connected to a different pod through an aggregation switch. Within a pod, the aggregation switches are connected to all edge switches.
Finally, each edge switch is connected to a different set of servers. Unlike conventional DC topologies, Fat-tree is built using links and switches of the same capacity.

All switches have $n$ ports. Hence, the network has $n$ pods, and each pod has $\frac{n}{2}$ aggregation
switches connected to $\frac{n}{2}$ edge switches. The edge switches are individually connected to $\frac{n}{2}$ different servers. Thus, using $n$-port switches, a Fat-tree can have $\frac{n}{2}*\frac{n}{2}*n=\frac{n^3}{4}$ servers.
Figure~\ref{fig:fatTree} shows a Fat-tree for $n=4$. Note that Fat-tree employs a more redundant core than the Three-layer topology.

\subsection{BCube}
\label{bcube}

\begin{figure}
\centering
\includegraphics[width=0.7\textwidth]{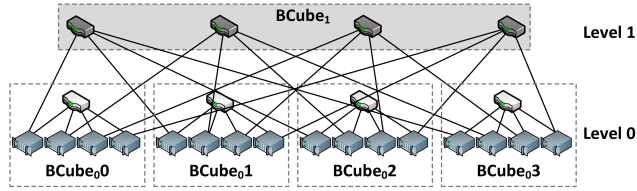}
\caption{BCube with 4-port switches ($n=4$) and 2-port servers ($l=1$).}
\label{fig:bcube}
\end{figure}

The BCube topology was designed for Modular Data Centers (MDC) that need a high network robustness~\cite{guo2009bcube}. A BCube is organized in layers of commodity mini-switches and servers, which participate in packet forwarding. The main module of a BCube is $BCube_0$, which consists of a single switch with $n$ ports connected to $n$ servers.
A $BCube_1$, on the other hand, is constructed using $n$ $BCube_0$ networks and $n$ switches. Each switch is connected to all $BCube_0$ networks through one server of each $BCube_0$.
Figure~\ref{fig:bcube} shows a $BCube_1$. More generally, a $BCube_l$ ($l \geq 1$) network consists of $n$ $BCube_{l-1}$s and $n^l$ switches of $n$ ports.
To build a $BCube_l$, the $n$ $Bcube_{l-1}$s are numbered from 0 to $n-1$ and the servers of each one from 0 to $n^l - 1$.
Next, the level $l$ port of the $i$-th server ($i \in [0,n^l - 1]$) of the $j$-th $BCube_l$ ($j \in [0,n-1]$) is connected to the $j$-th port of the $i$-th level $l$ switch.
A $BCube_l$ can have $n^{l+1}$ servers.
In BCube, servers participate in packet forwarding but are not directly connected.

\subsection{DCell}
\label{dcell}

\begin{figure}
\centering
\includegraphics[width=.4\textwidth]{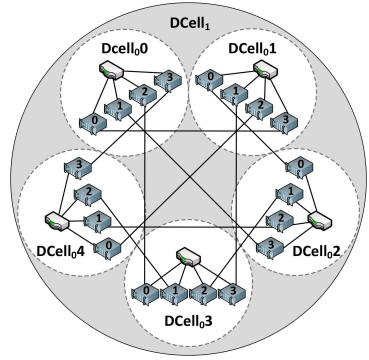}
\caption{DCell with 4-port switches ($n=4$) and 2-port servers ($l=1$).}
\label{fig:dcell}
\end{figure}

Similar to BCube, DCell is defined recursively and uses servers and mini-switches for packet forwarding.
The main module is $DCell_0$ which, as in $BCube_0$, is composed of a switch connected to $n$ servers.
A $DCell_1$ is built by connecting $n + 1$ $DCell_0$ networks, where a $DCell_0$ is connected to every other $DCell_0$ via a link connecting two servers.
A $DCell_1$ network is illustrated in Figure~\ref{fig:dcell}.

Note that in a DCell, unlike a BCube, switches are connected only to servers in the same DCell and the connection between different DCell
networks goes through servers. To build a $DCell_l$, $n + 1$ $DCell_{l-1}$ networks are needed. Each server in a $DCell_l$ has $l+1$ links, where the first link (level 0 link) is connected to the switch of its $DCell_0$, the second link connects the server to a node on its $DCell_1$, but in another $DCell_0$, and so on. Generalizing, the level $i$ link of a server connects it to a different $DCell_{i-1}$ in the same $DCell_{i}$. The procedure to build a DCell is more complex than that of a BCube, and is executed by the algorithm described in~\cite{guo2008dcell}.

The DCell capacity in a number of servers can be evaluated recursively, using the following equations: $g_l = t_{l-1} + 1$ and $t_l = g_l\times t_{l-1}$, where $g_l$ is the number of $DCell_{l-1}$ networks in a $DCell_{l}$, and $t_l$ is the number of servers in a $DCell_l$.
A $DCell_0$ network is a special case in which $g_0=1$ and $t_0=n$.

\section{Analysis Methodology}
\label{sec:methodology}
\begin{figure}
\centering
\includegraphics[width=0.64\textwidth]{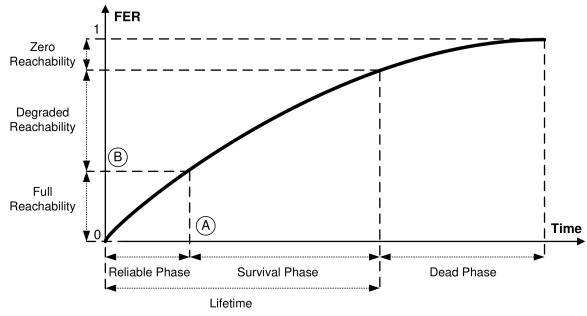}
\caption{Evolution of the DC reachability. As more network elements fail, more servers are disconnected and thus the reachability decreases. }
\label{fig:overview}
\end{figure}
\begin{figure}
\centering
{\includegraphics[width=0.23\textwidth]{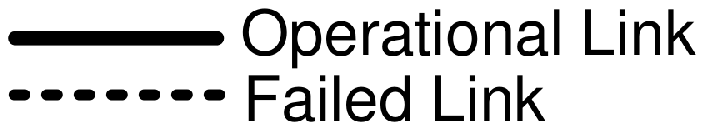}}\\
\subfigure[\hspace{-1mm}Reliable Phase.]
{\includegraphics[width=0.2\textwidth]{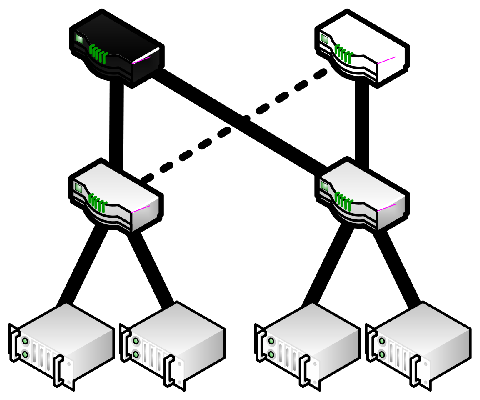}}
\subfigure[\hspace{-1.1mm}Survival Phase.]
{\includegraphics[width=0.2\textwidth]{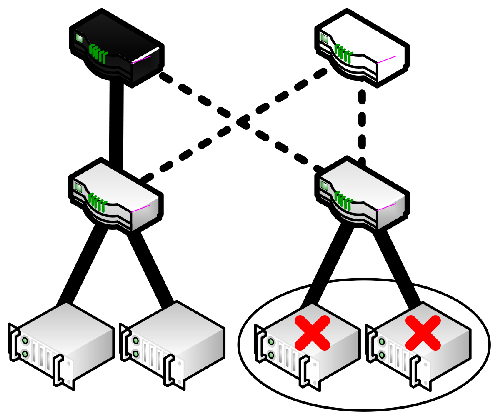}}
\subfigure[\hspace{-1mm}Dead Phase.]
{\includegraphics[width=0.2\textwidth]{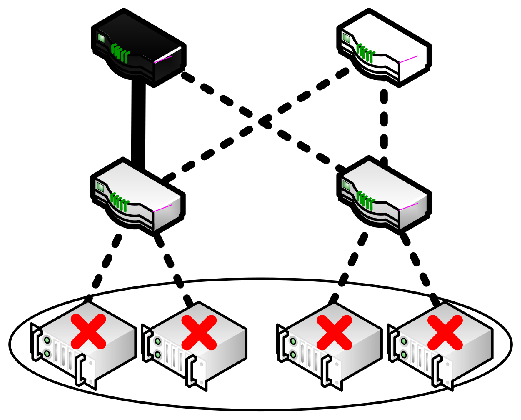}}
\caption{The different phases a network undergoes when facing link failures.}
\label{fig:phases}
\end{figure}

As the operating time of a DCN progresses, more network elements would fail and thus server reachability (i.e., number of connected servers and the connectivity between them) levels are expected to decrease. \textit{A server is considered disconnected when it has no paths to the DCN gateways, i.e., to the switches providing access to external networks like the Internet}. In this work, we evaluate DCNs, considering the failures of a given network element type, i.e., link, switch, or server. Each type of failure is evaluated \textit{separately} to analyze its particular influence. Independent of the element type, we define the \textit{lifetime as the amount of time until the disconnection of all DC servers}. 
Despite this theoretical definition in this work we do not analyze the DCN behavior for the whole lifetime, since it is not practical to have a DC with almost all its servers disconnected.
To quantify the failures, we define the \textit{Failed Elements Ratio} (FER), which is the fraction of failed elements of a given network element type (link, switch, or server). If no maintenance is performed on the DC, which is the case considered in this work, the FER for a given 
equipment type 
will increase as the time passes, meaning that more 
network elements are under failure. Figure~\ref{fig:overview} illustrates a hypothetical situation of the FER evolution according to the time. Starting the lifetime by the moment where a full maintenance was completed, a DC passes through a first phase in which failures do not cause server disconnection, defined here as the \textit{Reliable Phase}, and a second phase where at least one server is disconnected, that we define as the \textit{Survival Phase}. The lifetime period ends 
when the DC has no connected servers. After that, the DC enters the \textit{Dead Phase}. Figure~\ref{fig:phases} depicts each phase of a hypothetical network, only considering link failures. In this figure, each failed link is represented by a dashed line, an inaccessible server is represented with a cross, and the switch that acts as a gateway is colored in black. The disconnected fraction of the DC is circled in the figure. We can see that on the Reliable Phase the DCN can have failed links and on the Dead Phase it can have links that are still up.

Regarding the Reliable Phase, the circled letters in Figure~\ref{fig:overview} point out two metrics of interest. These are

\begin{itemize}
\item \textbf {A}: Indicates time elapsed until the first server is disconnected, called TTF (Time to Failure). In this work, we evaluate the mean value of this metric, called MTTF (Mean Time To Failure), which is the expected value of the TTF in a network (i.e., mean time elapsed until the first server disconnection).
\item \textbf {B}: Indicates the minimum value of FER that produces a server disconnection. In this work, we evaluate this metric as a mean value, called the \textit{Critical FER}. 
For example, a network with 100 switches that disconnects a server, on average, after the removal of 2 random switches, has a critical FER of $\frac{2}{100}=0.02$. The mean time to have a Critical FER is thus equal to the MTTF.
\end{itemize}

The Survival Phase deserves special attention if one is interested in quantifying the network degradation; for this phase, in Section~\ref{sec:survivalTime} we define and analyze the following in a set of representative metrics: Service Reachability and Path Quality.

\subsection{Link and Node Failures}
\subsubsection{Failure Model}
\label{sec:failureModel}

Our failure model is based on the following assumptions:
\begin{itemize}
\item \textbf{Failure isolation:} Each type of failure (link, switch or server) is analyzed separately. This is important to quantify the impact of a given element type on the considered topologies.
\item \textbf{Failure probability:} For the sake of simplicity, all the elements have the same probability of failure and the failures are independent of each other.
\item \textbf{Repairs:} The elements are not repairable. This is important to study how much time the network can operate without maintenance (e.g., Modular Data Center, where equipment repair is a difficult task).
\end{itemize}
\subsubsection{Failure Metrics}
\label{sec:simulationOverview}

We analyze failures from both a spatial and temporal perspective, using the two following metrics:

\textbf{Failed Elements Ratio (FER).} Defined before, this metric quantifies only the extension of the failures and does not depend on the probability distribution of the element lifetime. In the following, we also use the more specific term ``Failed Links/Switches/Servers Ratio'' to emphasize the failure type. 

\textbf{Elapsed Time.} As the time passes, more elements would fail. In this case, the elapsed time since the last full maintenance can indirectly characterize the failure state.
For a given FER, we have an expected time that this ratio will occur. 
It is worth mentioning that time can be defined in two ways: absolute and normalized. In the former, we measure the time in hours, days or months. In the latter, we normalize the time by the mean lifetime of an individual link or node, as detailed next. This measure is important to make the analysis independent of the mean lifetime, being agnostic to hardware characteristics.
\subsection{Failure Simulation}
\label{sec:failureSimMethodology}

A topology is modeled as an undirected, unweighted graph $G=(\mathcal{V},\mathcal{E})$, where $\mathcal{V}$ is the set of servers and switches, and $\mathcal{E}$ is the set of links. The set $\mathcal{V}$ is given by $\mathcal{V}=\mathcal{S} \cup \mathcal{C}$, where $\mathcal{S}$ is the server set and $\mathcal{C}$ is the switch set. To simulate the scenario of Section~\ref{sec:failureModel}, we randomly remove either $\mathcal{S'}$, $\mathcal{C}'$, or $\mathcal{E}'$ from $G$, where $\mathcal{S}'\subset \mathcal{S}$, $\mathcal{C}'\subset \mathcal{C}$ and $\mathcal{E}'\subset \mathcal{E}$, generating the subgraph $G'$. Note that we separately analyze each set of elements (switches, servers, and links). Finally, the metrics are evaluated by using the graph $G'$. Unless otherwise stated, all metrics are represented with their average values and confidence intervals, evaluated with a confidence level of 95\%. As can be seen next in this work, our results have a very narrow confidence interval, and thus, most of these intervals are difficult to visualize in the curves.\footnote{Topology generation, failure simulation, and metric evaluation are obtained using the graph manipulation tool NetworkX~\cite{hagberg2008exploring}.}

The evaluation starts by removing $f$ elements from $G$, where $0 \leq  f \leq F $ and $F$ is the total number of elements of a given type (link, switch or server) present on the original graph $G$. After that, we evaluate our metrics of interest as a function of $f$. The FER and Elapsed Time (Section~\ref{sec:simulationOverview}) are computed, respectively, by $\frac{f}{F}$ and by the mean time that $f$ elements fail, given that we have $F$ possibilities of failure (i.e., the total number of elements of a given type). To evaluate this amount of time, we first need to define a probability distribution for element failures. For simplicity, following a widely adopted approach, we consider that failures are independent and that the time $\tau$ that an element fails is random and follows an exponential distribution with mean $E[\tau]$~\cite{egeland2009availability,rahman2010survivable}. 
Hence, the mean time to have $f$ failed elements (Elapsed Time) is given by the following equation derived from Order Statistics~\cite{barlow1975statistical}: 
\begin{equation}
AT = E[\tau] \sum_{i=0}^{f-1} \frac{1}{F - i} \,, \text{ for } f \leq F.
\label{eq:absoluteTime}
\end{equation}
Equation~\ref{eq:absoluteTime} gives the Absolute Time defined in Section~\ref{sec:simulationOverview}. Note that we can make it independent of $E[\tau]$ by dividing the right term of Equation~\ref{eq:absoluteTime} by $E[\tau]$. The result is the Normalized Time given by
\begin{equation}
NT = \sum_{i=0}^{f-1} \frac{1}{F - i} \,, \text{ for } f \leq F.
\label{eq:normalizedTime}
\end{equation}
\subsection{Operational Subnetworks After Failures}
\label{sec:operational}

In our analysis, we first have to identify whether a network is operational to compute the metrics of interest. As failures may split the DCN, we define \textit{operational} as all the connected (sub)networks that have at least one gateway\footnote{We call \textit{gateway} in this work a switch that is responsible for the network access outside the DC. In practice, the gateway function is performed by a router connected to this switch.}. This node plays a fundamental role since it is responsible for interconnecting the DC with external networks, as the Internet. Hence, a subnetwork that has no gateway is not considered operational because it cannot receive remote commands to assign tasks to servers. A server in an operational network is considered as \textit{connected}.

As typical definitions of DCN topologies are not aware of gateway placement, we assume that all switches at the highest hierarchical level of each topology are in charge of such a task. For the topologies considered, we have the following possible gateways:
\begin{itemize}
\item \textbf{Three-layer:} The two core switches
\item \textbf{Fat-tree:} All core switches
\item \textbf{BCube:} For a BCube of level $l$, all the $l$-level switches
\item \textbf{DCell:} As there is no switch hierarchy in this topology, we consider that all switches are at the top level and therefore can be a gateway.
\end{itemize}

A possible issue with the above choices is that the comparison between topologies may be unfair depending on how many gateways we choose for each of them. We thus define a metric of reference, called the Gateway Port Density~(GPD):
\begin{equation}
GPD = \frac{n*g}{|\mathcal{S}|},
\label{eq:epd}
\end{equation}
where $n$ is the number of ports on the gateway, $g$ is the number of gateways and $|\mathcal{S}|$ is the total number of servers in the network.
The GPD gives an idea on the number of ports per server available in the gateways. As each gateway has $n$ ports, the DC has $n*g$ ports acting as the last access to the traffic before leaving the DC. Note that the number of ports connecting the gateway to outside the DCN is not accounted in $n$, since $n$ is the number of switch ports as given by each topology definition (Section~\ref{sec:topologies}). We assume that each gateway has one or more extra ports that provide external access. In addition, we do not consider the failure of these extra ports.
The maximum GPD (i.e., if we use all possible switches) for Fat-tree, BCube, and DCell is equal to 1. 
As the Three-layer topology uses only two core switches, its maximum GPD is very low (e.g., 0.007 for a network with 3456 servers).
Hence, unless stated otherwise, we use all the possible switches for all topologies in our evaluations.
We do not equalize all topologies with the maximum GPD of the Three-layer one to provide a better comparison between alternative DC topologies.
In addition, we show later in this work that this choice does not change our conclusions regarding the comparison between the Three-layer topology and the alternative ones.

\section{Reliable Phase}
\label{sec:reliableTime} 

The Reliable Phase corresponds to the period until the disconnection of the first server. It quantifies the amount of time a DC administrator can wait until the next network maintenance intervention makes the network fully reachable. We qualify the DCN performance in the Reliable Phase both theoretically and by simulation, as explained in this section.

\subsection{Theoretical analysis}

The MTTF can be evaluated as a function of the reliability, $R(t)$. $R(t)$ is defined as the probability that the network is on the Reliable Phase (i.e., all its servers are accessible) at time $t$. In other words, considering that the time spent in the Reliable Phase is a random variable $T$, the reliability is defined as $R(t) = P(T > t) = 1 - P(T \leq t)$. Note that $P(T \leq t)$ is the CDF (Cumulative Distribution Function) of the random variable $T$. As the MTTF is the expected value $E[T]$, we can use the definition of $E[T]$ for non-negative random variables as shown in the following:
\begin{equation}
MTTF = \int_{0}^{\infty} 1 - P(T \leq t) \, dt = \int_{0}^{\infty} R(t) \, dt.
\label{eq:mttfReliability}
\end{equation}

We evaluate $R(t)$ by using the Burtin-Pittel approximation~\cite{gertzbakh2009models} to network reliability given by
\begin{equation}
R(t) = 1 - \frac{t^rc}{{E[\tau]}^r} + O \left ({\frac{1}{E[\tau]}}^{r+1} \right) \approx e^{-\frac{t^rc}{{E[\tau]}^r}},
\label{eq:burtilPittel}
\end{equation}
where ${E[\tau]}$ is the expected (i.e., average) time that an element fails, considering that $\tau$ follows an exponential probability distribution. The parameters $c$ and $r$ are the number of min-cut sets and their size, respectively. A min-cut set is a set with the minimum number of elements that causes a server disconnection. For example, considering only link failures on the network of Figure~\ref{fig:conventional}, a min-cut set consists of a link between the server and the edge switch. Considering only switch failures in Figure~\ref{fig:conventional}, a min-cut set is an edge switch. 
The min-cut size is the number of elements (links, switches, or servers) in a single set (e.g., equal to 1 in the above mentioned examples). In Equation~\ref{eq:burtilPittel}, $\frac{t^rc}{{E[\tau]}^r}$ is the contribution of the min-cut sets to $R(t)$ and $O \left ({\frac{1}{E[\tau]}}^{r+1} \right)$ is an upper bound to the contribution of other cut sets. The idea behind the approximation is that if ${E[\tau]}$ is high (i.e., the failure rate of an individual element is low), $R(t)$ is mainly affected by the min-cut sets. This is valid for a DCN, since it is expected to have a large lifetime even for commodity equipment~\cite{gill2011understanding}.
The approximation is done by using the fact that the term $1 - \frac{t^rc}{{E[\tau]}^r}$ in Equation~\ref{eq:burtilPittel} coincides with the first two terms of the Taylor expansion of $e^{-\frac{t^rc}{{E[\tau]}^r}}$. Hence, considering that the contribution of the other cut sets is as small as the remaining terms of the Taylor expansion, we can write $R(t) \approx e^{-\frac{t^rc}{{E[\tau]}^r}}$. 

Combining Equations~\ref{eq:mttfReliability}~and~\ref{eq:burtilPittel}, as detailed in Appendix~\ref{app:compMTTFapprox}, we rewrite the MTTF as:
\begin{equation}
MTTF \approx \frac{E[\tau]}{r}\sqrt[r]{\frac{1}{c}} \Gamma \left (  \frac{1}{r} \right ),
\label{eq:mttfApprox}
\end{equation}
where $\Gamma(x)$ is the gamma function of $x$~\cite{abramowitz1970handbook}. With this equation, the MTTF is written as a function of $c$ and $r$ that, as we show later, depends on the topology employed and on its parameters.

\subsection{Simulation-based analysis}

The simulation is provided to measure the accuracy of the MTTF approximation stated before.
For each simulation sample, we find the minimum number of $f$ elements of a given type that disconnects a server from the network. This value is called the critical point.
The Normalized MTTF (NMTTF) in a sample can thus be evaluated by setting $f$ equal to the critical point in Equation~\ref{eq:normalizedTime}. The simulated value of MTTF ($NMTF_{sim}$) is thus the
average of the NMTTF values considering all samples. Algorithm~\ref{alg:SimulationBased} summarizes the simulation procedure.
The function \texttt{removeRandomElement} removes one random element of a given type (link, switch, or server) following the procedure described in Section~\ref{sec:failureSimMethodology}.
In addition, the function \texttt{allServersAreConnected} checks if all the servers in the network $G'$ (i.e., network with $f$ removed elements of a given type) are connected, as defined in Section~\ref{sec:operational}.
When the function \texttt{removeRandomElements} leads to a $G'$ with at least one disconnected server, the simulation stops and line 10 evaluates the Normalized MTTF (NMTTF) using Equation~\ref{eq:normalizedTime}, adding this measure to the \texttt{accNMTTF} variable. The \texttt{accNMTTF} is thus the sum of the NMTTF values found in all samples. At the end, this variable is divided by the total number of samples $nrSamples$ to achieve the average value of NMTTF ($NMTF_{sim}$) found on the simulation. Note that the simulated MTTF can be evaluated by multiplying $NMTF_{sim}$ by ${E[\tau]}$ , as indicated by Equation~\ref{eq:absoluteTime}.
The parameter \texttt{nrSamples} is set in this work in the order of thousands of samples to reach a small confidence interval.
\begin{algorithm}
\footnotesize
\SetAlgoLined
\SetKwFunction{allServersAreConnected}{allServersAreConnected}
\SetKwFunction{removeRandomElement}{removeRandomElement}
\KwIn{element type $type$, number of experimental samples $nrSamples$, total number of elements $F$, original network $G$.}
\KwOut{Simulated NMTTF $NMTTF_{sim}$.}
sample = 1\;
accNMTTF = 0\;
\While{$sample$ $\leq$ $nrSamples$ }{
  $G'$ = $G$\;
  $f$ = 0\;
  \While{ ($f$ $<$ $F$) and \allServersAreConnected{$G'$}}{
    $f$ += 1\;
    $G'$ = \removeRandomElement($type$,$G'$)\;
  }
  $accNMTTF$ += $\sum_{i=0}^{f-1} \frac{1}{F - i}$\;
  $sample$  += 1\;
}
$NMTTF_{sim} = \frac{accNMTTF}{nrSamples}$\;
\caption{NMTTF simulation}
\label{alg:SimulationBased}
\end{algorithm}

The comparison between the simulated and theoretical MTTF is done using the Relative Error (RE) defined as:
\begin{equation}
RE = \frac{|NMTTF_{sim} - NMTTF_{theo}|}{NMTTF_{sim}},
\label{eq:relativeError}
\end{equation}
where $NMTTF_{theo}$ is the normalized theoretical MTTF, obtained by dividing the MTTF by $E[\tau]$, and $NMTTF_{sim}$ is the value obtained in the simulation. It is important to note that, as shown in Equation~\ref{eq:mttfApprox}, the MTTF can be expressed by a first order term of $E[\tau]$. Consequently, we do not need, in practice, to use the value of $E[\tau]$ to normalize the theoretical MTTF, needing only to remove this term from the equation.
Using the results of RE, we show in Section~\ref{sec:reliableTimeResults} in which cases Equation~\ref{eq:mttfApprox} is an accurate approximation for the MTTF. In these cases, we show that the MTTF for each topology can be approximated as a function of the number of server network interfaces and the number of servers. 
\subsection{Results}
\label{sec:reliableTimeResults} 

In this section, we use the metrics detailed before to evaluate the topologies of Table~\ref{tab:configurations} in the Reliable Phase. We compare configurations with approximately the same number of connected servers. It is worth mentioning that although some of these topologies can be incrementally deployed, we only consider complete topologies where all servers' and switches' network interfaces are in use. Furthermore, for alternative DC topologies, the number of switch ports is not limited to the number of ports often seen in commercially available equipment (e.g., 8, 24, and 48) to produce a similar number of servers for the compared topologies. As one of the key goals of a DC is to provide processing capacity or storage redundancy, which increases with the number of servers, balancing the number of servers per topology is an attempt to provide a fair analysis. 
For the Three-layer topology, we fix $n_e=48$ and $n_a=12$, based on commercial equipment description found in~\cite{scotland}. Hence, for all configurations, each pair of aggregation switches provides connectivity to 576 servers.
As we employ a fixed number of ports in the aggregation and edge layers for the Three-layer topology, we specify in Table~\ref{tab:configurations} only the number of ports in a core switch connected to aggregation switches. 
We provide below the analysis according to each type of failure. We do not evaluate the reliability to server failures because a network failure is considered whenever one server is disconnected. Hence, a single server failure is needed to change from the Reliable to the Survival Phase.
\begin{table}
\caption{DCN topology configurations used in the analysis.}
\label{tab:configurations}
\begin{tabular}{lllllll}
\hline\noalign{\smallskip}
\hline \multirow{2}{*}{\bf Size}	&\multirow{2}{*}{\bf Name} 	& {\bf Switch} &{\bf Server} &\multirow{2}{*}{\bf Links} &\multirow{2}{*}{\bf Switches} &\multirow{2}{*}{\bf Servers}\\
& &  {\bf ports} & {\bf ports} &  & &\\ 
\noalign{\smallskip}\hline\noalign{\smallskip}
\hline\multirow{6}{*} {500} & Three-layer & $2$(core) & 1 & 605 & 16 & 576\\
& Fat-tree & 12 & 1 & 1296 & 180 & 432\\
& BCube2 & 22 & 2 & 968 & 44 & 484\\
& BCube3 & 8 & 3 & 1536 & 192 & 512\\
& DCell2 & 22 & 2 & 759 & 23 & 506\\
& DCell3 & 4 & 3 & 840 & 105 & 420\\
\noalign{\smallskip}\hline	
\multirow{7}{*} {3k} & Three-layer & $12$(core)  & 1 & 3630 & 86 & 3456\\
& Fat-tree & 24 & 1 & 10368 & 720 & 3456\\
& BCube2 & 58 & 2 & 6728 & 116 & 3364\\
& BCube3 & 15 & 3 & 10125 & 670 & 3375\\
& BCube5 & 5 & 5 & 15625 & 3125 & 3125\\
& DCell2 & 58 & 2 & 5133 & 59 & 3422\\
& DCell3 & 7 & 3 & 6384 & 456 & 3192\\
\noalign{\smallskip}\hline	
\multirow{7}{*} {8k} & Three-layer & $28$(core)  & 1 & 8470 & 198 & 8064\\
& Fat-tree & 32 & 1 & 24576 & 1280 & 8192\\
& BCube2 & 90 & 2 & 16200 & 180 & 8100\\
& BCube3 & 20 & 3 & 24000 & 1190 & 8000\\
& BCube5 & 6 & 5 & 38880 & 6480 & 7776\\
& DCell2 & 90 & 2 & 12285 & 91 & 8190\\
& DCell3 & 9 & 3 & 16380 & 910 & 8190\\
\noalign{\smallskip}\hline	
\end{tabular}
\end{table}

\subsubsection{Link Failures}
\label{sec:perfEvaluationLinkReliable}

To provide the theoretical MTTF for link failures, we use Equation~\ref{eq:mttfApprox} with the values $r$ and $c$ corresponding to each topology. Table~\ref{tab:cuts_link} shows these values for all considered topologies. For all topologies, the min-cut size is the number of server interfaces, which is always 1 for Three-layer and Fat-tree, and $l+1$ for BCube and DCell. Also, except for DCell with $l=1$, the number of min-cuts is equal to the number of servers. For DCell with $l=1$, we have another min-cut possibility, different from the disconnection of $l+1=2$ links from a single server. We call this possibility a ``server island'', which appears when the two connected servers lose the link to their corresponding switch. 
As an example, consider that in Figure~\ref{fig:dcell} Server 0 in $DCell_00$ and Server 3 in $DCell_01$ have lost the link with their corresponding switches. These two servers remain connected to each other but when disconnected from the network, they form a server island.
In DCell with $l=1$, each server is directly connected with only one server, since each one has two interfaces. Then, the number of possible server islands is $0.5|\mathcal{S}|$ and the number of min-cuts is given by $|\mathcal{S}| + 0.5|\mathcal{S}| = 1.5|\mathcal{S}|$. For a DCell with $l>1$, the number of link failures that produces a server island is greater than $l+1$ and therefore, this situation is not a min-cut.
\begin{table}
\caption{Min-cut size and number considering link failures.}
\label{tab:cuts_link}
\begin{tabular}{lll}
\hline\noalign{\smallskip}
\hline \textbf{Topology}			&\textbf{Min-cut size (r)} &\textbf{Number of min-cuts (c)}\\
\noalign{\smallskip}\hline\noalign{\smallskip}
Three-layer	&$1$ &$|\mathcal{S}|$\\
Fat-tree &$1$ &$|\mathcal{S}|$ \\
BCube &$l+1$ &$|\mathcal{S}|$ \\
DCell &$l+1$ &$1.5|\mathcal{S}|$ \ if \, $l=1$ , \, $|\mathcal{S}|$ \, otherwise\\
\noalign{\smallskip}\hline
\end{tabular}
\end{table}

Using the values of Table~\ref{tab:cuts_link} in Equation~\ref{eq:mttfApprox}, we get the following MTTF approximations, for link failures:
\begin{equation}
MTTF_{threeLayer} = MTTF_{fatTree} \approx \frac{E[\tau]}{|\mathcal{S}|};
\label{mttf_fatTree_link}
\end{equation}
\begin{equation}
    MTTF_{dcell} \approx \begin{cases}
               \frac{E[\tau]}{2}\sqrt{\frac{1}{1.5|\mathcal{S}|}} \Gamma \left (  \frac{1}{2} \right ),   &\text{ if }          l = 1;\\
               \frac{E[\tau]}{l+1}\sqrt[l+1]{\frac{1}{|\mathcal{S}|}} \Gamma \left (  \frac{1}{l+1} \right ) &\text{ otherwise }.
           \end{cases}
\end{equation}
\begin{equation}
	MTTF_{bcube} \approx \frac{E[\tau]}{l+1}\sqrt[l+1]{\frac{1}{|\mathcal{S}|}} \Gamma \left (  \frac{1}{l+1} \right ).
\end{equation}

\begin{figure}
\centering
%generate_criticalPoint_link_allTopologies_relError.gnu
\subfigure[Relative Error of MTTF approximation.]
{\includegraphics[width=0.48\textwidth]{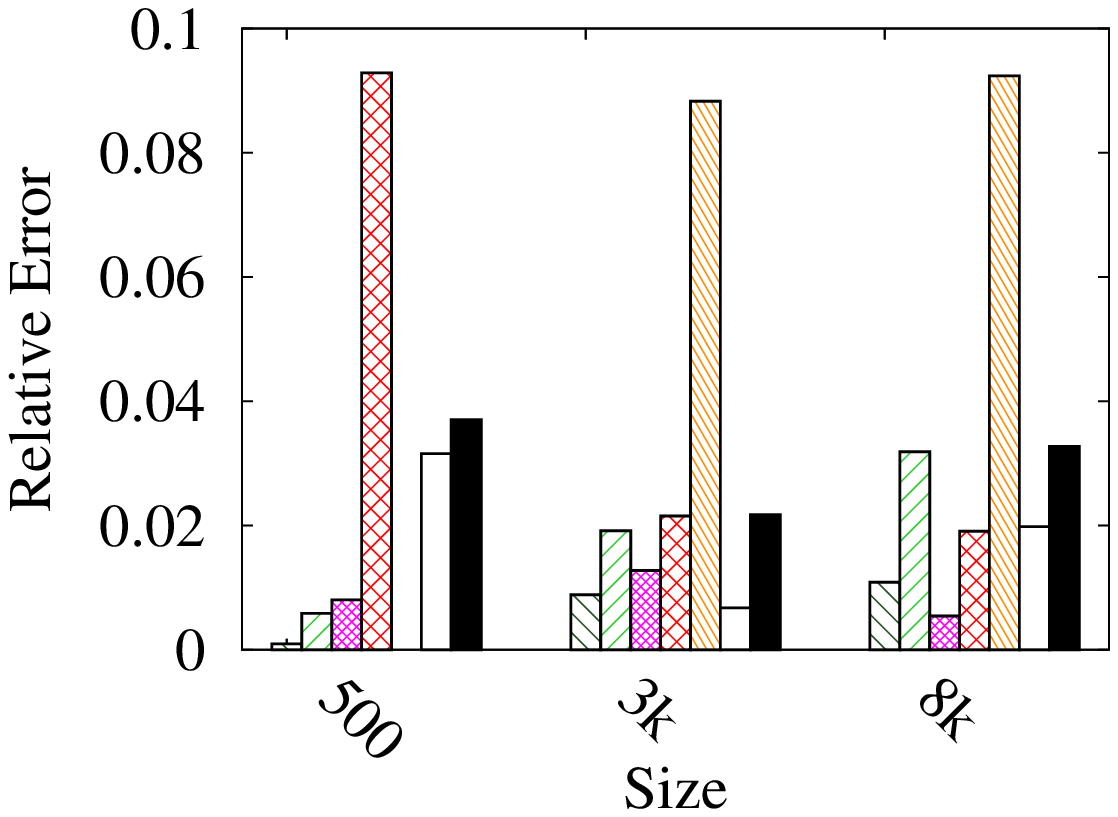}
\label{fig:link_mttf_relError}}
%gnuplot generate_criticalPoint_link_3k_unitMTTF.gnu
\subfigure[Elapsed Time and FER simulation for \mbox{3k-server} DCNs.]
{\includegraphics[width=0.36\textwidth]{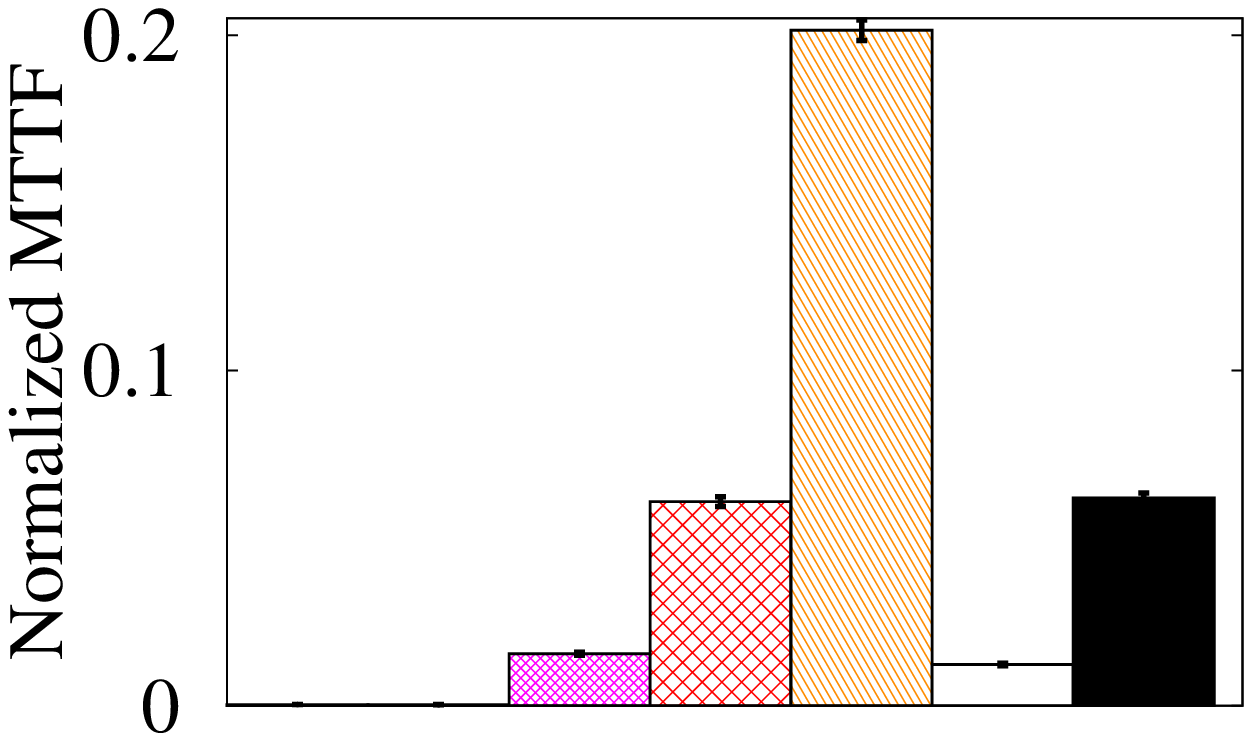}
%gnuplot generate_criticalPoint_link_3k_criticalPoint.gnu
\quad
\includegraphics[width=0.36\textwidth]{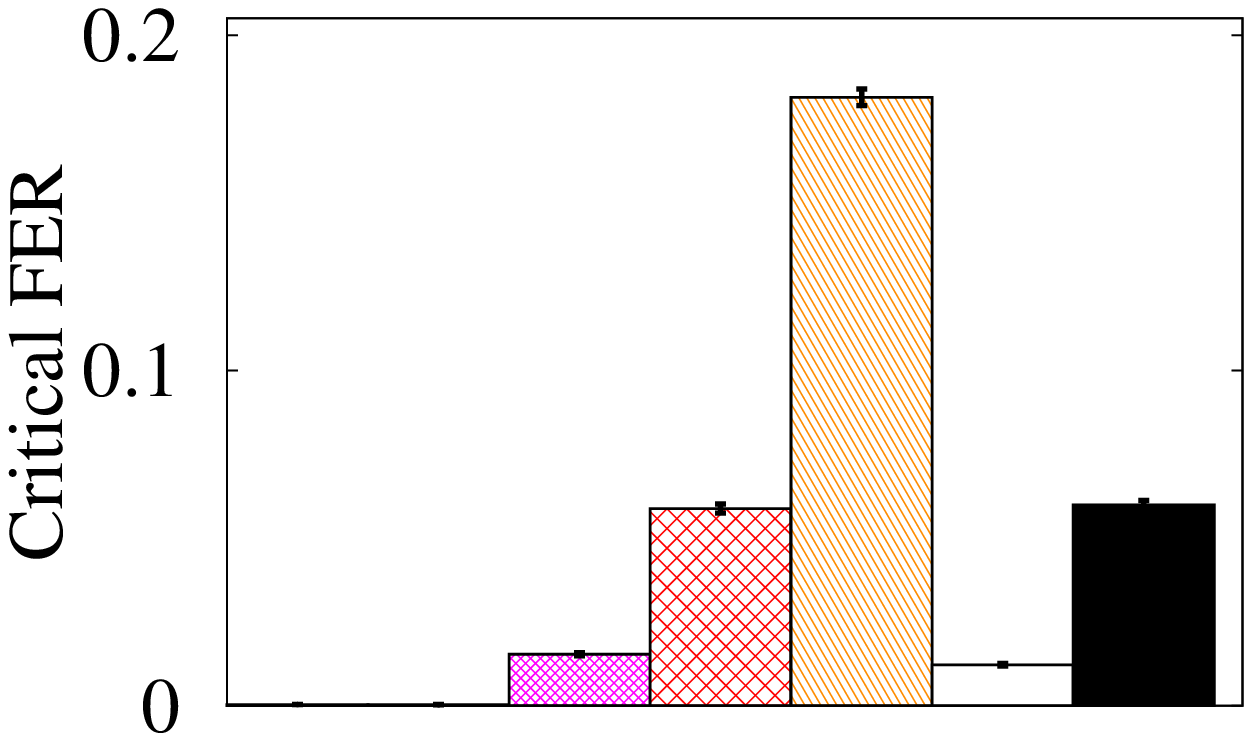}
\label{fig:link_mttf_unitmttf}
}
{\includegraphics[width=0.8\textwidth]{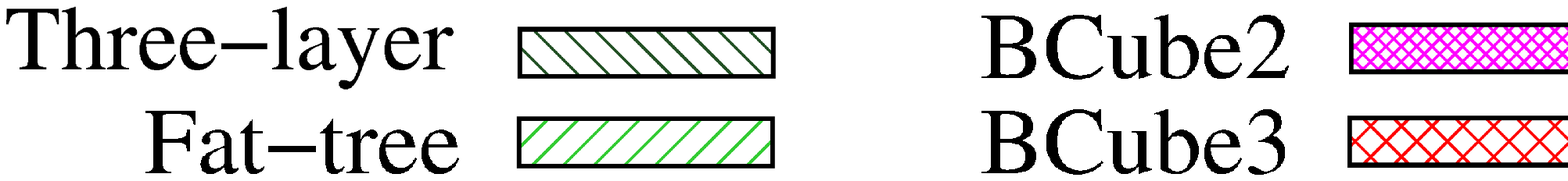}}
\caption{Reliable Phase analysis for link failures.}
\end{figure}

The results of Figure~\ref{fig:link_mttf_relError} show the RE (Equation~\ref{eq:relativeError}) for different network sizes. The figure shows that the MTTF estimation using min-cuts has less than a 10\% error.

Given the above equations and their comparison in Appendix~\ref{app:compMTTFEquations}, we can conclude that:\footnote{Hereafter, we split the result remarks in three items. The first one comments the performance of switch-centric topologies (Three-layer and Fat-tree), while the second one highlights the results of server-centric topologies (i.e., BCube and DCell). The last item, when available, indicates a general remark considering the three topologies.}

\begin{itemize}
\item \textit{Three-layer and Fat-tree Performance.} The two topologies have the same MTTF, presenting the lowest reliability considering link failures. According to the equations, the MTTF of Three-layer and Fat-tree is $\sqrt{\frac{|\mathcal{S}|\pi}{6}}$ lower than the worst case for a server-centric topology (DCell2). Hence, for a DCN with 3400 servers, the MTTF of Three-layer and Fat-tree is at least 42 times lower than that of server-centric topologies.
\item \textit{BCube and DCell Performance.} BCube has the same MTTF as DCell, except for two server interfaces where BCube performs better. However, as given by the equations, BCube2 is merely $1.23$ times better than DCell2 for any $|\mathcal{S}|$. In BCube and DCell, the increase in the number of server interfaces increases the MTTF.
\item \textit{General Remarks.} A higher number of servers $|\mathcal{S}|$ leads to a lower MTTF. This result emphasizes the importance of caring about reliability in large DCs, where $|\mathcal{S}|$ can be in the order of thousands of servers.
\end{itemize}
Figure~\ref{fig:link_mttf_unitmttf} shows the simulation of the Normalized MTTF and Critical FER for 3k-server topologies as an example. 
Note that the reliability of Three-layer and Fat-tree is substantially lower than that of the other topologies, and as a consequence their corresponding boxes cannot be seen in Figure~\ref{fig:link_mttf_unitmttf}.
\subsubsection{Switch Failures}
\label{sec:perfEvaluationSwitchReliable}

We employ the same methodology of Section~\ref{sec:perfEvaluationLinkReliable} to verify if we can use min-cuts to approximate the reliability when the network is prone to switch failures. Table~\ref{tab:cuts_switch} shows $r$ and $c$ values for this case. In Three-layer and Fat-tree, a single failure of an edge switch is enough to disconnect a server. Hence, the size of the min-cut is 1 and the number of min-cuts is the number of edge switches. 
In Three-layer, the number of edge switches is simply $\frac{|\mathcal{S}|}{n_e}$, where $n_e$ is the number of edge ports.
In a Fat-tree of $n$ ports, each edge switch is connected to $\frac{n}{2}$ servers, and thus, the number of edge switches is $\frac{|\mathcal{S}|}{\frac{n}{2}}$. As $|\mathcal{S}|=\frac{n^3}{4}$, we can write $n=\sqrt[3]{4|\mathcal{S}|}$ and therefore, the number of min-cuts is $\sqrt[3]{2|\mathcal{S}|^2}$. For BCube, a switch min-cut happens when, for a single server, the $l+1$ switches connected to it fail. The number of possible min-cuts is thus equal to the number of servers $|\mathcal{S}|$, as each server has a different set of connected switches.
As in the case of DCell, the reasoning is more complex. A min-cut is the set of switches needed to form a server island.
Although min-cuts for link failures generate server islands only in DCell2, all min-cuts generate this situation in both DCell2 and DCell3 for switch failures. For DCell2, it is easy to see that a server island is formed if two servers that are directly connected lose their corresponding switches, therefore $r=2$. As observed in Section~\ref{sec:perfEvaluationLinkReliable}, the number of possible server islands is the number of pairs of servers, given by $0.5|\mathcal{S}|$. For DCell3, we obtain the values $r$ and $c$ by analyzing DCell graphs for different values of $n$ with $l=2$. We observe that $r$ is always equal to 8, independent of $n$. Also, we observe the formation of server islands. Every island has servers from 4 different DCell modules of level $l=1$. Moreover, each DCell with $l=1$ has 2 servers from the island. Obviously, these 2 servers are directly connected to each other, from different DCell modules with $l=0$. Based on the analysis of different graphs, we find that DCell3 has $c=\binom{
n+2}{4}$. Hence, we can formulate the min-cuts for DCell2 and DCell3 as $r=2l^2$ and $c=\binom{n+l}{2l}$. Note that, for DCell2 $c=\binom{n+1}{2}=0.5*[n(n+1)]=0.5|\mathcal{S}|$, corresponding to the value found before. 
For DCell3 we find $c=\binom{n+2}{4}=0.125(2|\mathcal{S}| - 3\sqrt{4|\mathcal{S}|+1} + 3 )$, by replacing $n$ with the solution of $|\mathcal{S}| = [n(n+1)][(n+1)n +1]$. We leave the evaluation of $r$ and $c$ for DCell with $l>2$ as a subject for future work.
%.
\begin{table}[hb]
\caption{Min-cut size and number considering switch failures.}
\label{tab:cuts_switch}
\begin{tabular}{lll}
\hline\noalign{\smallskip}
\textbf{Topology} &\textbf{Min-cut size (r)} &\textbf{Number of min-cuts (c)}\\
\noalign{\smallskip}\hline\noalign{\smallskip}
Three-layer &$1$ &$\frac{|\mathcal{S}|}{n_e}$ \\
Fat-tree &$1$ &$\sqrt[3]{2|\mathcal{S}|^2}$ \\
BCube &$l+1$ &$|\mathcal{S}|$ \\
DCell $(l \leq 2)$ & $2l^2$ \ &$\binom{n+l}{2l}$ \\
\noalign{\smallskip}\hline
\end{tabular}
\end{table}
%
%generate_criticalPoint_switch_allTopologies_relError.gnu
\begin{figure}
\centering
%generate_criticalPoint_switch_allTopologies_relError.gnu
\subfigure[Relative Error of MTTF approximation.]
{\includegraphics[width=0.48\textwidth]{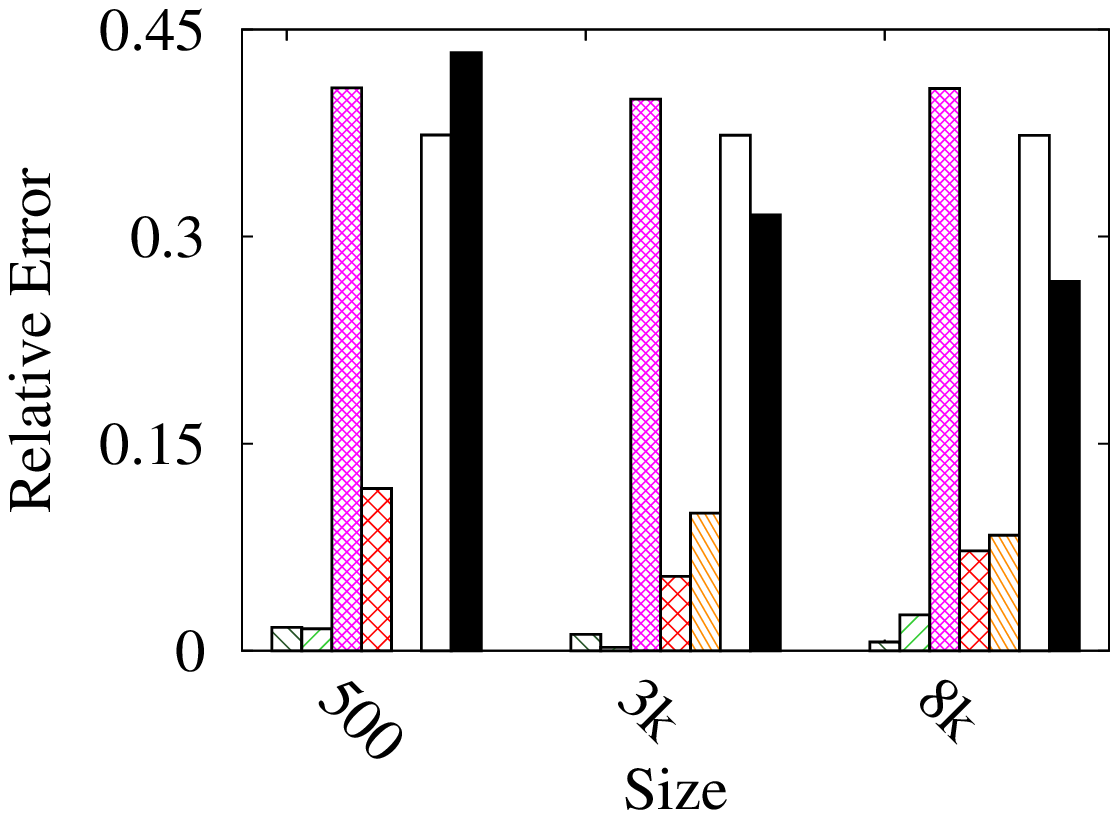}
\label{fig:switch_mttf_relError}}
%gnuplot generate_criticalPoint_switch_3k_unitMTTF.gnu
\subfigure[Elapsed Time and FER simulation for \mbox{3k-server} DCNs.]
{\includegraphics[width=0.36\textwidth]{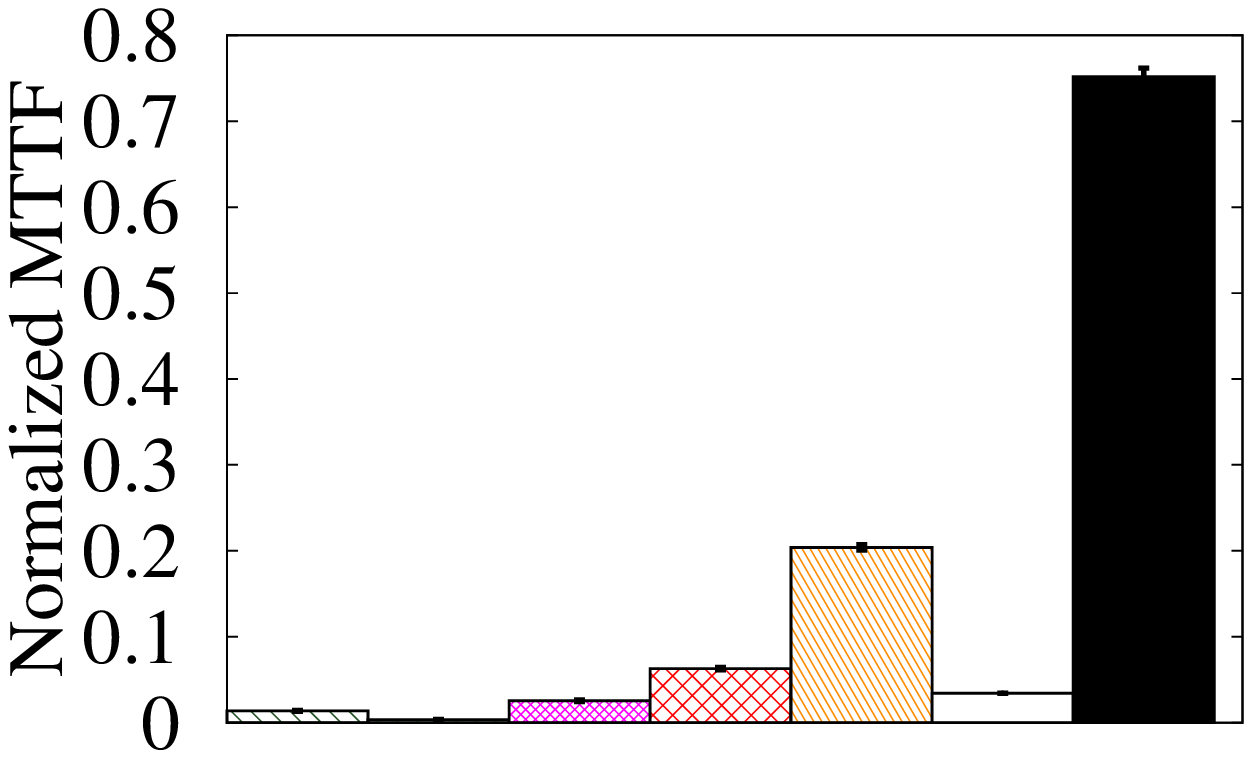}
%gnuplot generate_criticalPoint_switch_3k_criticalPoint.gnu
\quad
\includegraphics[width=0.36\textwidth]{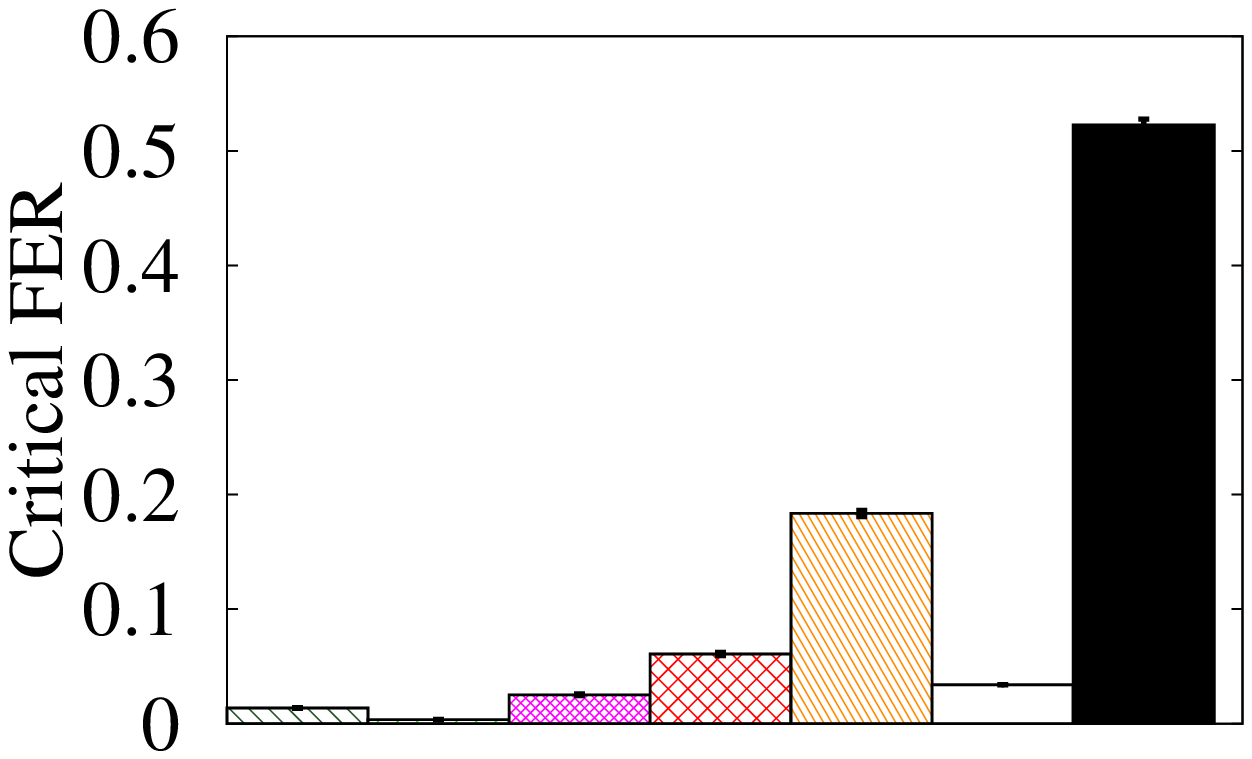}
\label{fig:switch_mttf_unitmttf}
}
{\includegraphics[width=0.8\textwidth]{legendAllCritical.eps}}
\caption{Reliable Phase analysis for switch failures.}
\end{figure}

Using Table~\ref{tab:cuts_switch} values in Equation~\ref{eq:mttfApprox}, we evaluate the theoretical MTTF for switch failures.
We compare these values with simulations using the same methodology as before, resulting in the RE shown in Figure~\ref{fig:switch_mttf_relError}. 
As the figure shows, the min-cut approximation is not well suited for switch failures in some topologies. The topologies that perform well for all network sizes of Table~\ref{tab:configurations} are Three-layer, Fat-tree, BCube5, and BCube3. 
The error for BCube2 is close to 40\%. The results for DCell show a bad approximation, since the minimum RE achieved was 27\%. 
However, we can write the exact MTTF for DCell2, since a failure in any two switches is enough to form a server island, as seen in Figure~\ref{fig:dcell}. 
Its MTTF is thus the time needed to have 2 switch failures, produced by doing $f=2$ and $F=n+1$ (i.e., total number of switches) in Equation~\ref{eq:absoluteTime}, 
and writing the number of switch ports as a function of the number of servers\footnote{The number of switch ports $n$ in function of $|\mathcal{S}|$ is evaluated by solving the equation $|\mathcal{S}|=n(n+1)$.} as $n=0.5(-1 + \sqrt{4|\mathcal{S}|+1})$: 
\begin{equation}
MTTF_{dcell} =  \frac{E[\tau] \sqrt{4|\mathcal{S}|+1}}{|\mathcal{S}|}, \text{ for } l=1.
\label{mttf_dcell2_switch}
\end{equation}
Based on the above analysis of RE, we have a low relative error when using the Burtin-Pittel approximation to estimate MTTF for Three-layer, Fat-tree, and BCube for $l > 1$. We can then write their MTTF using the following equations:
\begin{equation}
MTTF_{threeLayer} \approx \frac{E[\tau]n_e}{|\mathcal{S}|};
\label{mttf_conventional_switch}
\end{equation}
\begin{equation}
MTTF_{fatTree} \approx \frac{E[\tau]}{\sqrt[3]{2|\mathcal{S}|^2}};
\label{mttf_fatTree_switch}
\end{equation}
\begin{equation}
MTTF_{bcube} \approx \frac{E[\tau]}{l+1}\sqrt[l+1]{\frac{1}{|\mathcal{S}|}} \Gamma \left (  \frac{1}{l+1} \right ), \text{ for } l > 1 .
\end{equation}

Figure~\ref{fig:switch_mttf_unitmttf} shows the simulation of the Reliable Phase considering a 3k-server network. Since we do not have MTTF equations for all topology configurations, we compare the topologies using these results. It is important to note that this same comparison holds for the network with sizes 500 and 8k. In summary, we conclude that:
\begin{itemize}
\item \textit{Three-layer and Fat-tree Performance.} Three-layer and Fat-tree have very low reliability compared with other topologies, because a single failure in an edge switch disconnects the network. The MTTF of Fat-tree for 3k-server topologies is approximately $7.3$ times lower than that of BCube2, which is the server-centric topology with the lowest MTTF.
\item \textit{BCube and DCell Performance.} The number of server interfaces increases the MTTF, as in the case of link failures. Also, DCell has a higher reliability than BCube. This is due to less dependence on switches in DCell, as in DCell, each server is connected to 1 switch and $l$ servers while on BCube, only switches are attached to the servers. Although the performance of DCell2 is close to BCube2, the MTTF and Critical FER are much higher in DCell3 than in BCube3. The results show that, for 3k-server topologies, DCell3 is still fully connected when 50\% of the switches are down, and its MTTF is 12 times higher than that of BCube3.
\end{itemize}

\section{Survival Phase}
\label{sec:survival} 

After the first server disconnection, if no repair is done, the DC enters a phase that we call the Survival Phase, during which it can operate with some inaccessible servers. In this phase, we would like to analyze other performance metrics, such as the path length, which is affected by failures. This can be seen as a survivability measurement of the DCN, defined here as the DCN performance after experiencing failures in its elements~\cite{liew1994framework}.

We evaluate the survivability using the performance metrics for a given FER and Elapsed Time that corresponds to the Survival Phase. For example, we can measure the expected number of connected servers when 10\% of the links are not working. Also, we can measure this same metric after 1 month of DC operation. The survivability is evaluated by simulation using the methodology of Section~\ref{sec:simulationOverview}. The metrics used in the evaluation are detailed next.
\subsection{Metrics}
\label{sec:survivalTime} 
\subsubsection{Service Reachability}
The Service Reachability quantifies at what level DC servers are reachable to perform the desired tasks, by evaluating the number of accessible servers and their connectivity.
This measure is important to quantify the DC processing power, as it depends on the number of accessible servers.
Also, it can represent the DC capacity to store VMs in a cloud computing environment.
The Service Reachability can be measured by the following two metrics:

\textbf{Accessible Server Ratio (ASR).} This metric is the ratio between the number of accessible servers and the total number of servers of the original network, considering the current state of the network (i.e., a given FER). The ASR is defined by
\begin{equation}
ASR=\frac{\sum_{k \in \mathcal{A}}s_k}{|\mathcal{S}|},
\end{equation}
where $s_k$ and $|\mathcal{S}|$ are, respectively, the number of servers on the $k$ accessible subnetwork ($k \in \mathcal{A}$) and on the original network (i.e., without failures). The set of accessible subnetworks is given by $\mathcal{A}$. The ASR metric is based on the metric proposed in~\cite{albert2000error} to evaluate the robustness of complex networks. In that work, the robustness is measured as the fraction of the total nodes that after a random failure remains on the subnetwork with the largest number of nodes. However, their metric is not suitable for DCNs since we must take into account the existence of gateways and the existence of multiple operational subnetworks, as highlighted in Section~\ref{sec:operational}. 

\textbf{Server Connectivity (SC).} The ASR is important to quantify how many servers are still accessible in the network. Nevertheless, this metric, when used alone, does not represent the actual DC parallel processing capacity or redundancy. Accessible servers are not necessarily interconnected inside the DC. For example, a network with 100 accessible servers in 2 isolated subnetworks of 50 servers each performs better when executing a parallel task than a network with 100 accessible servers in 100 isolated subnetworks. As a consequence, we enrich the ASR metric with the notion of connectivity between servers. This connectivity is measured by evaluating the density of an auxiliary undirected simple graph, where the nodes are the accessible servers (i.e., servers that still have a path to a gateway) and an edge between two nodes indicates that they can communicate with each other inside the DC. Note that the edges of this graph that represent the reachability between servers are not related to the 
physical links. In other words, the 
proposed metric is the density of the graph of logical links between accessible servers. The density of an undirected simple graph with $\mathcal{|E|}$ edges and $S_a$ nodes is defined as~(\cite{coleman1983estimation}):
\begin{equation}
\frac{2\mathcal{|E|}}{S_a(S_a-1)}.
\label{eq:densityGraph}
\end{equation}
In our case, $\mathcal{|E|}$ is the number of logical links and $S_a=\sum_{k \in \mathcal{A}}s_k$ is the number of accessible servers. Note that in a network without failures the density is equal to 1 because every server can communicate with each other. In addition, a network with failures presenting only one accessible subnetwork also has this density equal to 1. The above evaluation can be simplified using the fact that, after a failure, the graph of logical links in each isolated subnetwork is a complete graph. Also, as subnetworks are isolated from each other, the value $\mathcal{|E|}$ is the sum of the number of edges of each subnetwork. As the subnetwork is a complete graph, it has $\frac{s_k(s_k-1)}{2}$ edges (i.e., pairs of accessible servers). Hence, we replace the value $\mathcal{|E|}$ of Equation~\ref{eq:densityGraph} according to the above reasoning, and define SC as:
\begin{equation}
SC= \begin{cases} 
	\frac{\sum_{k \in \mathcal{A}}s_k(s_k-1)}{S_a(S_a-1)}, &\text{ if } S_a > 1;\\
	0, &\text{ otherwise. }
    \end{cases}
\end{equation}

Our SC metric is similar to the A2TR (Average Two Terminal Reliability)~\cite{neumayer2010Network}. The A2TR is defined as the probability that a random chosen pair of nodes is connected in a network, and is also computed as the density of a graph of logical links. However, SC differs from A2TR since in our metric, we consider only the accessible servers, while A2TR considers any node. Hence, if applied in our scenario, A2TR would consider switches, accessible servers, and inaccessible servers.
\subsubsection{Path quality}
\label{sec:pathQuality}
We measure the Path Quality by evaluating the shortest paths of each topology.
The shortest path length is suitable to evaluate the behavior of the quality of paths in the network, since it is the basis of novel routing mechanisms that can be used in DCs, such as TRILL~\cite{rfc5556}, IEEE~802.1aq~\cite{spb}, and SPAIN~\cite{mudigonda2010spain}.
Hence, we define the following metric:

\textbf{Average Shortest Path Length.} This metric is the average of the shortest path lengths between the servers in the network. Note that in this analysis we do not consider paths between servers of different isolated subnetworks, since they do not have a path between them. The Average Shortest Path Length captures the latency increase caused by failures.
\subsection{Results}
\label{sec:survivalTimeResults} 

\begin{figure}
\centering
\includegraphics[width=0.48\textwidth]{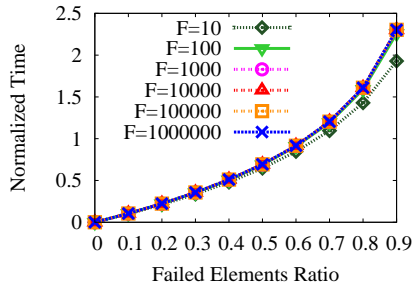}
\caption{Normalized Time as a function of the Failed Elements Ratio.}
\label{fig:FER_elapsedTime}
\end{figure}

As stated in Section~\ref{sec:simulationOverview}, failures can be characterized by using the FER and Elapsed Time. The FER does not depend on the probability distribution of the element lifetime, while the Elapsed Time assumes an exponential probability distribution. Due to space constraints, most of the results in this section are shown as a function of the FER, since they do not depend on the probability distribution. However, the survivability comparison between the topologies using the FER produces the same conclusions if we use the Elapsed Time. This is because, using Equation~\ref{eq:normalizedTime}, the Normalized Time for a given FER is almost independent on the total number of elements $F$, being agnostic to a specific topology and failure type.
For example, we use Equation~\ref{eq:normalizedTime} to plot in Figure~\ref{fig:FER_elapsedTime} the Normalized Time as a function of the FER (i.e $\frac{f}{F}$), for different total number of elements $F$ (e.g., total number of links). This figure shows that, for a large range of the FER, the relationship between Normalized Time and the FER is independent of $F$.

As done in Section~\ref{sec:reliableTimeResults}, we compare topologies that have approximately the same number of servers. For the sake of conciseness, the results are provided for the $3k$-server topologies detailed in Table~\ref{tab:configurations}. On the other hand, we observe that this number is sufficiently large to disclose the differences between the investigated topologies. Furthermore, as these topologies have a regular structure, our conclusions can be extrapolated to a higher number of servers~\cite{bilal2013Characterization}.
Finally, in this phase we provide results for a large range of the FER (i.e., from 0 to 0.4). Although this high failure ratio could be unrealistic for traditional data centers, we choose to use this range 
to provide a generic analysis, suitable for different novel scenarios. For example, Modular Data Centers present some challenges regarding their maintenance, which could make the DC operator wait for a high number of element failures before repairing the network~\cite{guo2009bcube}.
%
%gnuplot generate_NTvsFER.gnu
\subsubsection{Link Failures}
\label{sec:perfEvaluationLinkSurvival}

\begin{figure}
\centering
%generated by generate_3k_failureRatio_link_asr.gnu
\subfigure[Accessible Server Ratio.]
{\includegraphics[width=0.48\textwidth]{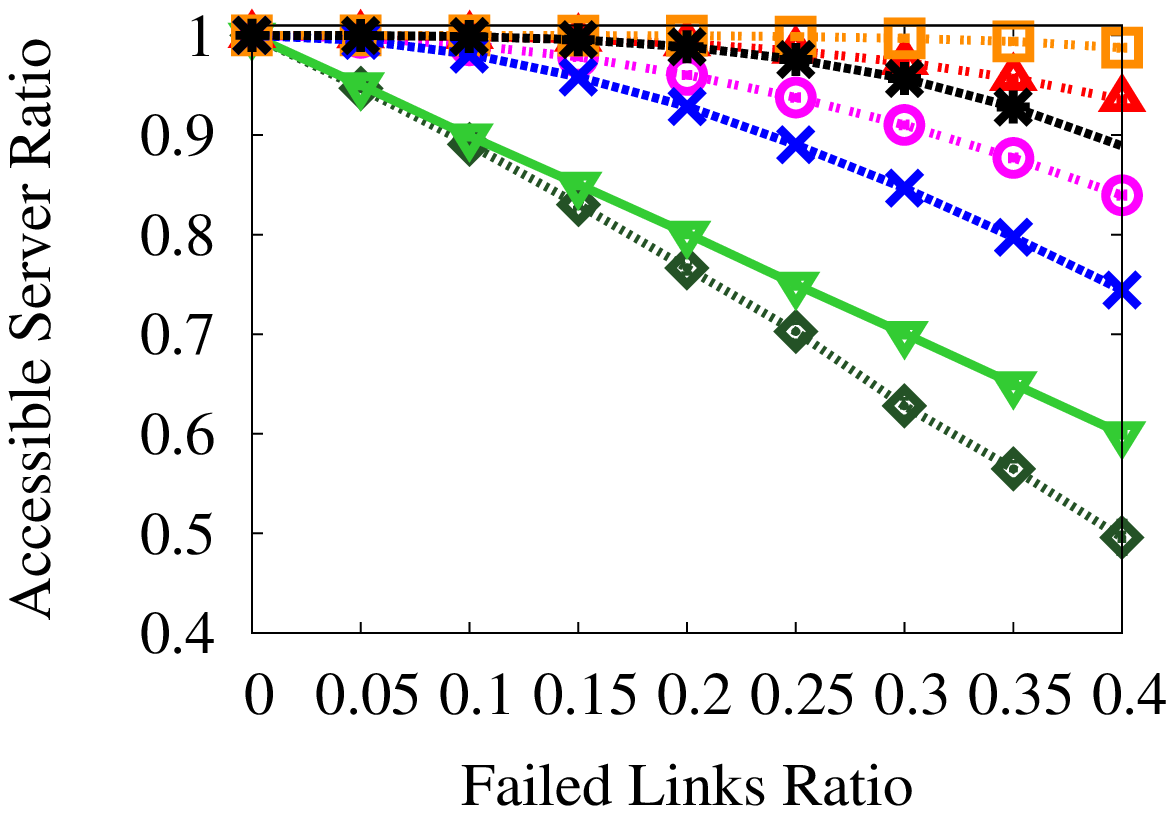}
\label{fig:linkAsr_fer}}
%generate_3k_failureRatio_link_sc.gnu
\subfigure[Server Connectivity.]
{\includegraphics[width=0.48\textwidth]{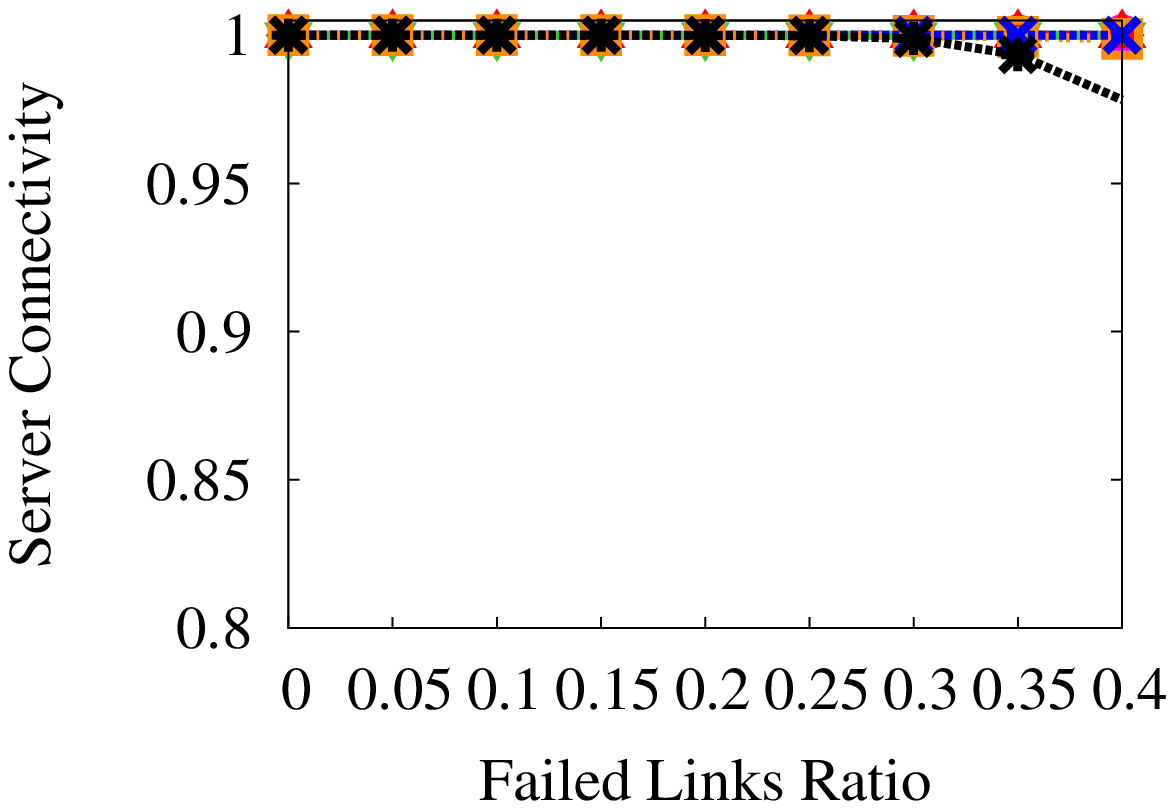}
\label{fig:link_sc_fer}}
%generate_3k_failureRatio_link_avgPathLength.gnu
\subfigure[Average Shortest Path Length.]
{\includegraphics[width=0.48\textwidth]{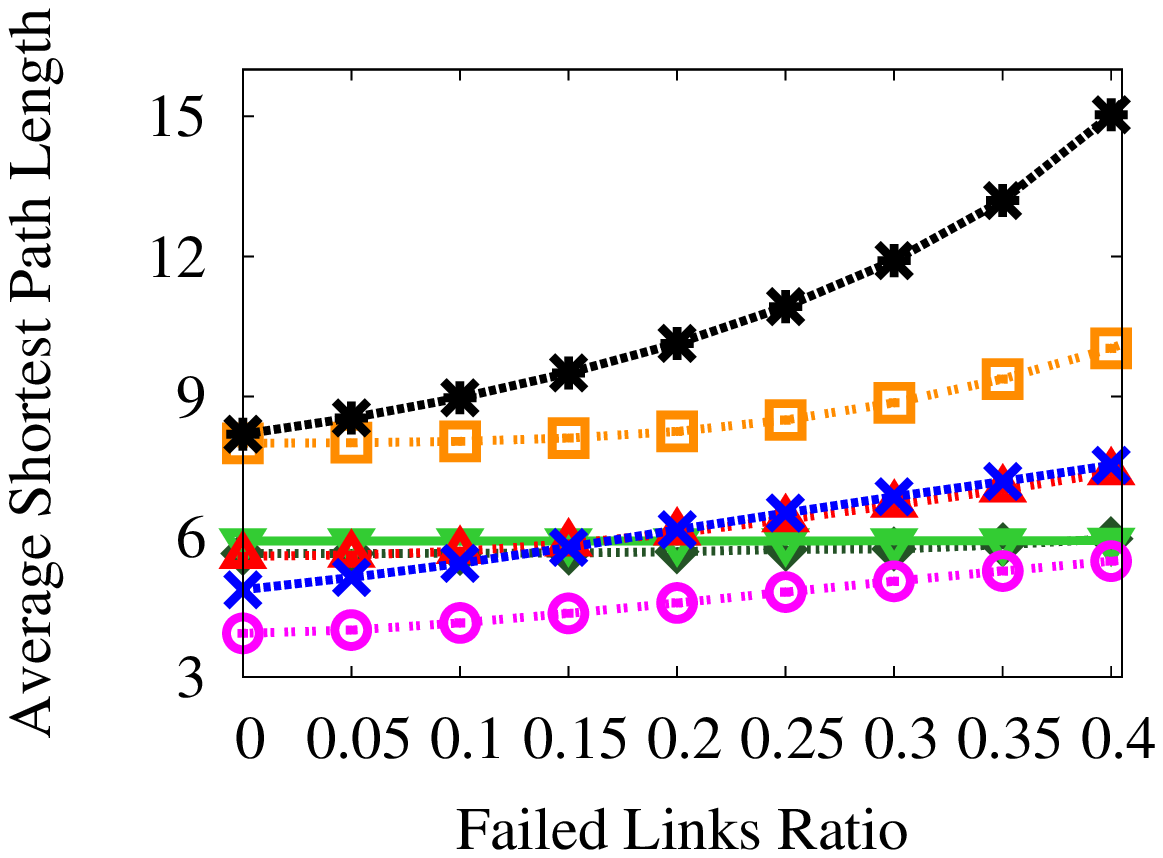}
\label{fig:link_avgPathLength_fer}}
%generate_3k_time_link_asr
\subfigure[Accessible Server Ratio along the time.]
{\includegraphics[width=0.48\textwidth]{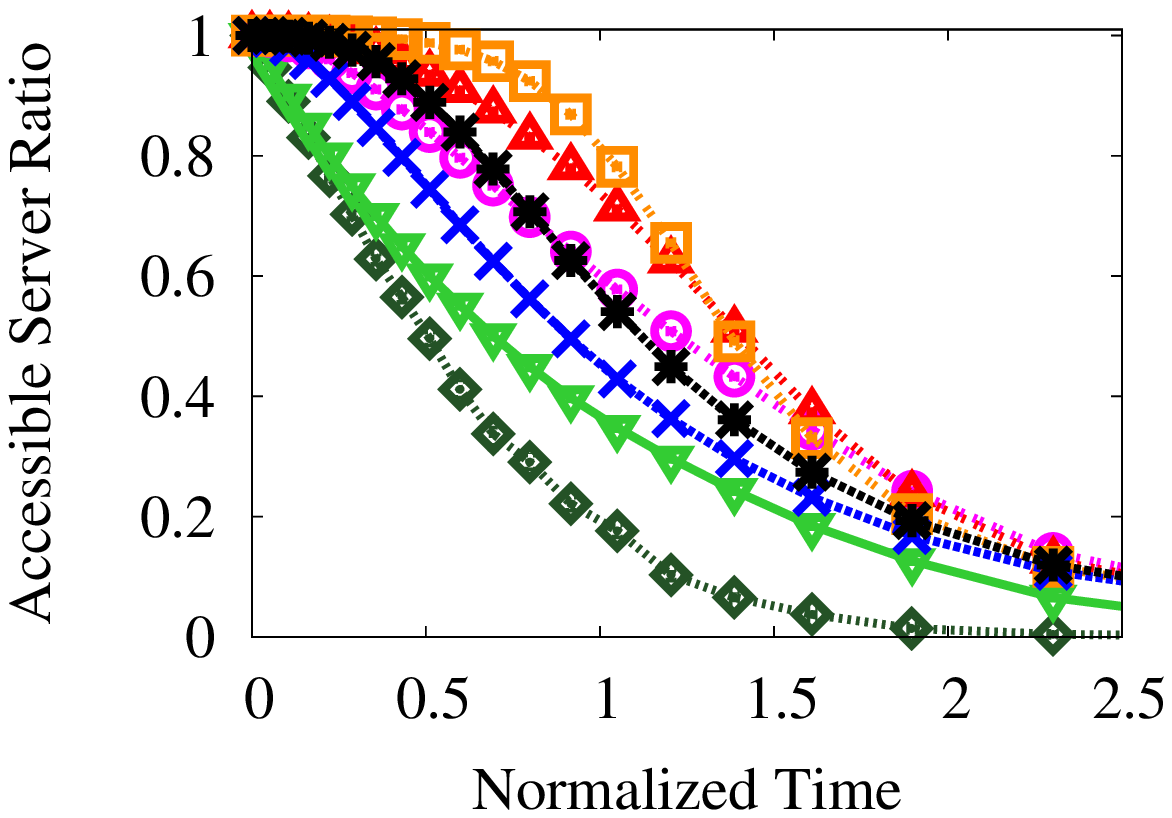}
\label{fig:link_asr_time}}
{\includegraphics[width=0.8\textwidth]{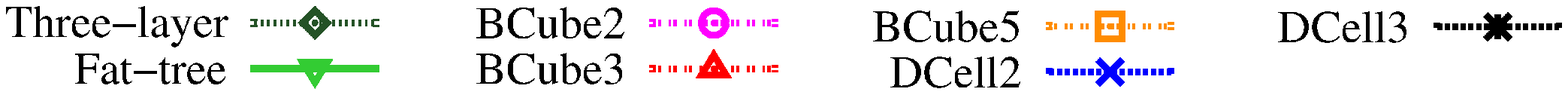}}
\caption{Survival Phase analysis for link failures.}
\end{figure}

Figures~\ref{fig:linkAsr_fer}~and~\ref{fig:link_sc_fer} plot, respectively, the ASR and SC as a function of the FER. We observe that:

\begin{itemize}
\item \textit{Three-layer and Fat-tree Performance.} Three-layer and Fat-tree have the worst performance values because the servers are attached to the edge switches using only one link. Hence, the failure of this link totally disconnects the server. In opposition, server-centric topologies have a slower decay on ASR since servers have redundant links. The results for Fat-tree show that a given Failed Links Ratio corresponds to a reduction in ASR by the same ratio (e.g., a FER of $0.3$ produces an ASR of $0.7$), showing a fast decay in Service Reachability. As Three-layer has a less redundant core and aggregation layers than Fat-tree, its ASR tends to decay faster than in the case of Fat-tree. As an example, Table~\ref{tab:configurations} shows that for a network with 3k servers, Fat-tree has almost three times the number of links than the Three-layer topology.
\item \textit{BCube and DCell Performance.} For the same type of server-centric topology, the survivability can be improved by increasing the number of network interfaces per server. As servers have more interfaces, their disconnection by link failures will be harder and thus a given FER will disconnect less servers. For example, considering a FER of $0.4$, the ASR is improved by $11\%$ in BCube and by $19\%$ in DCell if we increase the number of server interfaces from two to three. For the same number of server interfaces, the survivability of BCube is better than DCell. For instance, BCube maintains at least an ASR of $0.84$ when 40\% of its link are down, while in DCell this lower bound is $0.74$.
In DCell, each server is connected to 1 switch and $l$ servers, while in BCube the servers are connected to $l+1$ switches. As a switch has more network interfaces than a server, link failures tend to disconnect less switches than servers. Consequently, the servers in BCube are harder to disconnect from the network than in DCell. Obviously, the better survivability comes at the price that BCube uses more wiring and switches than DCell~\cite{guo2009bcube}.
\item \textit{General Remark.} For all topologies, the SC is very close to 1, meaning that link failures produce approximately only one subnetwork.
\end{itemize}

Figure~\ref{fig:link_avgPathLength_fer} shows the Average Shortest Path Length as a function of the FER. We can draw the following remarks:
\begin{itemize}
\item \textit{Three-layer and Fat-tree Performance.} Three-layer and Fat-tree keep their original length independent of the FER, showing a better Path Quality than other topologies as the FER increases.
\item \textit{BCube and DCell Performance.} The path length of server-centric topologies increases with the FER. BCube maintains a lower Average Shortest Path Length than DCell, by comparing configurations with the same number of server interfaces. Moreover, for a high FER ($0.4$) DCell has an increase of up to 7 hops in Average Shortest Path Length, while in BCube, the maximum increase is 2 hops. Also, for a given topology, the Average Shortest Path Length is greater when it has more server interfaces, even when there are no failures. As more server interfaces imply more levels in BCube and DCell, the paths contain nodes belonging to more levels and thus have a greater length.
\end{itemize}

Analyzing the above results, we observe a tradeoff between Service Reachability and Path Quality. On the one hand, the higher the number of server interfaces, the better the network survivability regarding the number of accessible servers. On the other hand, the higher the number of server interfaces, the higher the Average Shortest Path Length. Hence, \textit{increasing the Service Reachability by adding server interfaces implies a more relaxed requirement on the Path Quality}. 

Figure~\ref{fig:link_asr_time} illustrates how the survivability evolves in time, by plotting ASR as a function of the Normalized Time.
This is the same experiment shown in Figure~\ref{fig:linkAsr_fer}, but using the $X$-axis as given by Equation~\ref{eq:normalizedTime}, instead of $\frac{f}{F}$. 
Note that although Figure~\ref{fig:linkAsr_fer} shows the ASR up to a Failed Links Ratio of $0.4$, the last experimental point in Figure~\ref{fig:link_asr_time} is $2.3$, which corresponds approximately to a Failed Links Ratio of $0.9$.
The Normalized Time gives an idea of how the survivability is related to the individual lifetime of a single element, which is a link in this case. As a consequence, a Normalized Time equal to 1 represents the mean lifetime of a link given by $E[\tau]$. As shown in Figure~\ref{fig:link_asr_time}, most of the topologies present a substantial degradation of ASR when the Elapsed Time is equal to the mean link lifetime (Normalized Time of 1). Also, all topologies have very small reachability when the elapsed time is twice the link lifetime (Normalized Time equal to 2).
\subsubsection{Switch Failures}
\label{sec:perfEvaluationSwitchSurvival}

\begin{figure}
\centering
%generated by generate_3k_failureRatio_switch_asr.gnu
\subfigure[Accessible Server Ratio.]
{\includegraphics[width=0.48\textwidth]{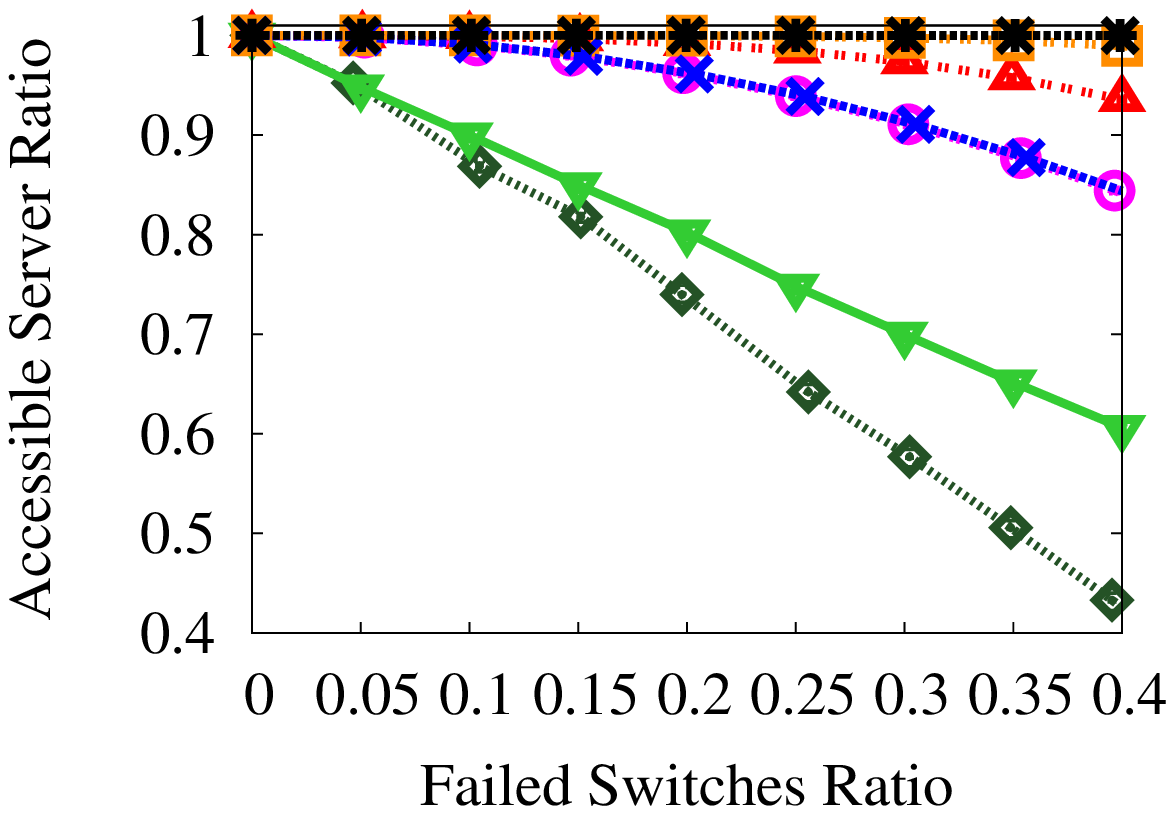}
\label{fig:switchAsr_fer}}
%generate_3k_failureRatio_switch_sc.gnu
\subfigure[Server Connectivity.]
{\includegraphics[width=0.48\textwidth]{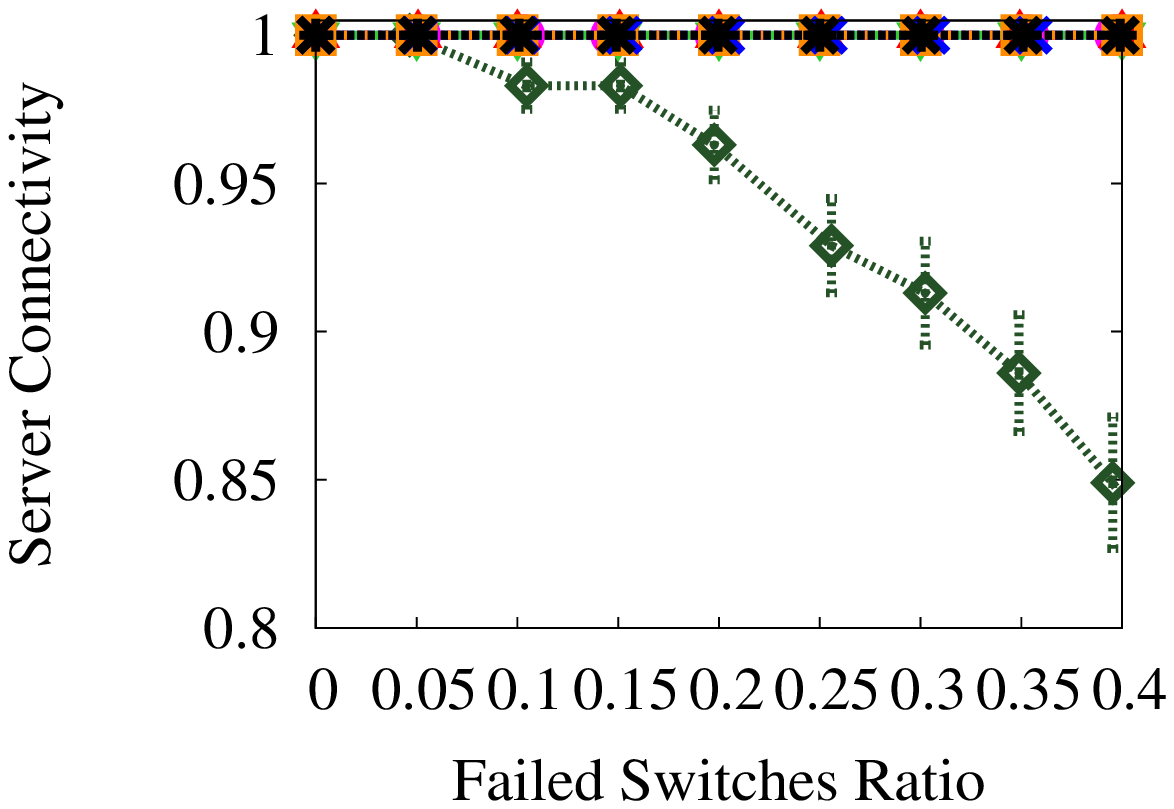}
\label{fig:switch_sc_fer}}
%generate_3k_failureRatio_switch_avgPathLength
\subfigure[Average Shortest Path Length.]
{\includegraphics[width=0.48\textwidth]{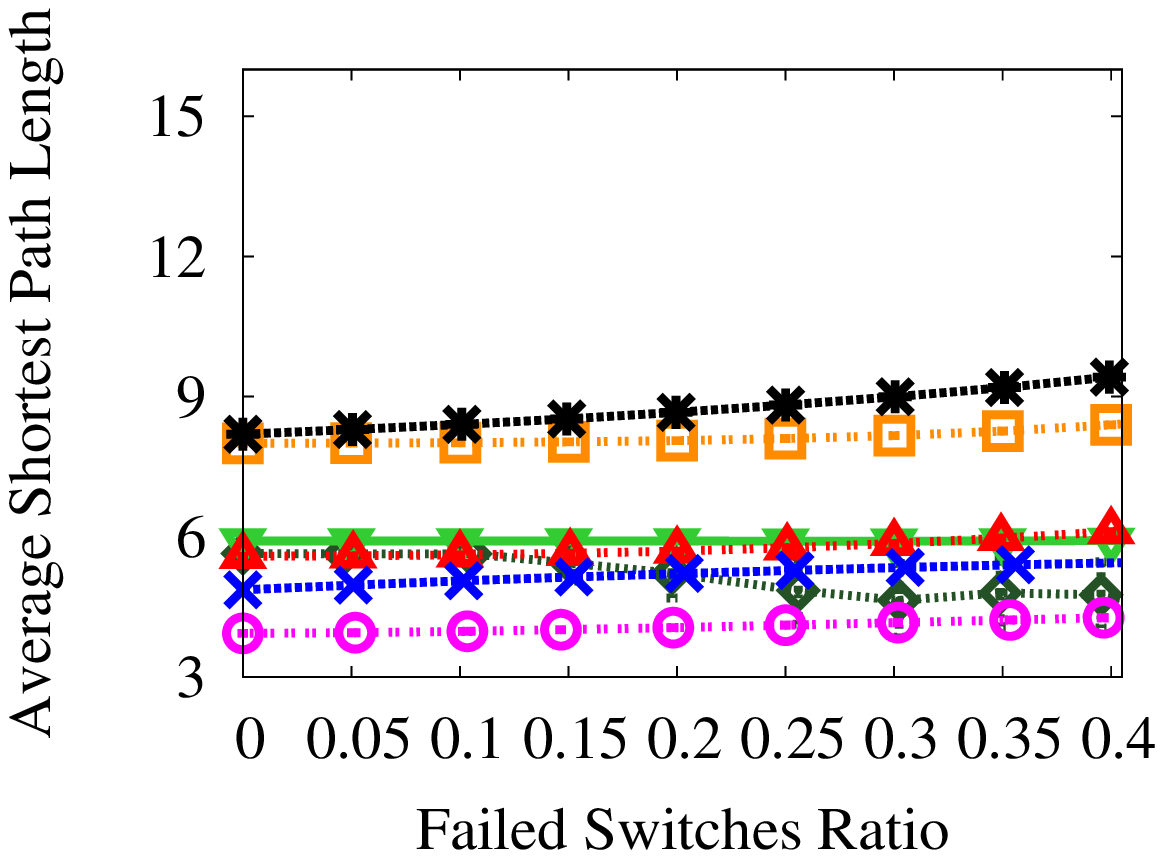}
\label{fig:switch_avgPathLength_fer}}
%generate_3k_time_switch_asr
\subfigure[Accessible Server Ratio along the time.]
{\includegraphics[width=0.48\textwidth]{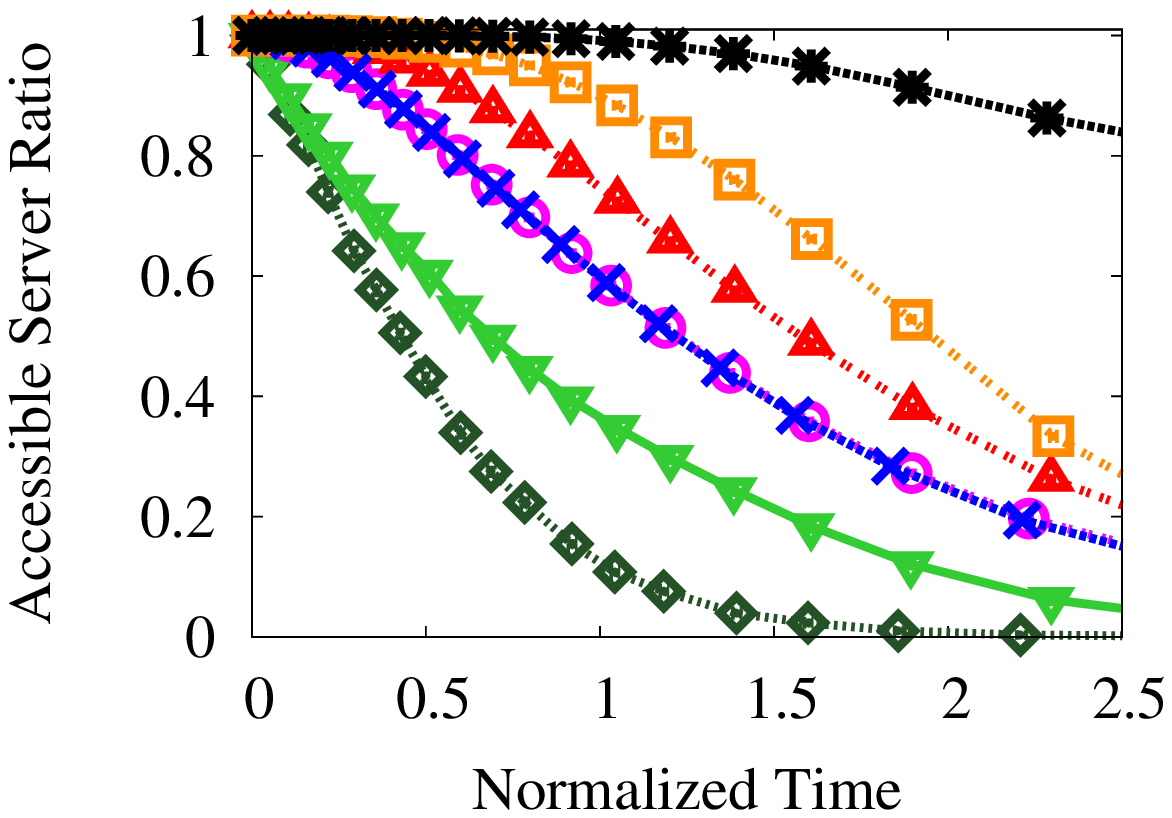}
\label{fig:switch_asr_time}}
{\includegraphics[width=0.8\textwidth]{legendAll.eps}}
\caption{Survival Phase analysis for switch failures.}
\end{figure}

Figures~\ref{fig:switchAsr_fer}~and~\ref{fig:switch_sc_fer} plot, respectively, the ASR and SC according to the Failed Switches Ratio. We observe that:
\begin{itemize}
\item \textit{Three-layer and Fat-tree Performance.} Three-layer and Fat-tree present the worst behavior due to the edge fragility. For Three-layer, a single failure on an edge switch is enough to disconnect 48 servers, which is the number of ports in this switch. For Fat-tree, a single failure on an edge switch disconnects $\frac{n}{2}$ servers, where $n$ is the number of switch ports, as seen in Figure~\ref{fig:fatTree}. Hence, for a 3k-server configuration, Fat-tree loses $\frac{24}{2}=12$ servers for a failure in an edge switch. Note that this number is four times lower than that of a Three-layer topology.
In addition, the Three-layer topology relies on only high-capacity gateways (i.e., core switches) to maintain all DC connectivity, while Fat-Tree has 24 smaller core elements acting as gateways. Although the comparison between Three-layer and Fat-tree is not necessarily fair, since they have a different GPD (Section~\ref{sec:operational}), the results show how much relying on a small number of high-capacity aggregate and core elements can decrease the topology performance.
As in the case of links, for Fat-tree, a given Failed Switches Ratio reduces the ASR by the same ratio, while in Three-layer the performance impact is more severe.
\item \textit{BCube and DCell Performance.} As in the case of link failures, increasing the number of server interfaces increases the survivability to switch failures. Considering a FER of $0.4$ for BCube and DCell, the ASR is increased respectively by $11\%$ and $19\%$ if we increase the number of server interfaces from two to three. In the case of BCube, a higher number of server interfaces represents a higher number of switches connected per server. Consequently, more switch failures are needed to disconnect a server. For DCell, a higher number of server interfaces represents less dependence on switches, as each server is connected to 1 switch and $l$ servers. We can also state that the survivability in DCell3 is slightly greater than in BCube3, showing an ASR $6\%$ higher for a FER of $0.4$, while BCube2 and DCell2 have the same performance. The first result is due to less dependence on switches in DCell, as explained in Section~\ref{sec:perfEvaluationSwitchReliable}. In the particular case of two 
server interfaces, this reasoning is not valid. Considering that the survivability is highly affected by min-cuts, each min-cut in DCell2 disconnects two servers; whereas in BCube2, each min-cut disconnects only one server. On the other hand, each Failed Switches Ratio in BCube2 represents approximately twice the absolute number of failed switches in DCell2. This relationship can be seen in Table~\ref{tab:configurations} where the total number of switches in BCube2 is approximately twice the number of switches in DCell2. For that reason, as the min-cuts have the same size in both topologies (Table~\ref{tab:cuts_switch}), a given Failed Switches Ratio in BCube2 will produce failures in approximately twice the number of min-cuts as in DCell2. Hence, BCube2 has twice the number of affected min-cuts, whereas DCell2 has twice the number of server disconnections per min-cut. Consequently, the number of disconnected servers is approximately the same in both topologies for a given Failed Switches Ratio.
\item \textit{General Remark.} SC is very close to 1 for all topologies, except for Three-layer. For a single experimental round in Three-layer, we can only have two possible SC values. In the first one, at least one gateway (core switch)  is up and we have one accessible subnetwork, and thus $SC=1$. In the second one, the two gateways are down (i.e., randomly chosen to be removed) and thus $SC=0$. As Figure~\ref{fig:switch_sc_fer} plots values averaged over all experimental rounds, the SC measure is simply the percentage of the round that outcomes $SC=1$. As can be seen, the outcome $SC=1$ is more frequent since $SC>0.8$ for the considered FER range. Hence, even in the case of Three-layer which only has 2 gateways, we have a low probability that the network is completely disconnected after the removal of random switches.
\end{itemize}

The results for Average Shortest Path Length on Figure~\ref{fig:switch_avgPathLength_fer} show that, \textit{for all topologies, switch failures do not lead to a significant increase in path length}.

Figure~\ref{fig:switch_asr_time} shows the evolution of ASR as a function of time, considering switch failures. As for link failures, the last experimental point is approximately $2.3$, corresponding to a Failed Switches Ratio of $0.9$. Compared with the results of link failures in Figure~\ref{fig:link_asr_time}, we can see that the topologies degrade slower under switch failures than under link failures, except for the Three-layer topology. Also, we note the high survivability of DCell3, which maintains a high ASR for a long time for switch failures. As stated before, this behavior shows its low dependence on switches.

\subsubsection{Server Failures}
\label{sec:perfEvaluationServerSurvival}

\begin{figure}
\centering
%generated by generate_3k_failureRatio_server_asr.gnu
\subfigure[Accessible Server Ratio.]
{\includegraphics[width=0.48\textwidth]{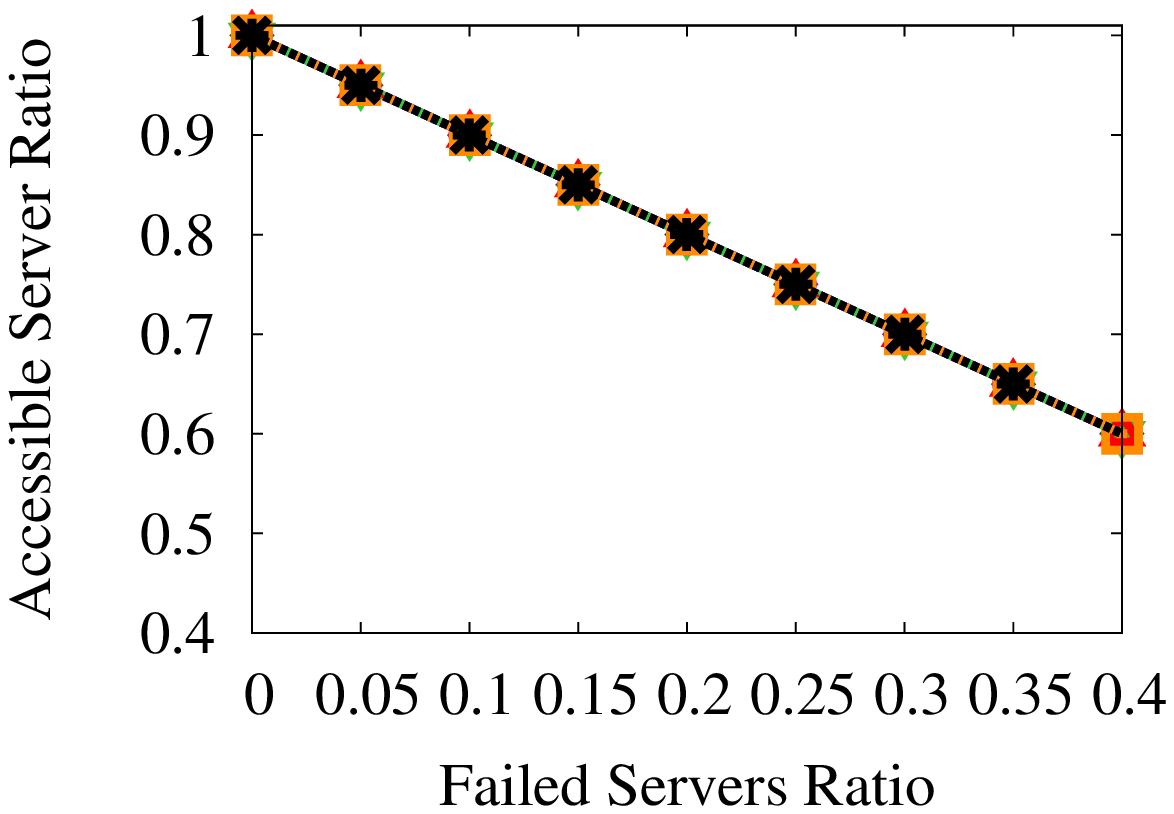}
\label{fig:serverAsr_fer}}
%generate_3k_failureRatio_server_sc.gnu
\subfigure[Server Connectivity.]
{\includegraphics[width=0.48\textwidth]{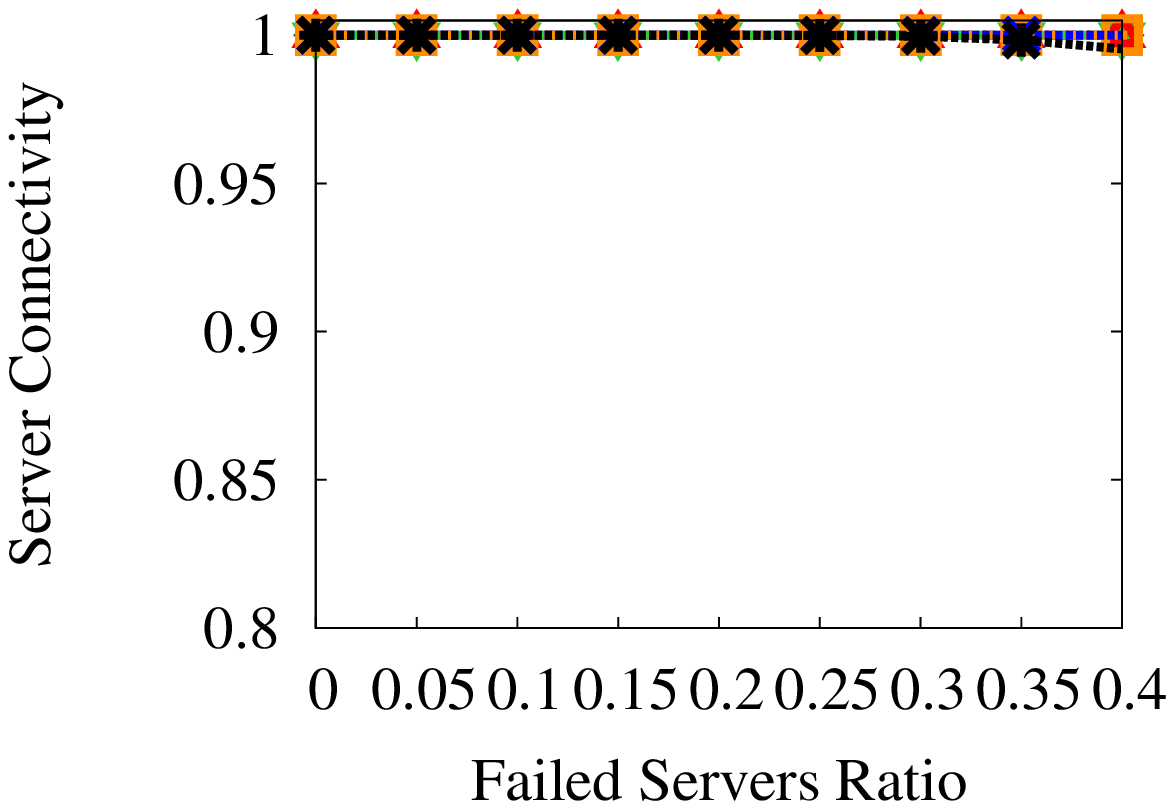}
\label{fig:server_sc_fer}}
%generate_3k_failureRatio_server_avgPathLength
\subfigure[Average Shortest Path Length.]
{\includegraphics[width=0.48\textwidth]{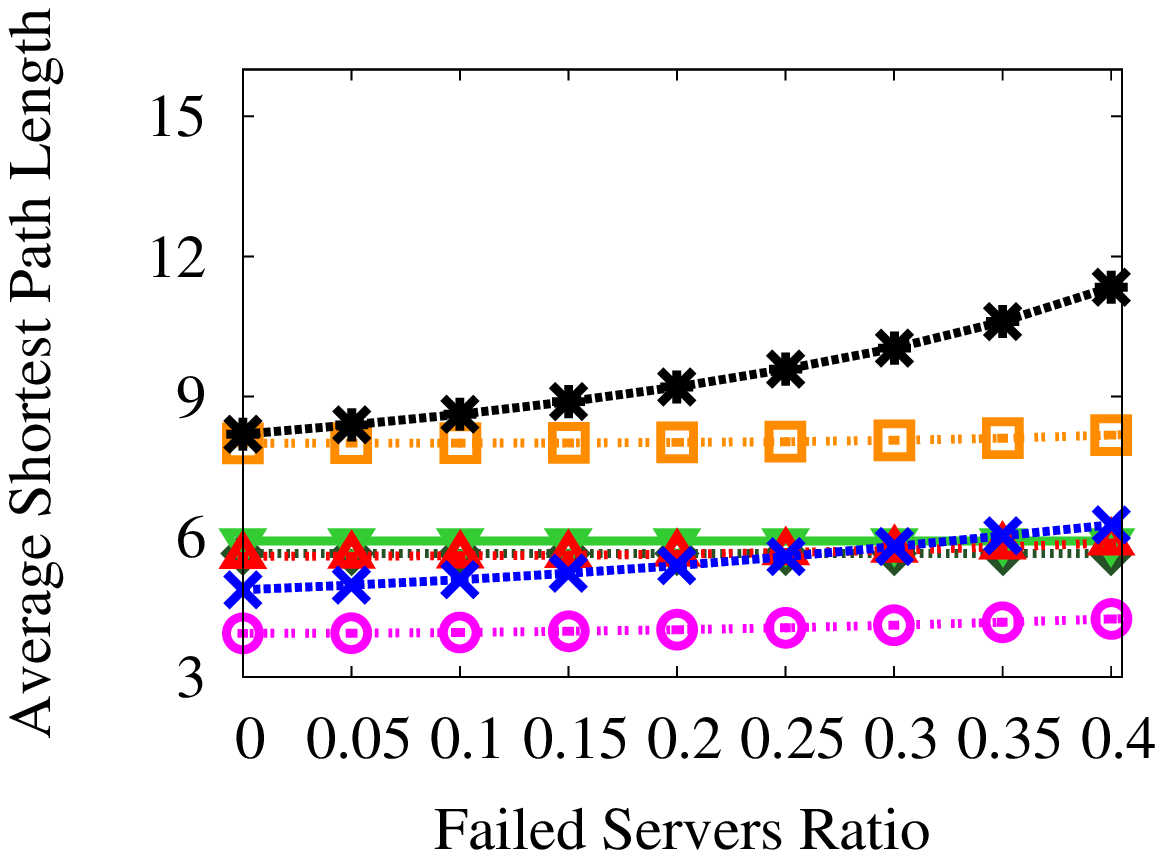}
\label{fig:server_avgPathLength_fer}}
%generate_3k_time_server_asr
\subfigure[Accessible Server Ratio along the time.]
{\includegraphics[width=0.48\textwidth]{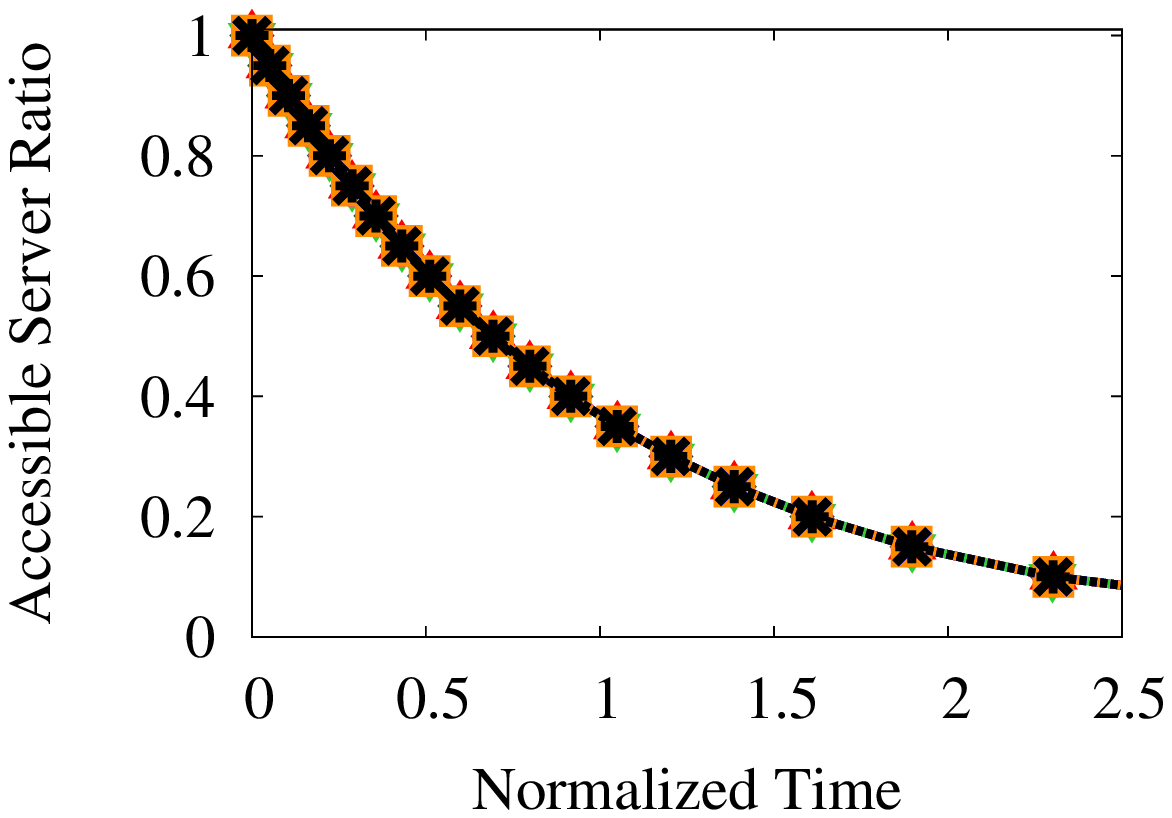}
\label{fig:server_asr_time}}
{\includegraphics[width=0.8\textwidth]{legendAll.eps}}
\caption{Survival Phase analysis for server failures.}
\end{figure}

Figure~\ref{fig:serverAsr_fer} shows that, for all topologies, the ASR decreases linearly with the Failed Servers Ratio. Although BCube and DCell depend on server forwarding, their Service Reachability is equal to that of Fat-tree and Three-layer when servers are removed. It means that \textit{a server failure does not lead to a disconnection of other servers in the network}. For all topologies, the SC is always very close to 1 for the considered range of the Failed Servers Ratio.

Despite the favorable results of Service Reachability under server failures, Figure~\ref{fig:server_avgPathLength_fer} shows that the path length in DCell slightly increases with failures (up to 3 hops for a FER of $0.4$), because DCell is more dependent on server forwarding than the other topologies. 

The evolution of the ASR in time is shown in Figure~\ref{fig:server_asr_time}. This result indicates that the Service Reachability of the remaining servers is not affected by server failures for a long period.

\subsubsection{Link and Switch Failures}
\label{sec:perfEvaluationLinkSwitchSurvival}

In the previous results we isolate each failure type to provide a more accurate comparison between the topologies. However, in a real data center environment, different failure types may coexist.
Hence, in this section we analyze the ASR of each topology by combining both link and switch failures. 
We focus on the ASR metric since, as shown before, it is more affected by failures than the other metrics. In addition, we do not consider server failures because it does not have a significant impact in ASR, as shown in Section~\ref{sec:perfEvaluationServerSurvival}.

\begin{figure}
\centering
\subfigure[Three-layer.]
{\includegraphics[width=0.48\textwidth]{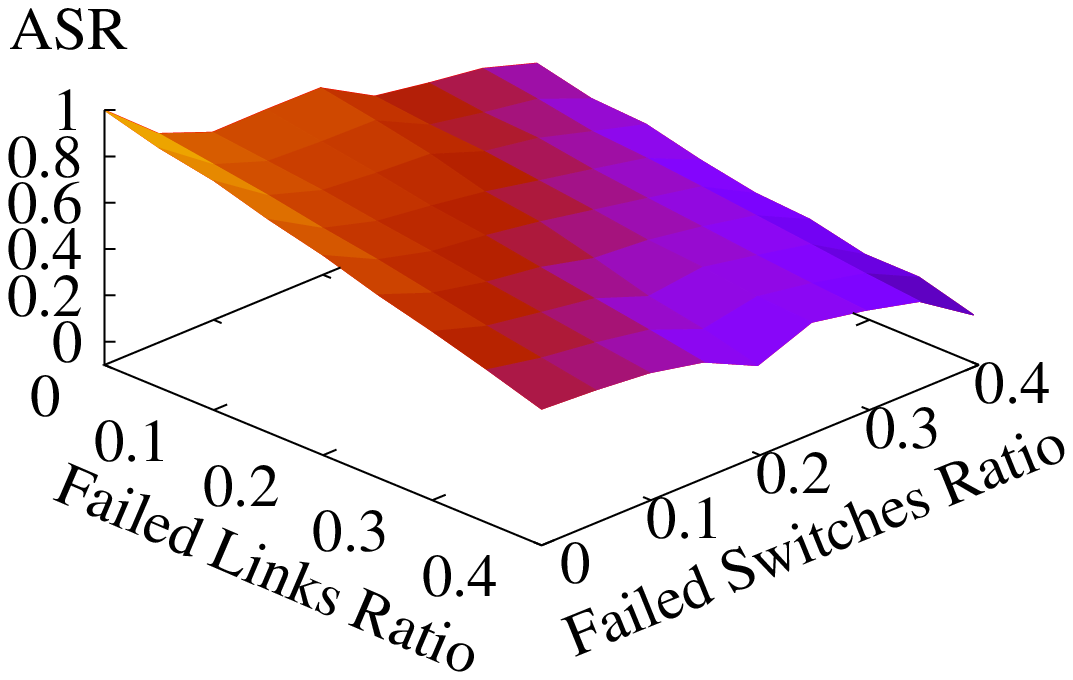}
\label{fig:allRangelinkSwitch_fatTree_24}}
\subfigure[Fat-tree.]
{\includegraphics[width=0.48\textwidth]{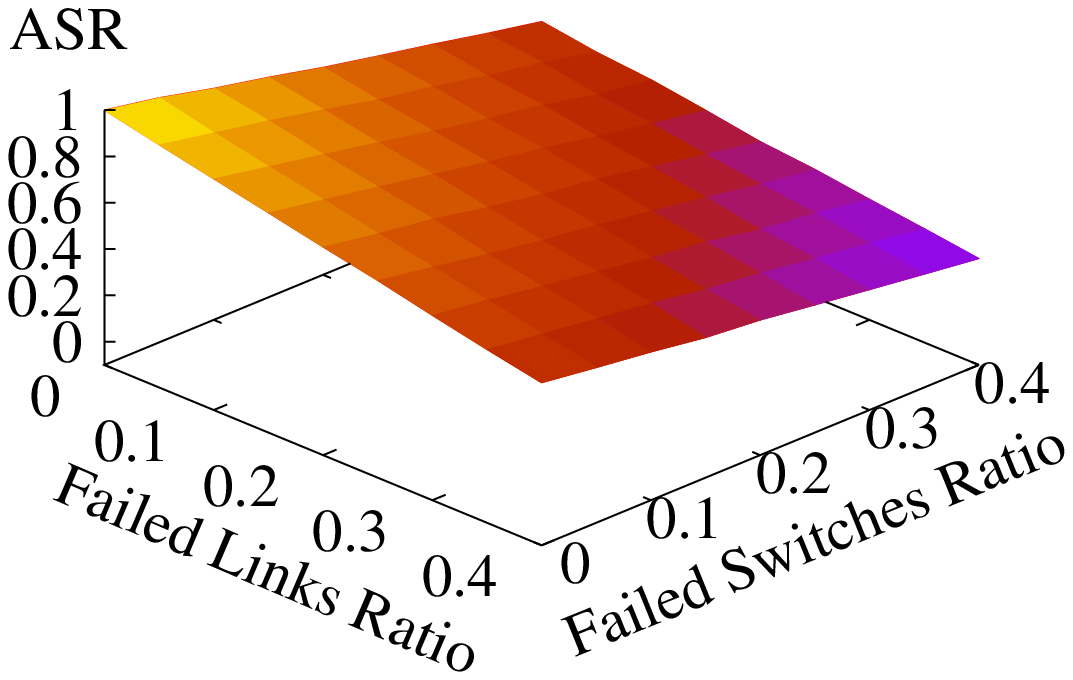}
\label{fig:allRangelinkSwitch_3tier_3456}}
\subfigure[BCube2.]
{\includegraphics[width=0.48\textwidth]{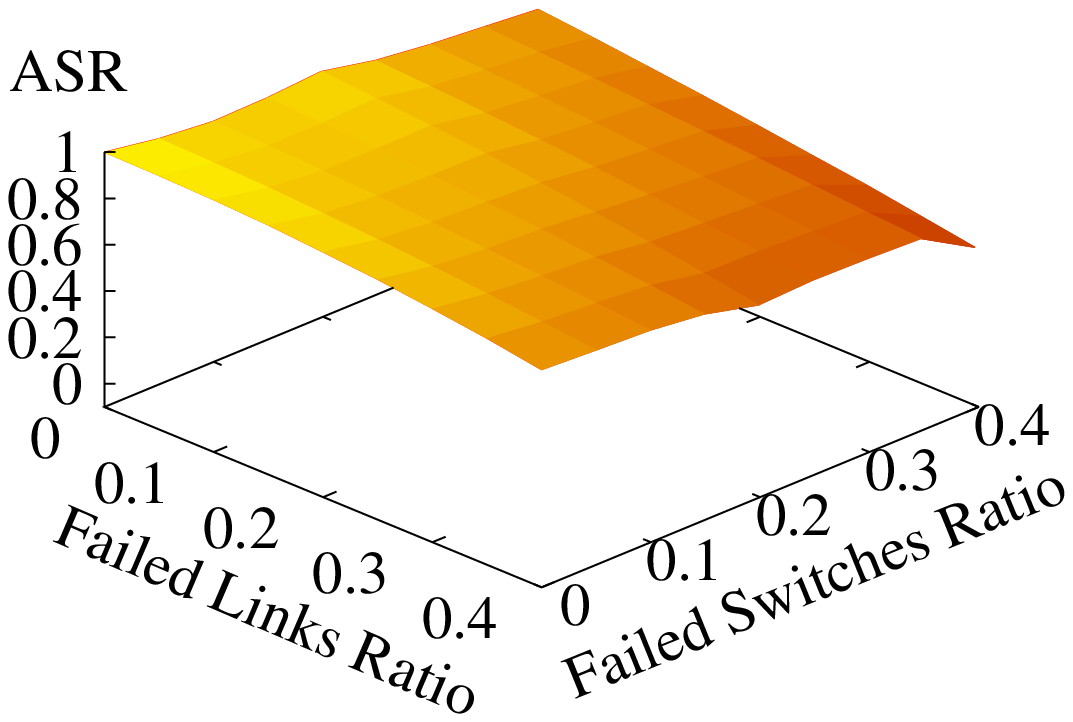}
\label{fig:allRangelinkSwitch_bcube_2_58}}
\subfigure[DCell2.]
{\includegraphics[width=0.48\textwidth]{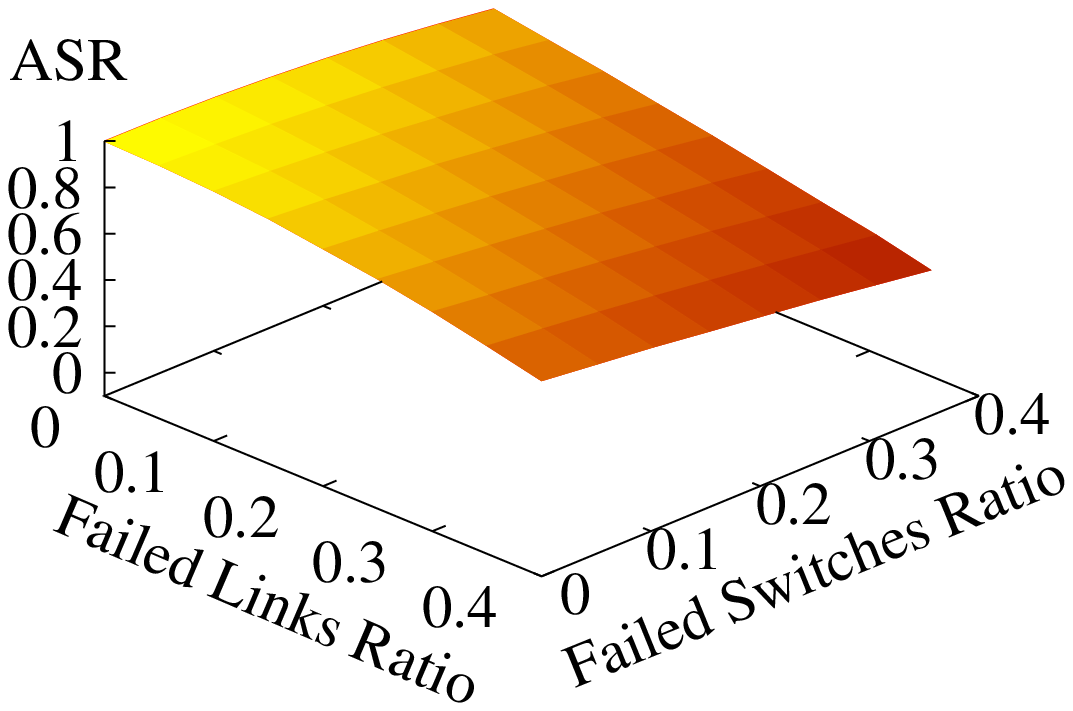}
\label{fig:allRangelinkSwitch_dcell_2_58}}
\subfigure[BCube3.]
{\includegraphics[width=0.48\textwidth]{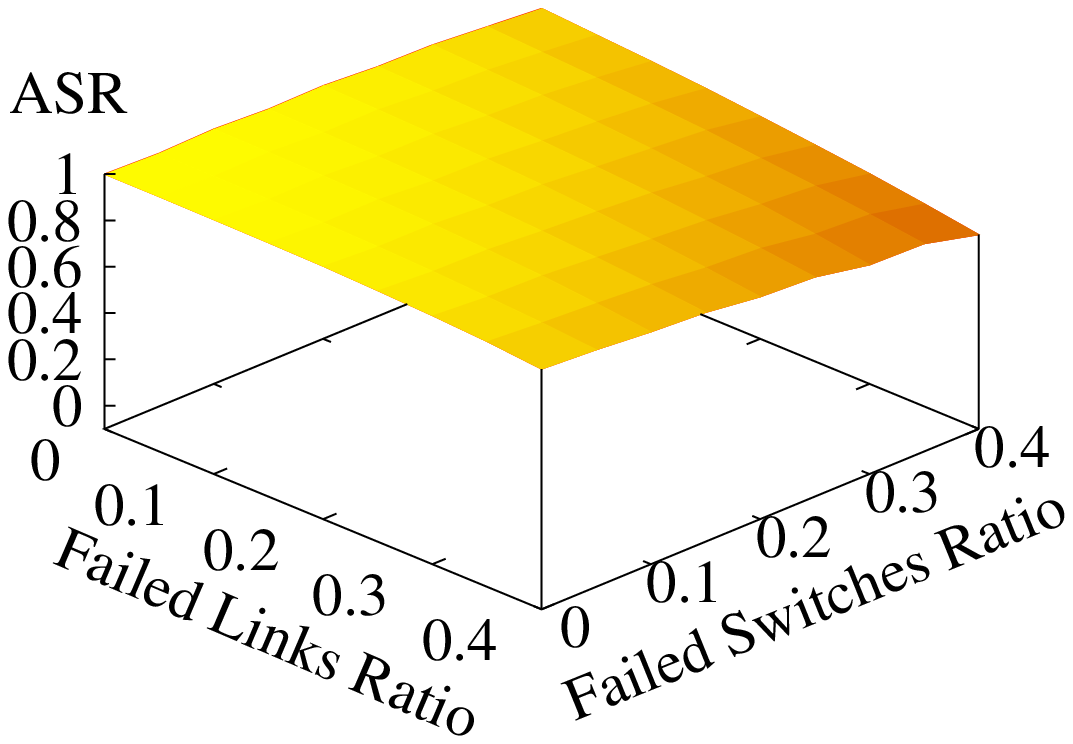}
\label{fig:allRangelinkSwitch_bcube_3_15}}
\subfigure[DCell3.]
{\includegraphics[width=0.48\textwidth]{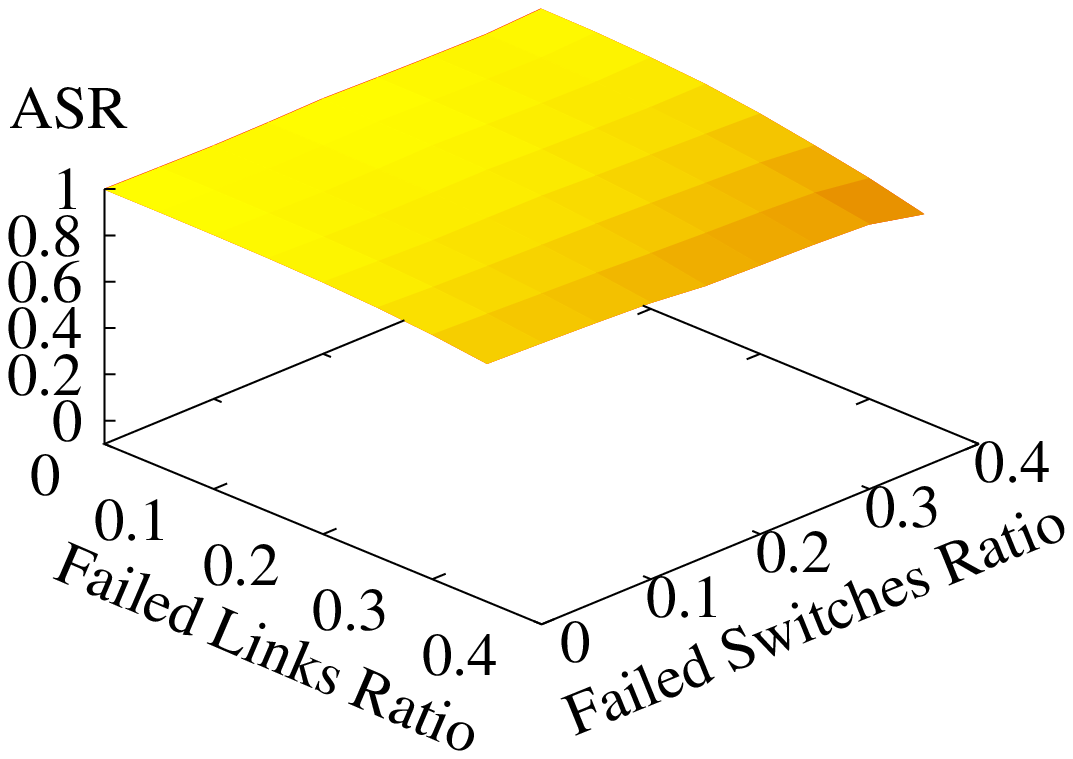}
\label{fig:allRangelinkSwitch_dcell_3_7}}
\subfigure[BCube5.]
{\includegraphics[width=0.48\textwidth]{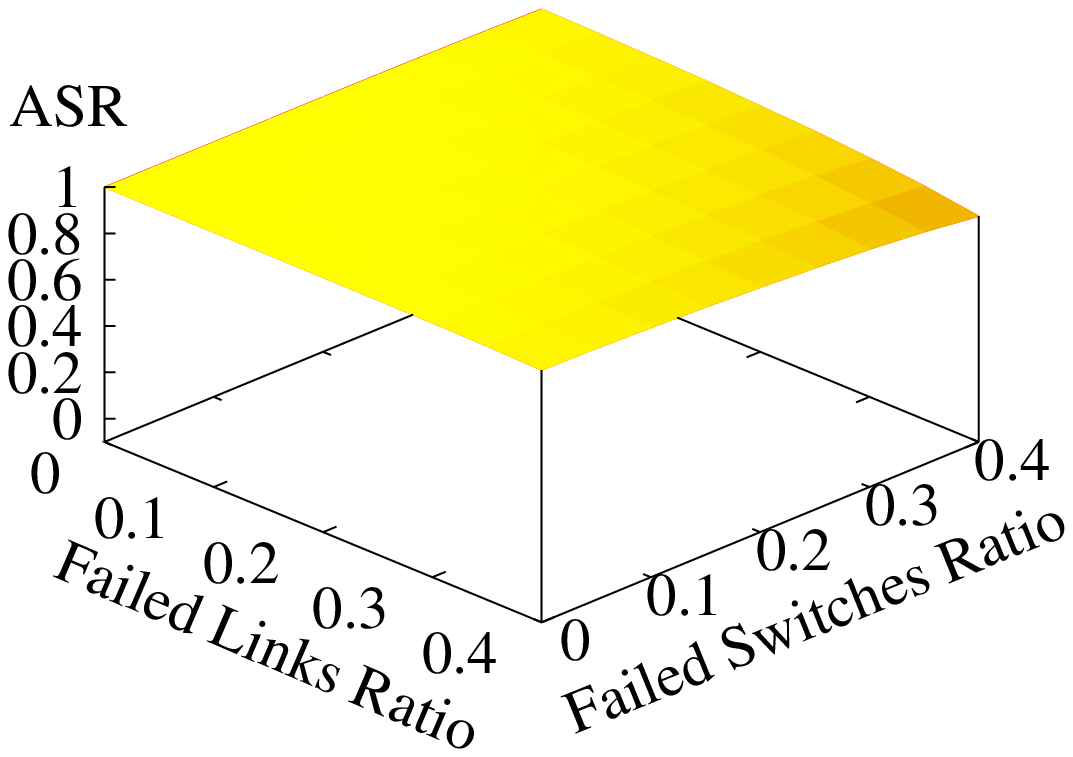}
\label{fig:allRangelinkSwitch_bcube_5_5}}\\
{\includegraphics[width=0.48\textwidth]{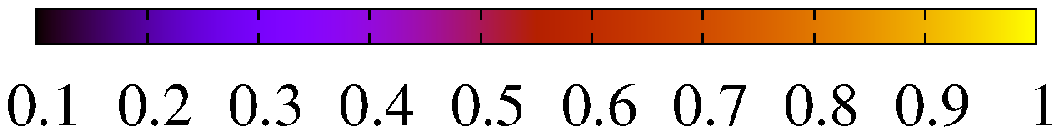}}
\caption{Variation of Failed Links Ratio and Failed Switches Ratio in all considered range.}
\label{fig:allRangelinkSwitch}
\end{figure}

The results are shown in Figure~\ref{fig:allRangelinkSwitch}. Each sub-figure represents the ASR for a given topology. For better visualization, we omit the confidence intervals. However, they are very narrow in this experiment. We observe that:

\begin{itemize}
\item \textit{Three-layer and Fat-tree Performance.} Three-layer and Fat-tree present the worst degradation in the ASR due to the fragility outlined in the previous results, when the failures are isolated. Note that Fat-tree performs better than Three-layer, because it employs more redundancy of links and switches.
\item \textit{BCube and DCell Performance.} BCube2 presents a slightly better survivability than DCell2, since for switch failures they perform equally, but BCube2 has a better survivability considering link failures. However, in the case of 3 interfaces, DCell3 performs slightly better than BCube3, since DCell3 has a very high survivability considering switch failures that compensates a worse performance to link failures. Note that BCube5 is almost unaffected by failures in the considered FER range.
\end{itemize}
%
%Generated by the script generateAll3d.sh at the directory generateAllRangeLinkSwitch 

\section{Qualitative Performance Analysis}
\label{sec:qualitative}

Considering our results for the Reliable and Survival phases, Table~\ref{tab:qualitative} provides a qualitative comparison of DCN topologies in terms of Reachability and Path Quality. 
The Reachability criterion combines the MTTF and the Service Reachability (i.e., ASR), since these metrics are closely related (i.e., a good MTTF implies a good Service Reachability). 
The topologies are evaluated considering five qualitative levels: bad, poor, fair, good, and excellent. 
The methodology used in this classification is detailed next in Section~\ref{subsec:defQualitative}.
Note that switch failures do not incur severe performance degradation in server-centric topologies. Hence, even if DCell performs better than BCube to switch failures, BCube still has a better overall performance since it is not classified as ``bad'', ``poor'' or ``fair'' in any criterion. Also, the Path Quality is not affected considerably by any failure type.
\begin{table}
\caption{Qualitative performance of DCN topologies considering both Reliable and Survival phases.}
\label{tab:qualitative}
\begin{tabular}{llllll}
\hline\noalign{\smallskip}
\hline {\bf Failure Type}	&{\bf Criterion} &{\bf Three-layer} &{\bf Fat-tree} &{\bf BCube} &{\bf DCell}\\
\noalign{\smallskip}\hline\noalign{\smallskip}
\multirow{2}{*} {\bf Link}	 &{\bf Reachability} &bad &poor  &good  &fair  \\
&{\bf Path Quality} &excellent &excellent &good &fair \\
\noalign{\smallskip}\hline
\multirow{2}{*} {\bf Switch}	 &{\bf Reachability} &bad &poor  &good  &excellent  \\
&{\bf Path Quality} &excellent &excellent &excellent &good \\
\noalign{\smallskip}\hline
\multirow{2}{*} {\bf Server}	 &{\bf Reachability} &excellent &excellent  &excellent  &excellent  \\
&{\bf Path Quality} &excellent &excellent &excellent &good \\
\noalign{\smallskip}\hline
\end{tabular}
\end{table}

\subsection{Methodology employed in the Qualitative Analysis}
\label{subsec:defQualitative}

We next detail the methodology employed in the analysis of Table~\ref{tab:qualitative}, regarding Reachability and Path Quality.

\subsubsection{Reachability}

For the Reachability analysis we use the following methodology:

\begin{itemize}
\item all topologies are considered as ``excellent'' in the server failure analysis since, on average, server failures no not lead to the disconnection of the remaining servers;
\item Three-layer topology is the only one classified as ``bad'' for link and switch failures, since the simulations show that it presents the worst performance;
\item for link and switch failures, we use the performance of DCell3 for a Failed Switches Ratio of $0.4$ as a reference value for ``excellent''. This topology has an ASR very close to 1 for a high Failed Switches Ratio ($0.4$) and also a high MTTF;
\item for link and switch failures, we use the performance of Fat-tree for a Failed Switches Ratio of $0.4$ as a reference value for ``poor''. In this topology, the ASR decreases linearly according to the FER, and its MTTF is significantly lower than in the other topologies, for both failure types;
\item the performance of BCube5 is not considered, since we do not employ a DCell with the same number of network interfaces;
\item for a given failure type (link or switch), a topology is classified as ``excellent'' if, for a FER value of $0.4$, \textbf{at least} one of its configurations (i.e., number of network interfaces) has a performance near (difference of $0,01$ in the ASR) the reference value for ``excellent'', and \textbf{all} configurations have an ASR greater than $0.8$; 
\item for a given failure type (link or switch), a topology is classified as ``poor'' if, for a FER value of $0.4$, \textbf{at least} one of its configurations (i.e., number of network interfaces) has a performance near (difference of $0,01$ in the ASR) the reference value for ``poor'',  and \textbf{all} configurations have an ASR less than $0,8$;
\item if a topology does not meet the requirements to be classified as ``poor'' or ``excellent'', it is classified as  ``good'' if, for \textbf{all} configurations, the topology has an ASR greater than $0.8$ for a FER value of $0.4$. Otherwise, it is classified as ``fair''.
\end{itemize}

\subsubsection{Path Quality}

For the Path Quality analysis we use the following methodology:

\begin{itemize}
\item as the Path Quality does not change significantly for all failure types, no topology is considered as ``bad'' or ``poor'' using this criterion;
\item the performance of BCube5 is not considered, since we do not employ a DCell with the same number of network interfaces;
\item for a given failure type, a topology is considered as ``excellent'' if, for a FER value of $0.4$, \textbf{all} its configurations have an Average Shortest Path Length less than or equal to 6. 
This reference value for ``excellent'' is the metric evaluated for Fat-tree, which does not change when the failure increases;
\item for a given failure type, a topology is considered as ``fair'' if, for a FER value of $0.4$, \textbf{at least} one of its configurations has an Average Shortest Path Length greater than 12. This reference value for ``fair'' is twice the value for ``excellent'';
\item for a given failure type, a topology is considered as ``good'' if it does not meet the requirements to be classified as ``fair'' or ``excellent''. Note that for Path Quality, the requirements to consider a topology as ``excellent'' are looser than for the Reachability case. This approach is adopted since, as shown by the results of Section~\ref{sec:survival}, the ASR variates more than the Average Shortest Path Length if we increase the FER.
\end{itemize}

\section{Gateway Port Density Sensibility Analysis} 
\label{sec:edpAnalysis}

In this section, we study how the choice of the number of gateways, or Gateway Port Density (GPD), influences the reliability and survivability of the DC.
In the case of survivability, we only evaluate the Service Reachability. The Path Quality concerns the paths between servers inside the DC, thus not depending on the choice of gateways. 

The results of Sections~\ref{sec:reliableTime}~and~\ref{sec:survival} were obtained with the maximum GPD (Section~\ref{sec:operational}), which is 1 for Fat-Tree, BCube and DCell, and 0.007 for Three-layer. In this section, we start by evaluating the metrics of each topology by setting the minimum GPD of each one. In other words, we choose in each topology only one switch to act as a gateway. As the network has only one gateway in this experiment, we only evaluate the ASR since the SC is always 1 when at least one server is reachable. Our experiments show that the MTTF and Service Reachability, considering link and server failures, are not substantially affected when we set a minimum GPD, as compared to the case of a maximum GPD. Hence, we only show in this section the results for switch failures.

\begin{figure}
\centering
%generate_criticalPoint_switch_3k_epd_0.0_unitMTTF.gnu
\includegraphics[width=0.36\textwidth]{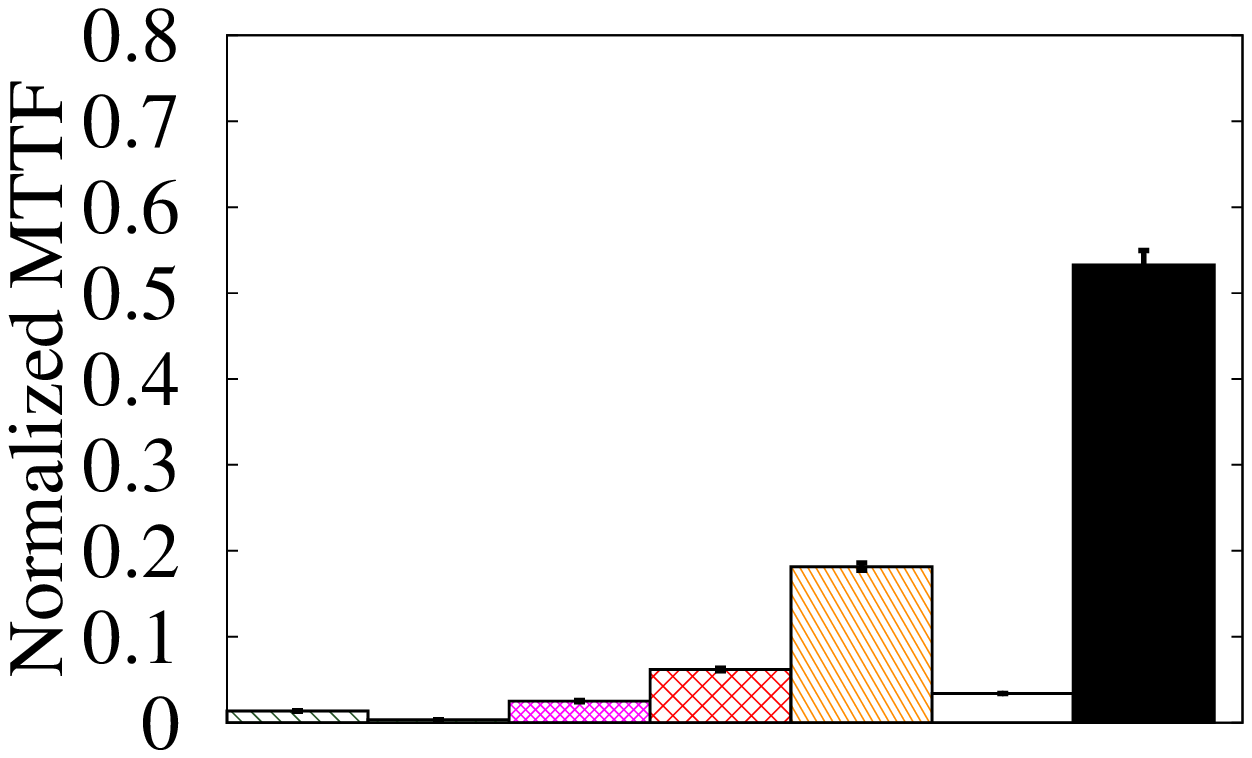} \quad
%gnuplot generate_criticalPoint_switch_3k_epd_0.0_criticalPoint.gnu
\includegraphics[width=0.36\textwidth]{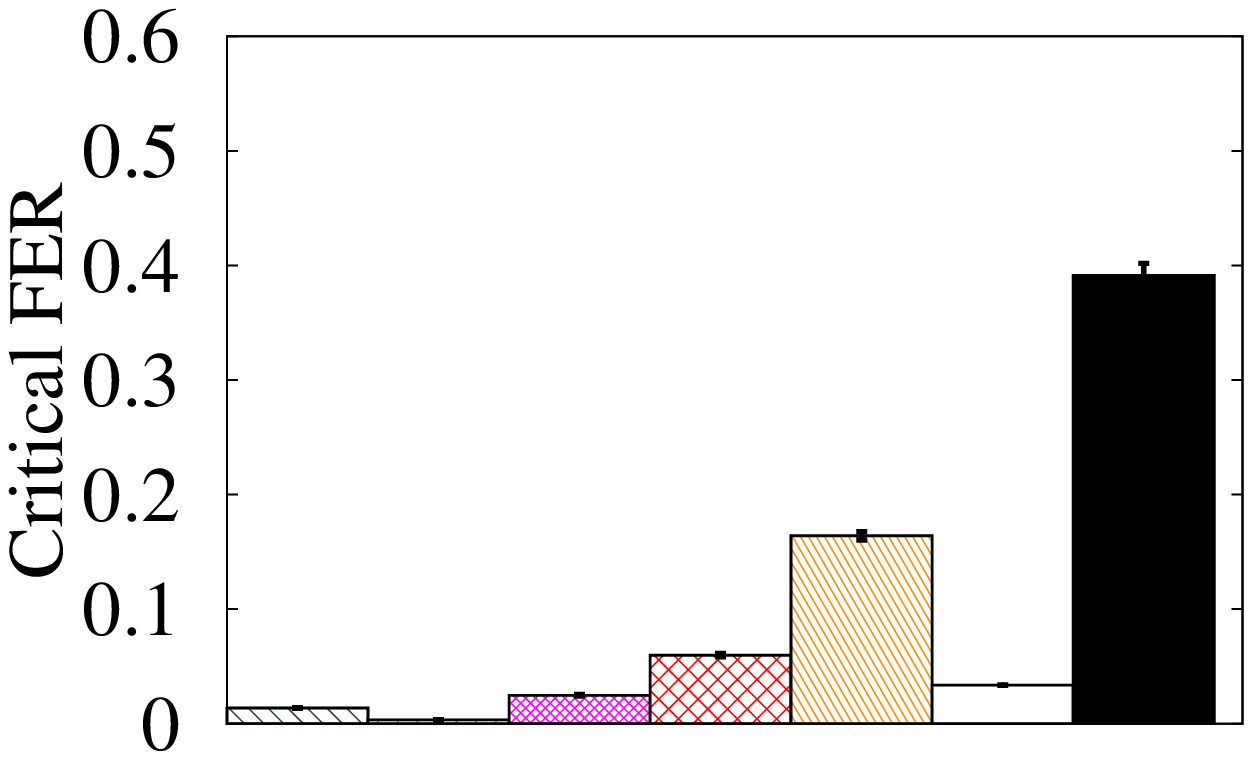}
\includegraphics[width=0.8\textwidth]{legendAllCritical.eps}
\caption{Reliable Phase analysis for switch failures, using the minimum GPD.}
\label{fig:switch_min_epd_mttf_unitmttf}
\end{figure}

Figure~\ref{fig:switch_min_epd_mttf_unitmttf} shows the results of the Reliable Phase for a minimum GPD.
Except for the case of DCell3, the reduction of MTTF and Critical FER is small when compared with Figure~\ref{fig:switch_mttf_unitmttf}. The results of DCell3 show that:
\begin{itemize}
\item Although DCell3 is highly reliable to switch failures, the choice of a minimum GPD produces a single point of failure that reduces to $29\%$ in its MTTF.
\item Even with a minimum GPD, the reliability of DCell3 is still higher than the one achieved by the other topologies with the maximum GPD, shown in Figure~\ref{fig:switch_mttf_unitmttf}.
\end{itemize}

\begin{figure}
\centering
\subfigure[Accessible Server Ratio using minimum GPD.]
{
%generate_3k_epd_0.0_failureRatio_switch_asr.gnu
\includegraphics[width=0.48\textwidth]{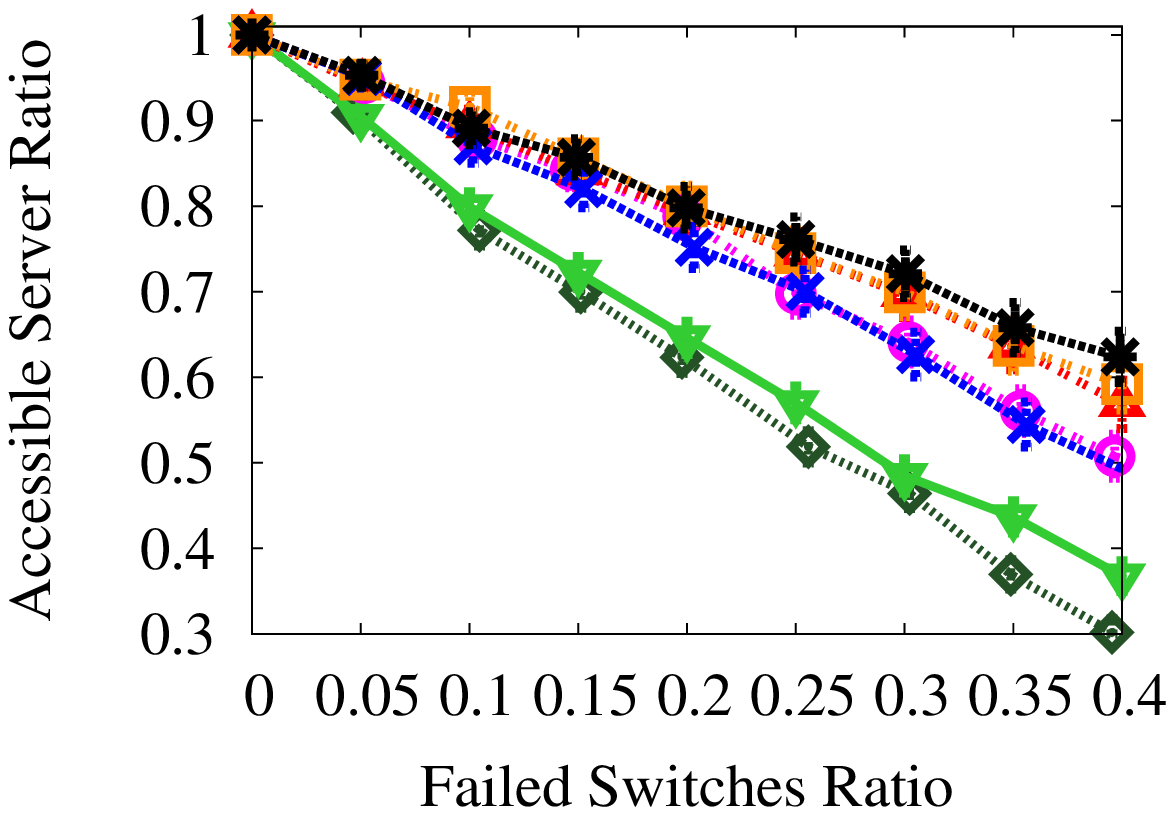}
\label{fig:switch_min_epd_asr_failureRatio}
}
\subfigure[ASR as function of the GPD for a Failed Switches Ratio of 0.05.]
{
%generate_3k_epdVariation_0.05_switch_asr.gnu
\includegraphics[width=0.48\textwidth]{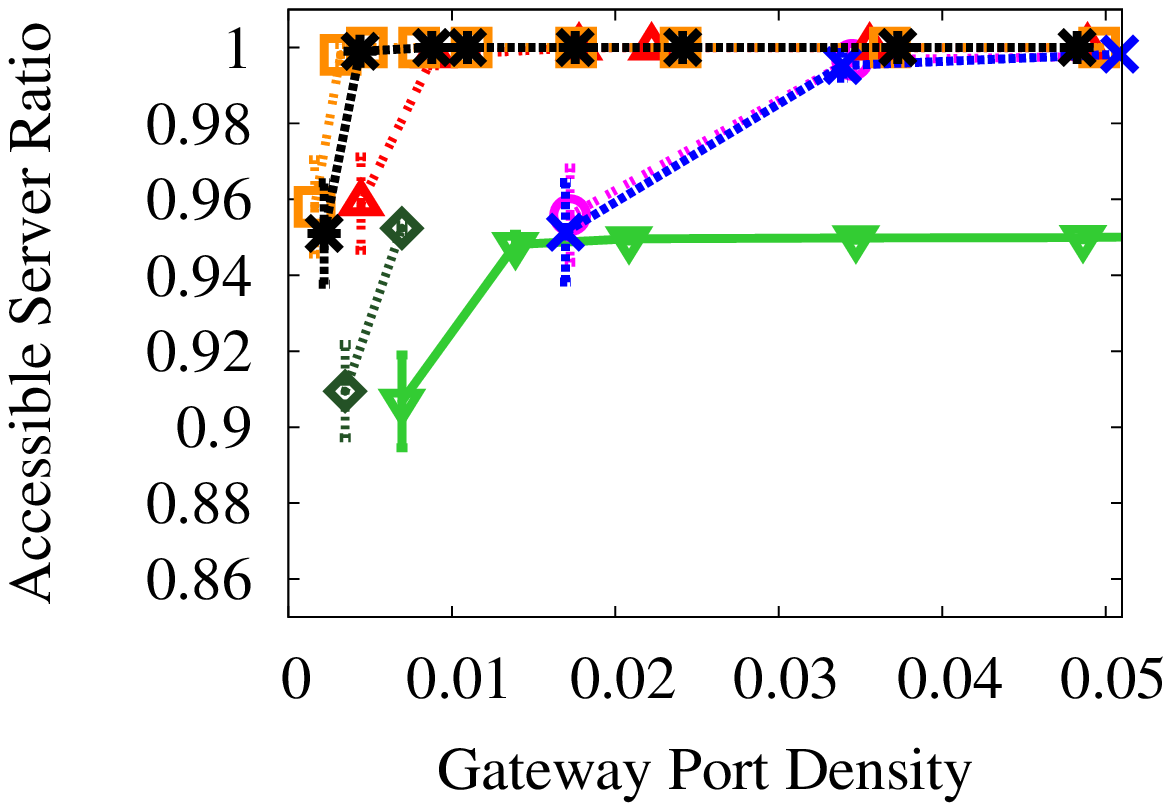}
\label{fig:switch_epdVariation_0.05_asr}
}
{\includegraphics[width=0.8\textwidth]{legendAll.eps}}
\caption{GPD sensibility analysis in Survivable Phase for switch failures.}
\end{figure}

Figure~\ref{fig:switch_min_epd_asr_failureRatio} shows that the survivability considering switch failures is highly affected by the minimum GPD. Also, comparing the ASR between topologies of the same type (i.e., switch-centric or server-centric), the results show that their performance is very close.
With a minimum GPD, a high decrease on survivability is expected since the topologies have a single element responsible for maintaining the connectivity of the whole network.
Hence, the network becomes totally disconnected if the gateway is down. Moreover, as the FER increases, the probability of failure of this switch increases, reducing the ASR on average. 

To complete the above analysis, Figure~\ref{fig:switch_epdVariation_0.05_asr} shows the ASR according to the GPD choice. This result is obtained by fixing the Failed Switches Ratio to 0.05 and varying the GPD of each topology from its minimum GPD to an approximate value of 0.05. Note that, although the curves are shown with continuous lines to facilitate visualization, for each topology the two lowest GPDs in the figure correspond to the utilization of 1 and 2 gateways. The ASR reduces significantly only when the GPD is at a minimum, showing the high robustness of the topologies regarding this value. Thus, we conclude that: 
\begin{itemize}
\item The robustness to GPD is more related to the probability of gateway failure than to the indirect loss of access to this element. This is explained by the fact that the failure of links does not significantly reduce the reliability and survivability, whereas switch failures do.
\item The choice of the number of gateways has little influence on Service Reachability. Severe performance degradation, according to this parameter, is only observed when a single switch is chosen and the network is prone to switch failures. 
\end{itemize}

It is important to note that this result does not address reliability and survivability according to the failure of the external access (i.e., ports connected to the outside world) itself. The results shown here just prove that the access to the gateways is not substantially affected by failed network elements, except the gateway itself. Obviously, considering failures of the external access, the reachability of the entire DC will increase as we increase their redundancy. However, we do not analyze this type of failure, since we are interested in evaluating the characteristics inside the DCN. Also, external accesses are generally easier to monitor and repair as they are less numerous than other network elements.

\section{Heterogeneous Elements} 
\label{sec:het}

In this work, we consider that all elements of a given type are equal and assume that all servers have the same hardware characteristics. In this section, we analyze the impact of these two assumptions in our results. First, we analyze how the failure of different types of switches and links impact the results for the Three-layer topology. Next, we redefine a Reachability metric to consider heterogeneous servers. For both cases we focus on the ASR metric, since its analysis in the previous results explain better the differences between topologies.

\subsection{Equipment Heterogeneity in Three-layer} 
\label{sec:het3Layer}

The methodology described in Section~\ref{sec:reliableTime} and employed in all the results of Section~\ref{sec:survival}, considers that all elements of a given type are equal. 
This is true for Fat-tree, BCube and DCell since the main goal of their design is to use homogeneous low cost switches. However, the Three-layer topology employs different switch types in each layer. Hence, we analyze this topology by considering three different types of switches and links. We perform this analysis by choosing a different Failure Element Ratio depending on the switch or link type.

\begin{figure}[h!]
\centering
\subfigure[Edge and Aggregation switch failures.]
{
%generate3TierHetNetHelements/generate_3tier_3456_3k_Aggswitch_failureRatio_asr.gnu
\includegraphics[width=0.48\textwidth]{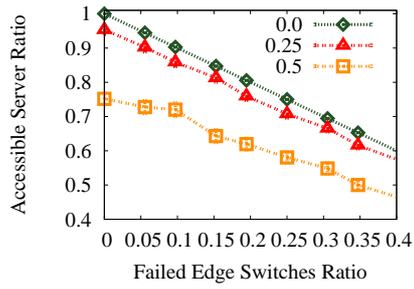}
\label{fig:aggswitch_failureRatio_asr}
}
\subfigure[Edge and Core switch failures.]
{
%generate3TierHetNetHelements/generate_3tier_3456_3k_Coreswitch_failureRatio_asr.gnu
\includegraphics[width=0.48\textwidth]{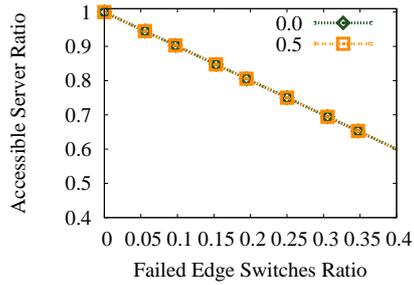}
\label{fig:coreswitch_failureRatio_asr}
}
\subfigure[Aggregation and Core switch failures.]
{
%generate3TierHetNetHelements/generate_3tier_3456_3k_AggCoreswitch_failureRatio_asr.gnu
\includegraphics[width=0.48\textwidth]{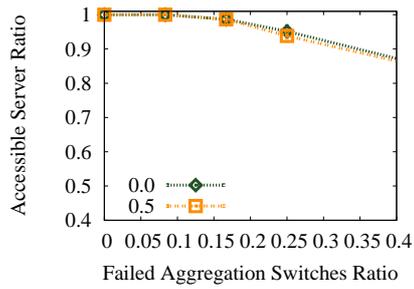}
\label{fig:aggscoreswitch_failureRatio_asr}
}
\caption{ASR considering different switch failure types in Three-layer with 3456 servers.}
\end{figure}

Our analysis employs the three switch types specified in the Three-layer topology definition: Edge, Aggregation, and Core. For a given analysis, we combine the failures in two switch types. Figure~\ref{fig:aggswitch_failureRatio_asr} shows the results for the Three-layer when varying the Failed Edge Switches Ratio, while keeping the Failed Aggregation Switches Ratio fixed and considering that no Core switch fails. Hence, each curve of Figure~\ref{fig:aggswitch_failureRatio_asr} represents a given Failed Aggregation Switches Ratio. We choose three different FER values for Aggregation switches: 0.0, 0.25, and 0.5. The results show that the impact of differentiating these two switch types is only significant for a high Failed Aggregation Switches Ratio. In addition, note that the curves for the two lowest values of the Failed Aggregation Switches Ratio (i.e., 0.0 and 0.25) in Figure~\ref{fig:aggswitch_failureRatio_asr} are very close to the curve for the Fat-tree in Figure~\ref{fig:switchAsr_fer}. This happens because, when we have a low failure ratio in the aggregation and in the core, the survivability of the Three-layer is dominated by the effect of edge switches. In Fat-tree, even when considering failure of all switches, the ASR is dominated by the edge switches, since the aggregation and core layers of Fat-tree are highly redundant. As the edge of Three-layer is identical to the edge of Fat-tree, their survivability is close in this case. Figure~\ref{fig:coreswitch_failureRatio_asr} shows the ASR according to the variation of the Failed Edge Switches Ratio for two values of Failed Core Switches Ratio, while considering that no Aggregation switch fails. Recall that, since we have only two Core switches, a failed ratio of 0.5 corresponds to one Core switch failure. The results show that, if only one core switch fails, the remaining Core switch is enough to maintain the DC connectivity. This can be easily confirmed by Figure~\ref{fig:conventional} that shows that each Core switch is connected to all Aggregation switches, allowing the network to operate with a single Core switch. Obviously, as core switches are the only gateways in the Three-layer, the ASR is zero if the two Core switches fail. Again, as the failure of one Core switch is negligible and no Aggregation switch fails, the ASR for the Three-layer in Figure~\ref{fig:coreswitch_failureRatio_asr} becomes close to the ASR for Fat-tree in Figure~\ref{fig:switchAsr_fer}. Finally, Figure~\ref{fig:aggscoreswitch_failureRatio_asr} shows the results when no edge switch fails and we vary the Failed Aggregation Switches Ratio, keeping the Failed Core Switches Ratio fixed in 0.0 or 0.25. The effect of one Core switch failure is negligible for the same reason as before. Note that, when the Failed Edge Switches Ratio is kept in zero, the Three-layer maintains high ASR values even for the high Failed Aggregation Switches Ratios. This result shows, as already mentioned in Section~\ref{sec:perfEvaluationSwitchSurvival}, that the edge has a major role on the survivability of the Three-layer.

\begin{figure}[h!]
\centering
\subfigure[Edge and Aggregation link failures.]
{
%generate3TierHetNetHelements/generate_3tier_3456_3k_Agglink_failureRatio_asr.gnu
\includegraphics[width=0.48\textwidth]{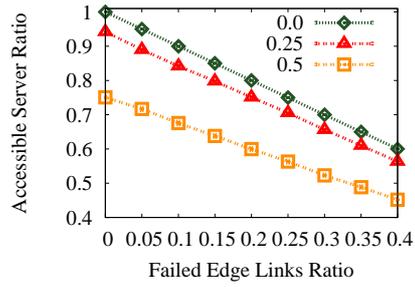}
\label{fig:agglink_failureRatio_asr}
}
\subfigure[Edge and Core link failures.]
{
%generate3TierHetNetHelements/generate_3tier_3456_3k_Corelink_failureRatio_asr.gnu
\includegraphics[width=0.48\textwidth]{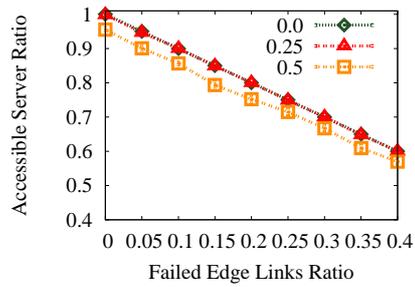}
\label{fig:corelink_failureRatio_asr}
}
\subfigure[Aggregation and Core link failures.]
{
%generate3TierHetNetHelements/generate_3tier_3456_3k_AggCorelink_failureRatio_asr.gnu
\includegraphics[width=0.48\textwidth]{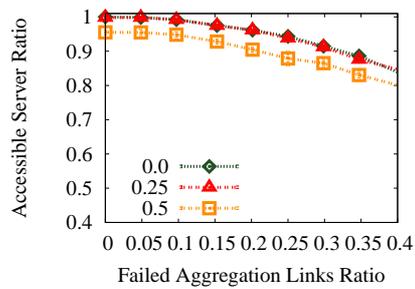}
\label{fig:aggscorelink_failureRatio_asr}
}
\caption{ASR considering different link failure types in Three-layer with 3456 servers.}
\end{figure}

To analyze the heterogeneity of the links in Three-layer, we define three link types: Edge, Aggregation, and Core. The first type corresponds to the links between the Edge switches and servers. The second one corresponds to the links between the Edge and Aggregation switches. The last type refers to the links between the Aggregation and Core switches. Note that Three-layer also has links between two Aggregation switches and between two Core switches, as shown in Figure~\ref{fig:conventional}. We disregard these link types in this analysis, because they do not affect the network survivability. Figure~\ref{fig:agglink_failureRatio_asr} shows the results for Three-layer when varying the Failed Edge Links Ratio, while keeping the Failed Aggregation Links Ratio fixed and considering that no Core link fails. As in the case of switch failures analyzed before, the survivability of Three-layer in Figure~\ref{fig:agglink_failureRatio_asr} becomes close to the survivability of Fat-tree in Figure~\ref{fig:linkAsr_fer} when the Failed Aggregation Links Ratio is low. This same behavior applies for the Failed Core Links Ratio, when we consider only failures of Edge and Core links in Figure~\ref{fig:corelink_failureRatio_asr}. Also from Figure~\ref{fig:corelink_failureRatio_asr}, note that the effect of Core links in the survivability is low. Finally, Figure~\ref{fig:aggscorelink_failureRatio_asr} shows that, if no Edge links fail, the ASR can be kept at high values. As in the case of switch failures, this result shows that Edge links play a major role in the survivability of Three-layer. Considering the analysis of switches and links, the results show that if a DC with the Three-layer topology employs high reliable equipment in the core and in the aggregation layers, its survivability can be close to that of Fat-tree.

\subsection{Server Heterogeneity} 
\label{sec:hetServers}

Another assumption made in our methodology is that all servers are equal, and thus the ASR accounts only the number of remaining servers in the network. However, as stated by Zhang~\textit{et. al}~\cite{zhang2014Dynamic}, the servers in a DC are very heterogeneous. In other words, different types of server hardware coexist in the infrastructure, and each type has its own CPU and memory capacity. Hence, the impact of disconnecting a high capacity server from the network is different from disconnecting a low capacity one. To show the impact of assuming homogeneous servers in the previous results, we propose another Reachability metric based on the ASR, called the Remaining Capacity Ratio (RCR), which takes into account the remaining capacity available after a failure.
The RCR metric is the ratio between the remaining capacity after a failure and the total capacity of the original DC. The RCR is defined as:
\begin{equation}
RCR=\frac{\sum_{i=1}^{|\mathcal{S}|}z_i a_i}{\sum_{i=1}^{|\mathcal{S}|}z_i},
\end{equation}
where $z_i$ and $a_i$ are, respectively, the capacity of the server $i$ and a binary variable indicating if this server is connected (i.e., it has a path to a gateway) after the failure. Hence, if server $i$ is connected, then $a_i=1$, and $a_i=0$ otherwise. The total number of servers on the original network is given by $|\mathcal{S}|$. The capacity value can be defined according to the DC application. In this section, we evaluate the RCR using CPU and memory capacity called, respectively, the Remaining CPU Ratio and Remaining Memory Ratio.

To evaluate the RCR metric in the considered DC topologies, we employ the information provided by Zhang~\textit{et. al}~\cite{zhang2014Dynamic}. In this article, Zhang~\textit{et. al} show the different types of servers employed in the DC, based on a real trace provided by Google~\cite{googleCusterData}. In addition, they show how many servers of a given type are installed in the DC, as well as their corresponding capacity. In our analysis, we use their information regarding the CPU and Memory capacity. These values are normalized in~\cite{zhang2014Dynamic}, so that the most powerful CPU or memory type has a capacity equal to 1. Their data shows ten different types of machines for a DC with approximately 12,000 servers. As our analysis comprises about 3,400 servers, we scale their number of servers to our DC size, by evaluating the fraction of servers from each type. Since six of their reported types together represent less than 1\% of the servers (i.e., less than 34 servers in our case), we consider these six types as one single type. This single type is the one with the highest number of servers among these six. Consequently, we have five machine types in our scenario, given by Table~\ref{tab:googleData}. In this table, we adopt the same type number specified in~\cite{zhang2014Dynamic}.
\begin{table}
\caption{Real dataset of server capacities, based on Google traces.}
\label{tab:googleData}
\begin{tabular}{llll}
\hline\noalign{\smallskip}
\hline \textbf{Type Number} &\textbf{CPU Capacity} &\textbf{Memory Capacity} &\textbf{Fraction of Servers}\\
\noalign{\smallskip}\hline\noalign{\smallskip}
1 &0.50 &0.50 &0.53 \\
2 &0.50 &0.25 &0.31 \\
3 &0.50 &0.75 &0.08 \\
4 &1.00 &1.00 &0.07 \\
5 &0.25 &0.25 &0.01 \\
\noalign{\smallskip}\hline
\end{tabular}
\end{table}

It is reasonable to expect that the survivability given by the RCR is higher in DCs where the capacity is uniformly distributed among the topology modules (e.g., the pods of Fat-Tree or a group of servers in Three-layer where the connectivity is maintained by the same pair of aggregation switches). When the capacity is uniformly distributed, if the entire module fails (e.g., if the pair of aggregation switches in Three-layer fails), the effect is lower than in the case were the failed module concentrates the most powerful servers. Hence, for each employed dataset, we choose two capacity distributions among the DC servers. In the first one, called \textit{Balanced}, we try to balance the total server capacity inside each topology module. For example, we try to assign, as much as possible, different server types in a Pod for Fat-tree, in a module for Three-layer, in lower level DCells for DCell and lower level BCubes for BCube. On the other hand, in the distribution called \textit{Unbalanced}, we try to put, as much as possible, servers of the same type together in the same module.
To analyze the impact of capacity distribution, we build a synthetic dataset where approximately 17\% of the servers concentrate 50\% of the capacity. We choose the value of 17\% since it corresponds to the fraction of servers inside a single module in Three-layer and, as we show next, we use this topology as a reference in our analysis.
\begin{table}
\caption{Synthetic dataset of server capacities.}
\label{tab:syntheticData}
\begin{tabular}{llll}
\hline\noalign{\smallskip}
\hline \textbf{Type Number} &\textbf{CPU Capacity} &\textbf{Memory Capacity} &\textbf{Fraction of Servers}\\
\noalign{\smallskip}\hline\noalign{\smallskip}
1 &1.00 &1.00 &0.16666666667 \\
2 &0.20 &0.20 &0.83333333333 \\
\noalign{\smallskip}\hline
\end{tabular}
\end{table}

\begin{figure}[h!]
\centering
\subfigure[Real data extracted from Google dataset.]
{
%generateGraphsHet/generate_3k_failureRatio_google_switch_3Tier.gnu
\includegraphics[width=0.47\textwidth]{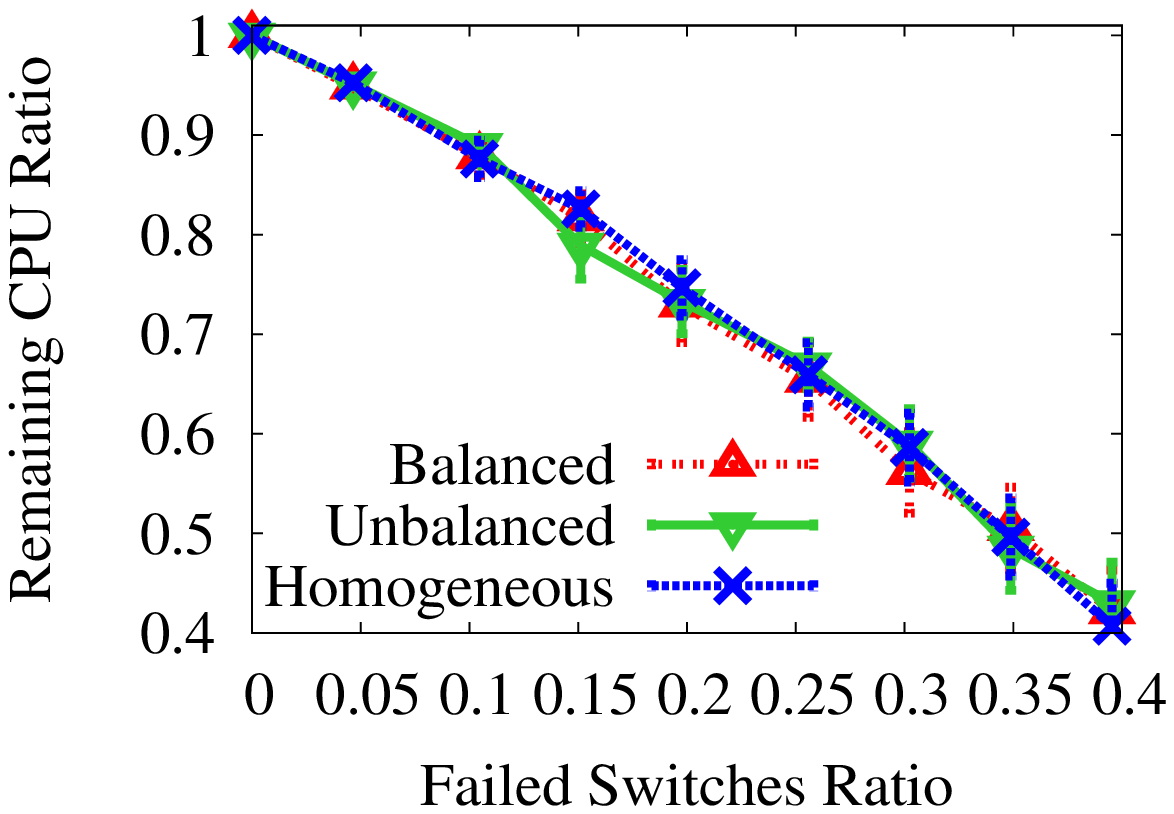}
\label{fig:3k_het_google_switch_3Tier}
}
\subfigure[Synthetic data.]
{
%generateGraphsHet/generate_3k_failureRatio_synthetic_switch_3Tier.gnu
\includegraphics[width=0.47\textwidth]{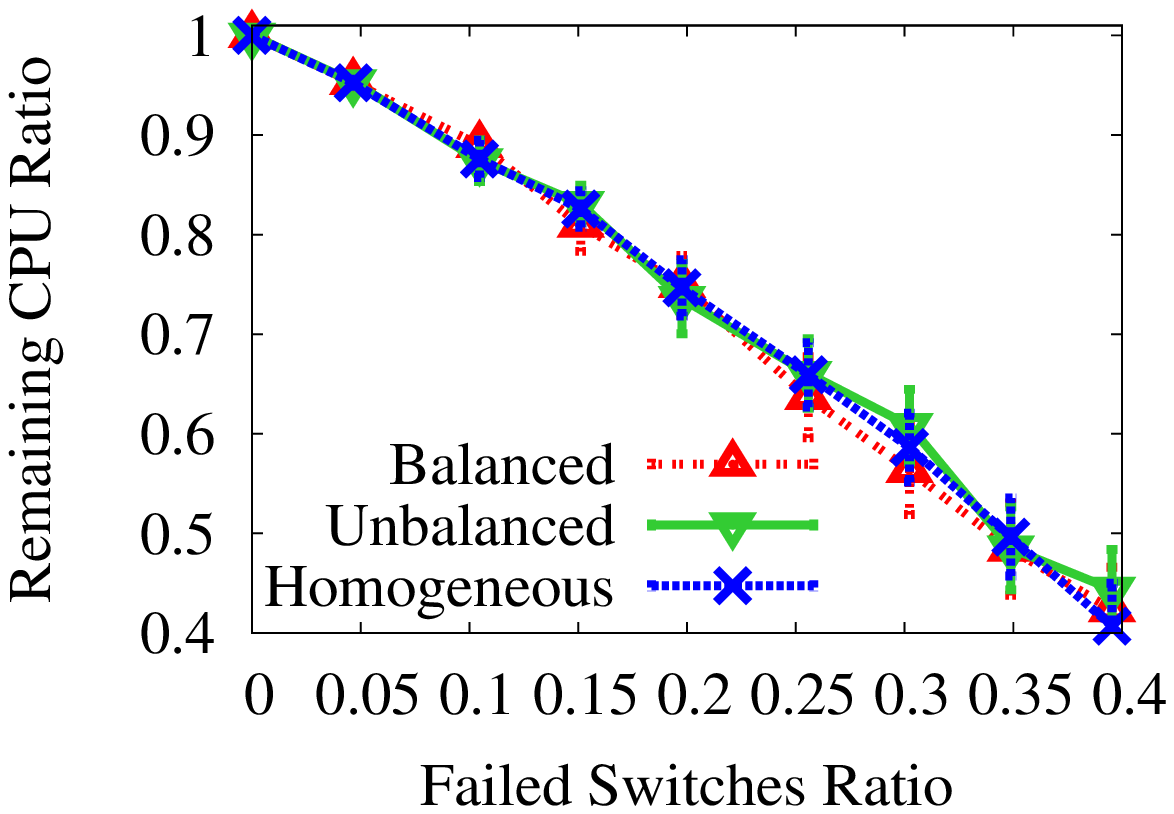}
\label{fig:3k_het_synthetic_switch_3Tier}
}
\caption{Remaining CPU Ratio considering different switch failure types in Three-layer with 3456 servers.}
\label{fig:3k_het_switch_3Tier}
\end{figure}

We first perform the experiments for RSR employing the same methodology of Section~\ref{sec:failureSimMethodology}, using both datasets and the two capacity distributions. Consequently, we remove from the network a random number of switch or links, and evaluate the RSR metric using its average value achieved in the simulation. Since the values are averaged, it is expected that the dataset and the capacity distribution play no major role on the DC survivability. It is true because in some simulation rounds the module with a high capacity may fail, but in other ones, the module with a low capacity fails. Hence, we do not expect substantial differences between the results for a heterogeneous DC and a homogeneous DC. For the sake of conciseness, in Figure~\ref{fig:3k_het_switch_3Tier} we show only the results for the CPU capacity of Three-layer when prone to switch failures. We choose this topology since it is the most fragile among the considered topologies, and thus, the heterogeneity tends to have a higher impact. The figure shows the results for each dataset employing the Balanced and Unbalanced distributions, as well as a reference curve for the homogeneous case (i.e., where all servers have the same capacity). As can be noted by Figure~\ref{fig:3k_het_switch_3Tier}, the heterogeneity has a very low impact in the RSR when considering average values. Hence, the results for the RSR metric from Figure~\ref{fig:3k_het_switch_3Tier} become close to those for the ASR metric in Figure~\ref{fig:switchAsr_fer}.

\begin{figure}[h!]
\centering
\subfigure[Remaining CPU Capacity.]
{
%generateSpecificFailureHet/generate_3Tier_specificResultHet_cpu.gnu
\includegraphics[width=0.45\textwidth]{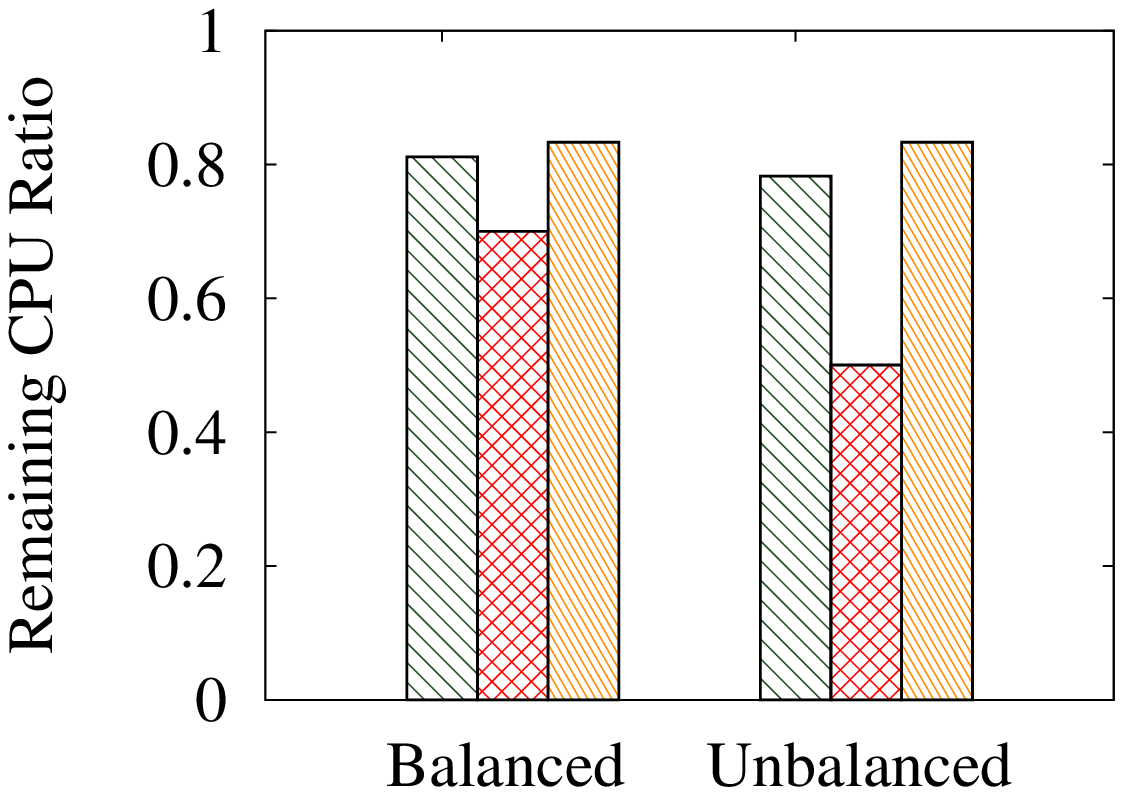}
\label{fig:3k_3Tier_het_specificCPU}
}
\subfigure[Remaining RAM Capacity.]
{
%generateSpecificFailureHet/generate_3Tier_specificResultHet_ram.gnu
\includegraphics[width=0.45\textwidth]{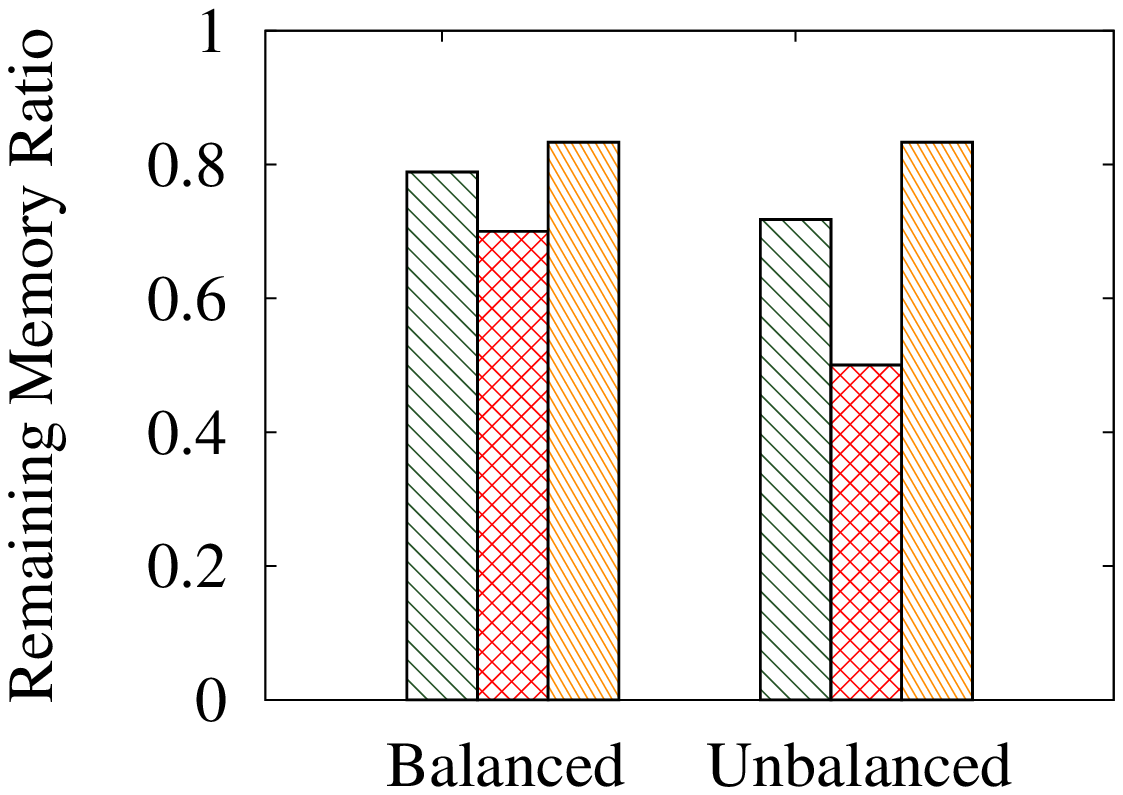}
\label{fig:3k_3Tier_het_specificRAM}
}
{\includegraphics[width=0.55\textwidth]{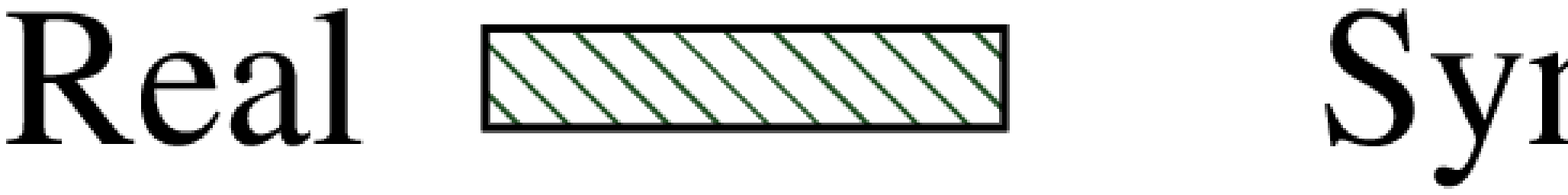}}
\caption{Remaining Capacity Ratio considering the removal of an entire module in Three-layer with 3456 servers.}
\label{fig:3k_3Tier_het_specific}
\end{figure}

As shown before, the average values of RCR do not capture the impact in the survivability caused by server heterogeneity.
Therefore, we perform an analysis in the Three-layer topology by removing a given pair of aggregation switches. In Three-layer topology, removing an aggregation switch pair disconnects an entire module, which is a group of 576 servers when the entire DC has 3,456 servers. In this experiment, we choose to remove the module that concentrates the highest CPU or RAM capacity. Figures~\ref{fig:3k_3Tier_het_specificCPU}~and~\ref{fig:3k_3Tier_het_specificRAM} show, respectively, the RCR results for CPU and RAM using the different datasets and capacity distribution. We also plot the results for the homogeneous case for reference\footnote{For the homogeneous case, the Balanced and Unbalanced results correspond to the same scenario, since all servers are equal}. The values of RCR are deterministic since we remove a specific pair of Aggregation switches. The results show that, as expected, the Balanced distributions lead to a higher survivability for both datasets. However, the impact of balancing server capacity is higher for the Synthetic dataset, since the two existent server types have very different capacity values, as shown in Table~\ref{tab:syntheticData} . For the Google DC case, we note that considering CPU capacity, the difference between the two capacity distributions is small. This happens since three types of servers in this dataset have the same CPU capacity, as shown in Table~\ref{tab:googleData}. Furthermore, these three types together correspond to 92\% of the servers. On the other hand, we can note that the difference between the Balanced and Unbalanced cases is significant for memory capacity. This happens since memory configurations are more heterogeneous in the real dataset; from five machine types, we have four memory capacities, as shown in Table~\ref{tab:googleData}. Finally, we can note that, considering a real dataset and a balanced capacity distribution, the performance of Three-layer in a heterogeneous scenario is close to the homogeneous scenario. The same methodology employed in these results applies to the other topologies considered in this work. However, as they have a more redundant network and thus higher survivability, the effect of heterogeneity is even lower.

\section{Related Work}
\label{sec:related} 
 
Our work provides an analysis of DC reliability and survivability considering the utilization of recently proposed DCN topologies. In some sense, it complements the existing study presented in~\cite{guo2009bcube} where the performance of Fat-tree, BCube, and DCell are compared considering switch and server failures. In that work, Guo~\textit{et al.} evaluated the survivability of those topologies by defining the ABT (Aggregate Bottleneck Throughput) metric. To evaluate this metric, they consider that every server is sending a packet flow to all other servers. They define the bottleneck throughput as the lowest throughput achieved among the flows. Hence, the ABT is defined as the number of existent flows times the throughput of the bottleneck flow. Their evaluation uses a single configuration for each topology (BCube and DCell with respectively 4 and 3 server interfaces and a Fat-tree with 5 switch levels) with a total number of 2,048 servers. Furthermore, they considered the utilization of the routing schemes originally proposed for each of the 
three architectures. They concluded that BCube performs well under both server and switch failures, and Fat-tree suffers from a high ABT drop when switches fail. On the other hand, the results showed that DCell has a low ABT even in the case of zero failures, but this value does not significantly change under failures. 
Our work differs from~\cite{guo2009bcube} in that our analysis is not restricted to specific traffic patterns and routing schemes, but is instead generic with focus on topological aspects. Also, we provide additional metrics that allow an evaluation of the reliability and the survivability of each topology, and analyze the relationship between their number of server network interfaces and their robustness to failures. Finally, we evaluate the Service Reachability, which was not addressed by Guo~\textit{et al.}. 

Bilal~\textit{et al.}~\cite{bilal2013Characterization} analyzed the robustness of Fat-tree, DCell, and a topology with three switch layers, demonstrating that classical robustness metrics derived from graph theory (e.g., average nodal degree) do not alone provide an accurate robustness factor for DCNs. Similarly to Guo~\textit{et al.} and to our work, they measure different metrics by varying the number of failed elements. They propose a metric that accounts for all metrics analyzed in their work, and use it to conclude that DCell outperforms the robustness of Fat-Tree and the topology with three switch layers, while this last one has the worst robustness performance. Different from their work, we analyze in more detail the behavior of Server Reachability according to failures, allowing us to highlight some topological characteristics that make a topology more robust to a given type of failures.
In addition, we provide an analysis for different number of server ports in BCube and DCell, while they focused on a specific DCell configuration with two different sizes and did not analyze BCube. Finally, we provide an evaluation of DC degradation according to the time, which also allows us to model and analyze the MTTF of the considered topologies.

Our previous work~\cite{couto2012Reliability} provides a comparison of the survivability (i.e., metrics in the Survivable Phase) of the above-mentioned topologies. In the present article, we extend the 
analysis by adding metrics such as those of the Reliable Phase and all the analysis considering the Elapsed Time. We also redefine the metrics of survivability used in that work to provide a more realistic analysis, by considering the existence of gateways. 

Still considering DC topologies, Ni~\textit{et al.}~\cite{ni2014provisioning} provided a theoretical analysis of bandwidth requirements for the switch-centric topologies Fat-tree and VL2, considering the failure of $k$ links in the network. They concluded that Fat-tree requires less link capacity than VL2 to support (i.e., provide full bandwidth communication between servers) $k$ failures when $k$ is small. For large values of $k$, VL2 outperforms Fat-tree.

There are also studies that have provided measurements in real DCs to investigate their reliability. Vishwanath~and~Nagappan~\cite{vishwanath2010characterizing} provided a characterization of server failures in DCs, by analyzing an environment with over 100,000 servers spread in different countries and continents. Among other observations, they concluded that the causes of most server failures are faulty hard disks. Gill~\textit{et al.}~\cite{gill2011understanding} measured the impact of network components on DC reliability. They used logs of failure events of some production DCs. Although they did not provide measurements using alternative DCN topologies, they stated that commodity switches are highly reliable. Consequently, a high degree of reliability can be achieved by using low-cost topologies such as Fat-tree, BCube and DCell. Also, they highlighted that legacy DCNs are highly reliable, presenting more than four 9`s of availability for about 80\% of the links and for about 60\% of the network devices. Nevertheless, as their study 
focused on legacy DCNs, this conclusion could not apply to emerging DCN scenarios such as Modular Data Centers (MDC) and low-cost architectures.

\section{Conclusions and Future Directions}
\label{sec:conclusion}

In this work, we evaluated the behavior of recently proposed DCN topologies considering that their different elements are prone to failures. The results allow us to conclude which topology behaves better for a given failure scenario. We can state that:
\begin{itemize}
\item The conventional Three-layer DC configuration has a lower redundancy level of links and switches than the alternative data center architectures. Hence, it shows the worst behavior when comparing to Fat-tree, BCube, and DCell to both link and switch failures. However, Three-layer can achieve survivability values close to those of Fat-tree, if it employs highly reliable equipment in the aggregation and core layers.
Fat-tree has a high redundant core but a vulnerable edge, which reduces its robustness to failures as compared to BCube and DCell. In Fat-tree, when a given fraction of the total links or switches fail, the same fraction of servers is disconnected. Consequently, Fat-tree shows a substantially lower performance than BCube and DCell, which lose no more than $26\%$ of their servers for a high percentage of failed elements ($40\%$). Also, Fat-tree achieves an MTTF at least $42$ times lower than other server-centric topologies for link failures and at least $7.2$ times lower for switch failures. On the other hand, Three-layer and Fat-tree maintain their original path length, while in BCube and DCell a high failure ratio can increase the average path length by 2 and 7 hops, respectively. Nevertheless, the increase in path length for server-centric topologies is generally not severe as compared with the server reachability degradation in switch-centric ones.
\item BCube performs better than the other topologies in environments with predominant link failures, maintaining at least 84\% of its servers when 40\% of the links are down, against 74\% in DCell. This is explained because, as a server-centric network, BCube employs redundant server interfaces. Also, the servers are directly connected only to switches. As switches in BCube have a higher degree than servers, the disconnection of the network by link removal will be harder in BCube than in DCell, since this last one employs servers directly connected to each other.
\item DCell presents the best performance under switch failures, being able to achieve an MTTF up to 12 times higher than BCube. This behavior is explained by the high dependence on servers to maintain a connected network.
\end{itemize}
 
By adding server interfaces, we have also shown that the improvement in reliability and survivability is upper bounded by the maximum tolerated path length. This happens because, even in the case without failures, increasing the number of servers interfaces in BCube and DCell increases the Average Shortest Path Length. Concerning server-centric topologies, we found that although they rely on servers to forward packets, a server failure does not lead to the disconnection of the remaining servers. 

Finally, we have also shown that the min-cut is an appropriate metric to approximate the reliability for link failures. Hence, we provided closed-form MTTF formulas for the considered topologies. For switch failures, the results show that the utilization of min-cuts is not well suited for some topologies.

In our future work, we will aim at evaluating the performance of DCN topologies considering correlated failures (e.g., failure of an entire rack), relaxing the assumption of independence between failures. Also, an interesting direction is to build more scenarios where all the three failure types (i.e., link, switch and server) coexist, complementing the study of Section~\ref{sec:perfEvaluationLinkSwitchSurvival}. 

\appendix

\section{MTTF Approximation}
\label{app:compMTTFapprox}

In this appendix we obtain Equation~\ref{eq:mttfApprox}, derived from the combination of Equations~\ref{eq:mttfReliability}~and~\ref{eq:burtilPittel}.
First, we replace the reliability $R(t)$ in Equation~\ref{eq:mttfReliability} by the reliability approximation given by 
Equation~\ref{eq:burtilPittel}, resulting in
\begin{equation}
MTTF = \int_{0}^{\infty} R(t) \, dt \approx \int_{0}^{\infty} e^{-\frac{t^rc}{{E[\tau]}^r}} \, dt.
\label{eqApp:mttfReliability}
\end{equation}
Hence, we find the MTTF by evaluating the integral in the rightmost term of Equation~\ref{eqApp:mttfReliability}.
The evaluation starts by performing the following variable substitution: 
\begin{equation}
t=x^{\frac{1}{r}} \Leftrightarrow  dt = \frac{1}{r}x^{\left (\frac{1}{r}-1\right )} \, dx.
\end{equation}
Note that the interval of integration in Equation~\ref{eqApp:mttfReliability} does not change after the variable substitution, since $t=0$ results in 
$x=0$ and $t \rightarrow \infty$ results in $x \rightarrow \infty$.
Hence, after the variable substitution, we can write Equation~\ref{eqApp:mttfReliability} as:
\begin{equation}
MTTF \approx \frac{1}{r}\int_{0}^{\infty} x^{\left ( \frac{1}{r} - 1\right )} e^{\frac{-xc}{{E[\tau]}^r}} \, dx.
\label{eqApp:mttfVariableSubst}
\end{equation}
The integral of Equation~\ref{eqApp:mttfVariableSubst} is evaluated using the gamma function defined as~\cite{abramowitz1970handbook}:
\begin{equation}
\Gamma(z) =  k^z \int_{0}^{\infty} x^{z-1}e^{-kx} \, dx, (\Re z > 0 , \Re k > 0).
\label{eqApp:gammaDefinition}
\end{equation}
For better clarity, we rewrite the integral of Equation~\ref{eqApp:gammaDefinition} as:
\begin{equation}
\int_{0}^{\infty} x^{z-1}e^{-kx} \, dx = \frac{\Gamma(z)}{k^z}.
\label{eqApp:gammaDefinitionRew}
\end{equation}
We make $z = \frac{1}{r}$ and $k=\frac{c}{{E[\tau]}^r}$ in Equation~\ref{eqApp:gammaDefinitionRew} and multiply its both sides by $\frac{1}{r}$, obtaining
\begin{equation}
\frac{1}{r} \int_{0}^{\infty} x^{\left (\frac{1}{r}-1\right )}e^{-\frac{xc}{{E[\tau]}^r}} \, dx = \frac{1}{r}\frac{\Gamma \left (\frac{1}{r} \right)}{{\frac{c}{{E[\tau]}^r}}^{\frac{1}{r}}} = \frac{E[\tau]}{r}\sqrt[r]{\frac{1}{c}} \Gamma \left (  \frac{1}{r} \right ).
\label{eqApp:gammaDefinitionSubst}
\end{equation}

Note that the leftmost term in Equation~\ref{eqApp:gammaDefinitionSubst} is the MTTF approximation given by Equation~\ref{eqApp:mttfVariableSubst}. 
Hence, we can write the MTTF as:
\begin{equation}
MTTF \approx \frac{E[\tau]}{r}\sqrt[r]{\frac{1}{c}} \Gamma \left (  \frac{1}{r} \right ).
\label{eqApp:mttfFinal}
\end{equation}

\section{Comparison of MTTF equations for link failures}
\label{app:compMTTFEquations}

In BCube we have $MTTF_{bcube} \approx \frac{E[\tau]}{l+1}\sqrt[l+1]{\frac{1}{|\mathcal{S}|}} \Gamma \left (  \frac{1}{l+1} \right )$. Hence, we will start by showing that if we have a new configuration with $l'=l+1$ (i.e., one more server interface) we can increase the MTTF. For simplicity, we consider that $|\mathcal{S}|$ is equal for the configurations using both $l$ and $l'$. Although it is not necessarily true, because the number of servers depends on the combination of $l$ and $n$, we can adjust $n$ to have a close number of servers for $l$ and $l'$, as done on the configurations of Table~\ref{tab:configurations}.
First, we need to state that

\begin{equation}
\frac{E[\tau]}{l'+1}\sqrt[l'+1]{\frac{1}{|\mathcal{S}|}} \Gamma \left (  \frac{1}{l'+1} \right ) > \frac{E[\tau]}{l+1}\sqrt[l+1]{\frac{1}{|\mathcal{S}|}} \Gamma \left (  \frac{1}{l+1} \right ).
\label{eq:statementLMTTF_link}
\end{equation}

Doing $l'=l+1$, and rearranging the terms we have the following requirements for the above formulation to be true:

\begin{equation}
|\mathcal{S}| > {\left ( \frac{l+2}{l+1} \frac{\Gamma \left (  \frac{1}{l+1} \right )}{\Gamma \left (  \frac{1}{l+2} \right )} \right)}^{(l+1)(l+2)}.
\label{eq:serverGreaterLink}
\end{equation}

The right term of Equation~\ref{eq:serverGreaterLink} is a decreasing function of $l$ over the considered region ($l \geq 1$). Hence, it is sufficient to prove that Equation~\ref{eq:serverGreaterLink} is true for $l=1$. Doing $l=1$ in Equation~\ref{eq:serverGreaterLink}, we have $|\mathcal{S}| > 0.955$, which is true for a feasible DC.

As DCell with $l>1$ has the same MTTF of a BCube with the same $l$, the above reasoning is valid for this topology. For DCell2 ($l=1$), the equation of the MTTF is the same of BCube2 ($l=1$) except that DCell2 has the value $\sqrt{\frac{1}{1.5|\mathcal{S}|}}$ instead of $\sqrt{\frac{1}{|\mathcal{S}|}}$. Consequently, the MTTF of BCube2 is greater than that of DCell2. We can thus conclude that DCell2 has the lowest MTTF among server-centric topologies. Hence, to show that the MTTF of Fat-tree is smaller than the MTTF of all server-centric topologies, we compare it to DCell2. We thus need to prove that

\begin{equation}
\frac{E[\tau]}{|\mathcal{S}|} < \frac{E[\tau]}{2}\sqrt{\frac{1}{1.5|\mathcal{S}|}} \Gamma \left (  \frac{1}{2} \right ).
\label{eq:statementLMTTF_fatTree_link}
\end{equation}

The solution of this equation is $|\mathcal{S}| > 1.909$, which is always true considering a real DC.

\begin{acknowledgements}
The authors would like to thank FAPERJ, CNPq, CAPES, CTIC research agencies and the Systematic FUI 15 RAVIR (\url{http://www.ravir.io}) project for their financial support to this research.
\end{acknowledgements}


\bibliographystyle{spmpsci}
%\bibliography{header,dcn}
\bibliography{jnsmReliabilitySurvivabilityDC_r2}

\noindent
\textbf{Rodrigo S. Couto} received his Doctor of Science degree in Electrical Engineering in 2015 and his \textit{cum laude} Electronics and Computing Engineer degree in 2011, both from Universidade Federal do Rio de Janeiro (UFRJ). Since March 2015 he has been an Associate Professor with the Universidade do Estado do Rio de Janeiro (UERJ). His major research interests include data center networks, cloud computing, network reliability and network virtualization.\\ 

\noindent
\textbf{Stefano Secci} is an Associate Professor at the University Pierre and Marie Curie (UPMC - Paris VI, Sorbonne Universites). He received a  \textit{Laurea} degree in Telecommunications Engineering from Politecnico di Milano, in 2005, and a dual Ph.D. degree in computer networks from the same institution and Telecom ParisTech, in 2009. His current research interests are about Internet resiliency and Cloud networking.\\

\noindent
\textbf{Miguel Elias M. Campista} is an associate professor with Universidade Federal do Rio de Janeiro (UFRJ), Rio de Janeiro, Brazil, where he works since 2010. He received his Telecommunications Engineer degree from the Fluminense Federal University (UFF), Niter{\'o}i, Brazil, in 2003 and his M.Sc. and D.Sc. degrees in Electrical Engineering from UFRJ, Rio de Janeiro, Brazil, in 2005 and 2008, respectively. His major research interests are in wireless networks, routing, home networks, and complex networks.\\

\noindent
\textbf{Lu\'is Henrique M. K. Costa} received his Electronics Engineer and M.Sc. degrees in Electrical Engineering from Universidade Federal do Rio de Janeiro (UFRJ), Rio de Janeiro, Brazil, respectively, and a Dr. degree from the Universit\'e Pierre et Marie Currie (Paris 6), Paris, France, in 2001. Since August 2004 he has been an associate professor with COPPE/UFRJ. His major research interests are in the areas of routing, wireless networks, and future Internet.

\end{document}